\UseRawInputEncoding
\documentclass[%
superscriptaddress,
preprint,
 amsmath,amssymb,
 aps,
prc,
]{revtex4-2}

\usepackage{graphicx}
\usepackage{xtab}
\usepackage{dcolumn}
\usepackage{bm}
\usepackage[utf8]{inputenc} 
\usepackage{hyperref}
\usepackage{isotope}

\usepackage{soul}
\usepackage{orcidlink}
\usepackage[normalem]{ulem}

\usepackage{xcolor}
\usepackage[normalem]{ulem}
\usepackage{orcidlink}

\usepackage{float}

\usepackage{subcaption}

\usepackage{titlesec}

\titleformat{\paragraph}
{\normalfont\normalsize\bfseries}{\theparagraph}{1em}{}
\titlespacing*{\paragraph}
{0pt}{3.25ex plus 1ex minus .2ex}{1.5ex plus .2ex}

\newcommand{ \ket }[ 1 ]{ \big\vert {#1} \big\rangle }
\newcommand{ \bra }[ 1 ]{ \big\langle {#1} \big\vert }
\newcommand{ \overlap }[ 2 ]{ \big\langle \, {#1} \, \big\vert \, {#2} \, \big\rangle }

\DeclareMathOperator*{\sumintegral}{%
\mathchoice{\ooalign{$\displaystyle\sum$\cr\hidewidth$\displaystyle\int$\hidewidth\cr}}
{\ooalign{\raisebox{.14\height}{\scalebox{.7}{$\textstyle\sum$}}\cr\hidewidth$\textstyle\int$\hidewidth\cr}}{\ooalign{\raisebox{.2\height}{\scalebox{.6}{$\scriptstyle\sum$}}\cr$\scriptstyle\int$\cr}}{\ooalign{\raisebox{.2\height}{\scalebox{.6}{$\scriptstyle\sum$}}\cr$\scriptstyle\int$\cr}}
}

\newcommand{ \nmax }{ N_{ \mathrm{max} } }

\newcommand{ \emninteraction }{ \mathrm{NN-N}^{4}\mathrm{LO}(500) + 3 \mathrm{N}_{ \mathrm{lnl} } }
\newcommand{ \emninteractionstar }{ \mathrm{NN}^{4}\mathrm{LO}(500) + 3 \mathrm{N}_{ \mathrm{lnl} }^{*} }

\newcommand{ \carbonfermidecay }{ {}^{10}\mathrm{C} \rightarrow {}^{10}\mathrm{B} }

\begin{document}

\preprint{Pre-print numbers: CERN-TH-2026-133, INT-PUB-26-028, LA-UR-26-25143,}
\preprint{LLNL-JRNL-2020513, N3AS-26-01}

\title{Future directions in nuclear $\beta$ decay at FRIB and beyond}

\author{Garrett B.~King \orcidlink{0000-0002-5553-868X}}
\email[]{kingg@lanl.gov}
\affiliation{Theoretical Division, Los Alamos National Laboratory, Los Alamos, NM 87545, USA}

\author{Ayala Glick-Magid \orcidlink{0000-0003-4499-8303}}
\affiliation{School of Physics and Astronomy, Tel-Aviv University, Tel-Aviv 69978, Israel}

\author{Grigor Sargsyan \orcidlink{0000-0002-3589-2315}}
\affiliation{Facility for Rare Isotope Beams, Michigan State University, East Lansing, MI 48824, USA}

\author{Mark A. Caprio \orcidlink{0000-0001-5138-3740}}
\affiliation{Department of Physics and Astronomy, University of Notre Dame, Notre Dame, IN 46556, USA}

\author{Kyle G. Leach\orcidlink{0000-0002-4751-1698}}
\affiliation{Arthur B. McDonald Canadian Research Institute for Astroparticle Physics, Queen's University, Kingston, ON K7L 3N6, Canada}
\affiliation{Department of Phyiscs, Engineering Physics \& Astronomy, Queen's University, Kingston, ON K7L 3N6, Canada}
\affiliation{Facility for Rare Isotope Beams, Michigan State University, East Lansing, MI 48824, USA}
\affiliation{TRIUMF, 4004 Wesbrook Mall, Vancouver BC Canada V6T 2A3}

\author{John A. Behr\orcidlink{0000-0003-2820-6767}}
\affiliation{TRIUMF, 4004 Wesbrook Mall, Vancouver BC Canada V6T 2A3}

\author{Francesca Bonaiti\orcidlink{0000-0002-3926-1609}}
\affiliation{Facility for Rare Isotope Beams, Michigan State University, East Lansing, MI 48824, USA}
\affiliation{Physics Division, Oak Ridge National Laboratory, Oak Ridge, Tennessee 37831, USA}

\author{Maxime Brodeur\orcidlink{0000-0001-9323-1198}}
\affiliation{Department of Physics and Astronomy, University of Notre Dame, Notre Dame, IN 46556, USA}

\author{Graham Chambers-Wall\orcidlink{0000-0002-9403-8450}}
\affiliation{Department of Physics, Washington University, Saint Louis, MO 63130, USA}

\author{Heather L. Crawford\orcidlink{0000-0002-7765-4235}}
\affiliation{Nuclear Science Division, Lawrence Berkeley National Laboratory, Berkeley, CA 94720, USA}

\author{Maria Dawid\orcidlink{0000-0003-4021-9368}}
\affiliation{Department of Physics, Washington University, Saint Louis, MO 63130, USA}

\author{Wouter Dekens\orcidlink{0000-0002-7850-5901}}
\affiliation{Institute for Nuclear Theory, University of Washington, Seattle, WA 98195, USA}

\author{Michael Gennari\orcidlink{0000-0001-5271-8784}}
\affiliation{Institut f{\"u}r Kernphysik and PRISMA++ Cluster of Excellence, Johannes Gutenberg-Universit{\"a}t Mainz, 55128 Mainz, Germany}

\author{Robert Grzywacz \orcidlink{0000-0002-0920-2587}}
\affiliation{Department of Physics and Astronomy, University of Tennessee, Knoxville, TN 37996}

\author{Peter Gysbers\orcidlink{0000-0002-5269-9170}}
\affiliation{Facility for Rare Isotope Beams, Michigan State University, East Lansing, MI 48824, USA}

\author{Heather S. Harrington\orcidlink{0000-0002-9127-874X}}
\affiliation{Center for Experimental Nuclear Physics and Astrophysics, University of Washington, Seattle, WA 98195, USA}

\author{Heiko Hergert\orcidlink{0000-0003-0520-0856}}
\affiliation{Facility for Rare Isotope Beams, Michigan State University, East Lansing, MI 48824, USA}
\affiliation{Department of Physics and Astronomy, Michigan State University, East Lansing, MI 48824, USA}

\author{Lotta Jokiniemi\orcidlink{0000-0002-9327-5868}}
\affiliation{Technische Universit\"at Darmstadt, Department of Physics, 64289 Darmstadt, Germany}
\affiliation{ExtreMe Matter Institute EMMI, GSI Helmholtzzentrum f\"ur Schwerionenforschung GmbH, 64291 Darmstadt, Germany}

\author{Brenden Longfellow\orcidlink{0000-0001-8743-962X}}
\affiliation{Lawrence Livermore National Laboratory, Livermore, CA 94550, USA}
\affiliation{Physics Division, Argonne National Laboratory, Argonne, IL 60439, USA}

\author{Rebeka S. Lubna \orcidlink{0000-0003-4976-3976}}
\affiliation{Facility for Rare Isotope Beams, Michigan State University, East Lansing, MI 48824, USA}

\author{Kelsey A.~Lund\orcidlink{0000-0003-0031-1397}}
\affiliation{Department of Physics, University of California, Berkeley, CA 94720, USA}
\affiliation{Institute for Nuclear Theory, University of Washington, Seattle, WA 98195, USA
}

\author{Giacomo Marocco \orcidlink{0000-0001-7325-8190}}
\affiliation{Physics Division, Lawrence Berkeley National Laboratory, Berkeley, CA 94720, USA}

\author{Anna E. McCoy \orcidlink{0000-0002-1033-1474}}
\affiliation{Physics Division, Argonne National Laboratory, Argonne, IL 60439, USA}

\author{Dan Melconian \orcidlink{0000-0002-0142-5428}}
\affiliation{Cyclotron Institute and Department of Physics \& Astronomy, Texas A\&M University, College Station, TX 77843, USA}

\author{Alexis Mercenne \orcidlink{0000-0002-2624-3911}}
\affiliation{Department of Physics and Astronomy, Louisiana State University, Baton Rouge, LA 70803, USA}

\author{David C. Moore \orcidlink{0000-0002-2358-4761}}
\affiliation{Wright Laboratory, Department of Physics, Yale University, New Haven, CT 06520, USA}

\author{Oscar Naviliat-Cuncic \orcidlink{0000-0001-5082-2131}}
\affiliation{Facility for Rare Isotope Beams, Michigan State University, East Lansing, MI 48824, USA}
\affiliation{Department of Physics and Astronomy, Michigan State University, East Lansing, MI 48824, USA}
\affiliation{International Laboratory for Nuclear Physics and Nuclear Astrophysics,
CNRS-MSU, East Lansing, MI 48824, USA}

\author{Samuel J. Novario\orcidlink{0000-0001-7836-1269}}
\affiliation{Department of Physics, Washington University, Saint Louis, MO 63130, USA}

\author{Timilehin H. Ogunbeku \orcidlink{0000-0002-4442-7757}}
\affiliation{Lawrence Livermore National Laboratory, Livermore, CA 94550, USA}

\author{Walter C.~Pettus \orcidlink{0000-0003-4947-7400}}
\affiliation{Department of Physics and Center for Exploration of Energy and Matter, Indiana University, Bloomington, IN 47401, USA}

\author{Ryan Plestid \orcidlink{0000-0003-0779-7289}}
\affiliation{Theoretical Physics Department, CERN, 1 Esplanade des Particules, CH-1211 Geneva 23, Switzerland}

\author{Bertis C. Rasco \orcidlink{0000-0002-4061-1178}}
\affiliation{Physics Division, Oak Ridge National Laboratory, Oak Ridge, Tennessee 37831, USA}

\author{Ante Ravli\'c \orcidlink{0000-0001-9639-5382}}
\affiliation{Facility for Rare Isotope Beams, Michigan State University, East Lansing, MI 48824, USA}
\affiliation{Department of Physics, Faculty of Science, University of Zagreb, Bijeni\v cka c. 32, 10000 Zagreb, Croatia}

\author{Yukiya Saito \orcidlink{0000-0003-1320-k8903}}
\affiliation{Department of Physics and Astronomy, University of Tennessee, Knoxville, TN 37996}
\affiliation{Facility for Rare Isotope Beams, Michigan State University, East Lansing, MI 48824, USA}
\affiliation{Department of Physics and Astronomy, University of Notre Dame, Notre Dame, IN 46556, USA}

\author{Dustin P. Scriven \orcidlink{0000-0002-3853-2200}}
\affiliation{Facility for Rare Isotope Beams, Michigan State University, East Lansing, MI 48824, USA}

\author{Chien-Yeah Seng \orcidlink{0000-0002-3062-0118}}
\affiliation{Department of Physics and Astronomy, University of Tennessee, Knoxville, TN 37996}

\author{Nadezda A. Smirnova \orcidlink{0000-0001-8944-7631}}
\affiliation{LP2IB (CNRS/IN2P3 -- University of Bordeaux), 33175 Gradignan cedex, France}

\author{Artemis Spyrou \orcidlink{0000-0002-8642-8352}}
\affiliation{Facility for Rare Isotope Beams, Michigan State University, East Lansing, MI 48824, USA}
\affiliation{Department of Physics and Astronomy, Michigan State University, East Lansing, MI 48824, USA}

\author{Vandana Tripathi \orcidlink{0000-0002-8850-1357}}
\affiliation{Department of Physics, Florida State University, Tallahassee, FL 32306}

\author{Xing Wu \orcidlink{0000-0002-6646-820X}}
\affiliation{Facility for Rare Isotope Beams, Michigan State University, East Lansing, MI 48824, USA}
\affiliation{Department of Physics and Astronomy, Michigan State University, East Lansing, MI 48824, USA}

\author{Zhengyu Xu \orcidlink{0000-0001-8626-1276}}
\affiliation{Department of Physics and Astronomy, University of Tennessee, Knoxville, TN 37996}

\date{\today}

\begin{abstract}
Motivated by the opportunities presented for studies relevant to nuclear structure, astrophysics, and fundamental symmetries with nuclear $\beta$ decay, the Facility for Rare Isotope Beams (FRIB) Theory Alliance topical program ``Future Directions in Nuclear $\beta$ Decays at FRIB'' was held in September of 2025. This white paper summarizes the main points of discussion over the two-week program, and it aims to provide a snapshot of the current status of the field while also highlighting important questions and opportunities for future work. We provide an overview of the experimental tools and techniques that enable modern $\beta$ decay studies, discuss the current state of nuclear many-body approaches used to study $\beta$ decays, and highlight the important science questions that can be addressed by weak decays. 
\end{abstract}

\maketitle

\tableofcontents

\section{Preface}
Nuclear $\beta$ decay occupies a unique position at the intersection of nuclear physics, astrophysics, particle physics, and neutrino science. Studies of $\beta$ decay provide essential insights into the structure and properties of nuclei at the limits of stability, play a central role in understanding the nucleosynthetic processes responsible for the observed elemental abundances in the universe, and enable precision tests of the Standard Model and searches for physics beyond it. Recent advances in radioactive beam facilities, detector technologies, and nuclear many-body theory have significantly expanded the scientific opportunities offered by $\beta$ decay studies.

Motivated by these opportunities, the Facility for Rare Isotope Beams (FRIB) Theory Alliance topical program "Future Directions in Nuclear $\beta$ Decays at FRIB" was held in September of 2025.
The program brought together theorists and experimentalists working across a broad range of topics related to weak decays, including precision tests of fundamental symmetries, nuclear structure studies, nuclear astrophysics, neutrino physics, and the development of modern many-body approaches for describing weak processes in nuclei. While particular attention was given to the capabilities that FRIB will provide for future $\beta$ decay studies, discussions also considered the broader experimental and theoretical landscape and the complementary roles of other facilities and research programs worldwide.


The first week of the program focused primarily on precision decay measurements and their interpretation. Discussions covered experimental techniques for decay spectroscopy, measurements relevant to searches for new physics, and the theoretical frameworks required to connect experimental observables to underlying nuclear and particle physics.
The second week shifted the focus to how FRIB will allow us to study weak decays far from stability. Discussions of lifetime measurements, $\beta$ decay strength functions, and $\beta$ decay spectroscopy far from stability made up the experimental program, while discussions of many-body computations and astrophysical simulations composed the theoretical section. 

Throughout the program, dedicated discussion sessions were held to identify key scientific opportunities and outstanding challenges in the field, and to determine how theory and experiment could most effectively leverage the capabilities of FRIB and other facilities to maximize what is learned about nuclear structure, astrophysics, and beyond the Standard Model physics. This white paper is a product of those discussions. Its goal is to provide a snapshot of the current status of nuclear $\beta$ decay research, highlight important open questions, and outline promising directions for future work. The document reflects the perspectives of the participants of the program and is intended as a resource for the broader community.

The white paper is organized around both the tools that enable modern $\beta$ decay studies and the broad range of scientific questions that these studies address. We first review the experimental and theoretical methods used to investigate $\beta$ decay, before discussing their applications to nuclear structure, astrophysics, precision tests of the Standard Model, searches for new interactions, and neutrino physics.

The remainder of this white paper is organized as follows. We begin in Sec.~\ref{sec:methods} 
with an overview of the experimental methods used to measure $\beta$ decay properties, including half-lives, branching ratios, $\beta$-delayed neutron emission, and spectroscopic observables. Sec.~\ref{sec:many_body} 
summarizes the theoretical many-body approaches that underpin modern descriptions of weak decays in nuclei and discusses future developments in the field. The role of $\beta$ decay as a probe of nuclear structure is examined in Sec.~\ref{sec:structure}, while Sec.~\ref{sec:astro} highlights the importance of weak decays in nuclear astrophysics and nucleosynthesis.

The subsequent sections focus on applications of $\beta$ decay to fundamental physics. Sec.~\ref{sec:V_ud} reviews the determination of $V_{ud}$ from nuclear $\beta$ decay and its implications for tests of Cabibbo-Kobayashi-Maskawa (CKM) unitarity. Sec.~\ref{sec:BSM} discusses searches for non-Standard-Model scalar and tensor interactions using $\beta$ decay observables, emphasizing both experimental efforts and the nuclear theory input required to interpret them. Finally, Sec.~\ref{sec:neutrino} explores the connections between $\beta$ decay and neutrino physics, including studies of neutrino mass, sterile neutrinos, neutrino interactions, and neutrinoless double-$\beta$ decay.

\section{Methods for Measuring $\beta$ Decay Properties\label{sec:methods}}
\label{sec:exp_methods}

The dedicated $\beta$ decay experimental setup at FRIB is the FRIB Decay Station initiator (FDSi) \cite{Allmond03042025}, a collaborative effort among multiple U.S. institutions. FDSi is a highly reconfigurable system with two focal planes designed for discrete and total absorption spectroscopy at the first and second focal planes, respectively.
The primary detection system includes $\beta$, $\gamma$, and neutron detectors to study $\beta$ decay properties along with other structural phenomena of exotic nuclei produced at FRIB. Its two–focal-plane design can maximize scientific output within a single experiment  \cite{Cox2024}. The first focal plane features a central implantation detector surrounded by the DEcay Germanium Array initiator (DEGAi-A) for $\gamma$-ray spectroscopy and the Neutron (Xn) Tracking initiator (NEXTi) \cite{Pete2016}. NEXTi can be replaced by a second $\gamma$ detector array, DEGAi-B, for studies in the neutron-deficient region. Various implantation detectors can be used depending on experimental goals, including XScintillator-Si DSSD (XSiSi) for heavy-mass region and YSO detectors (stationary or moving) \cite{YOKOYAMA201993} for improved timing measurements.
The second focal plane can house the Modular Total Absorption Spectrometer (MTAS) \cite{karny2016mo} (Sec.~\ref{TAS}). MTAS can be exchanged with an array of $^3$He neutron counters coupled with Ge detectors (S3HeNi) for $\beta$-Xn measurements. Both focal planes can accommodate user-provided devices as required.

 \begin{figure}[t!]
        \center\includegraphics[width=1.0\textwidth]{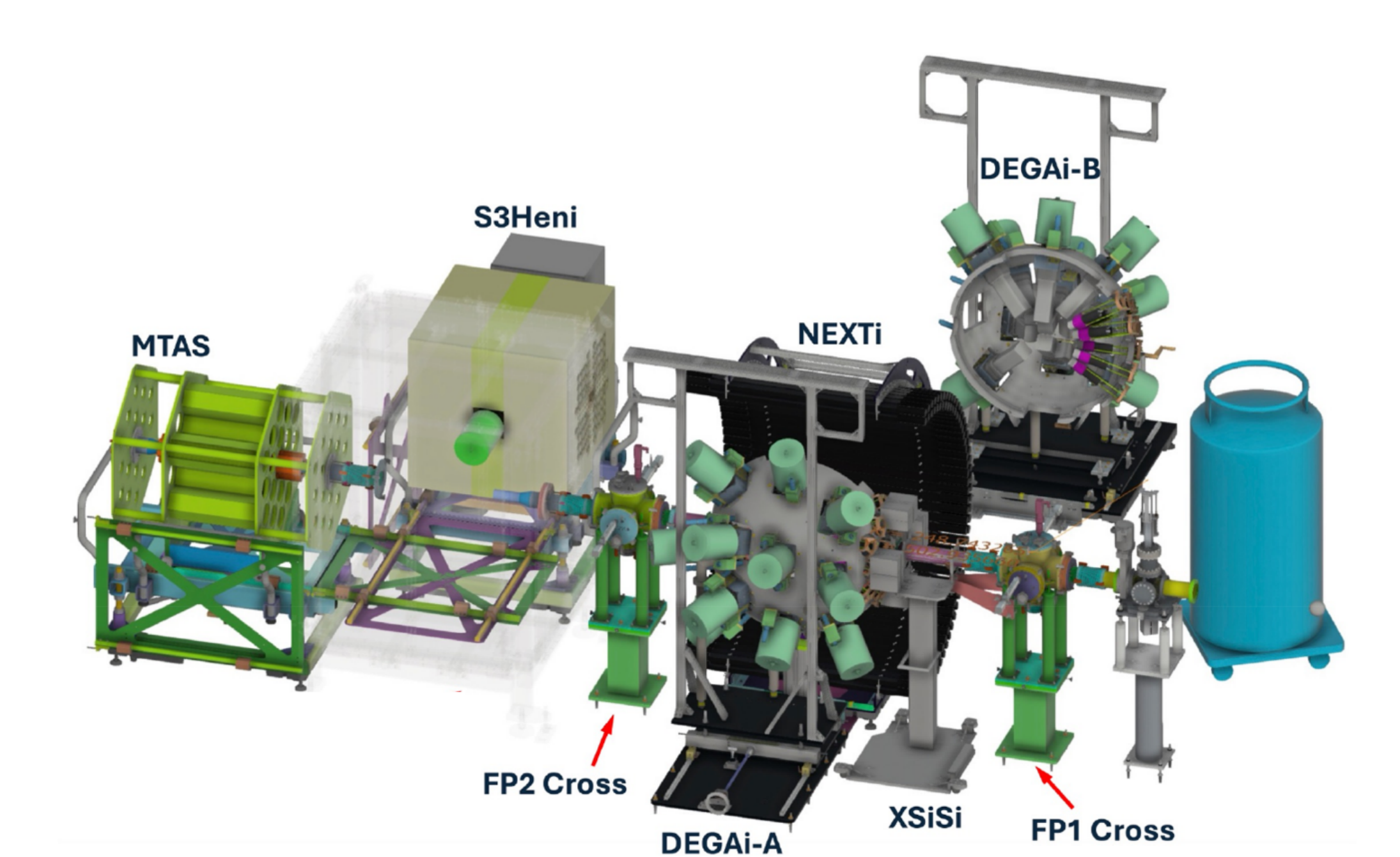}
        \caption{\label{section:exp:figure:bn}
        FRIB Decay Station Initiator is designed to perform comprehensive $\beta$ decay measurements in flexible configurations \cite{Allmond03042025}. Key elements of this systems are marked.}
    \end{figure}



\subsection{Half-Life Determination}

The half-life of a decaying system is among the first experimental observables which can be obtained, requiring minimal statistics for a first determination.  This basic observable can provide important first insights into nuclear structure, especially when considered systematically.  Half-lives are also one of the key inputs in predicting abundance patterns for astrophysics.

Half-life measurements can be challenging due to the large range of possible values.  Determination of long half-lives (days and longer) are not discussed here, but may have very different considerations experimentally, especially for the longest cases (billions of years and more).  At FRIB there are two main methods to measure $\beta$ decay half-lives: 
\begin{enumerate}
    \item For short half-lives (roughly less than 1~s) a fast beam consisting of a cocktail of rare isotopes is tracked through in-beam detectors and is then implanted into a position-sensitive detector. $\beta$ particles are also detected in the same detector and in this way the implanted ions can be correlated with the $\beta$ decays event-by-event. The implantation detector is located along the beam direction and can be surrounded by $\gamma$ detector array(s)  or neutron array, allowing for the $\beta$-delayed $\gamma$ and neutron detection, respectively. At FRIB, the main instrument used for fast-beam decay experiments is the FRIB Decay Station initiator (FDSi) \cite{Allmond03042025}, illustrated in Fig.~\ref{section:exp:figure:bn}.  
    Additional setups exist such as the SuN + mini DSSD setup for total absorption spectroscopy measurements \cite{dombos2021total, simon2013sun} and  GAseous Detector with GErmanium Tagging (GADGET) to study low-energy $\beta$-delayed proton decays. 
    
    \item For half-lives longer than roughly 1~s, the implant-decay correlation becomes more difficult, and experiments are typically done using a moving implant system, either a tape transport system or a mechanically moving implant detector. In such experiments a low-energy beam, with as close to 100\% purity as possible, is implanted onto a tape located at the center of a $\gamma$-array and/or neutron array. Beam implantation takes place for a fixed amount of time, the beam is turned off and $\beta$ decays are measured for a second time window, and then the tape is rotated such that any daughter/granddaughter decays take place outside of the detection system, and a fresh piece of tape is inserted at the implantation point. This sequence of events is typically called ``tape cycle'' and the amounts of time are selected based on the half-lives of the parent and the decay products. At FRIB the SuNTAN tape transport system was developed to be used together with the SuN detector at the low-energy beam area \cite{harris2025suntan,simon2013sun}. A novel development functionally, similar to tape but using a solid-state implantation detector, was recently implemented at FDSi. In this case, a moving implant detector swings into and away from the beam path. Combined with beam pulsing, it replicates a tape cycle. Here, the beam can be transported between two focal planes, enabling complementary experiments. This system can also be used with very short half-lives, when beam pulsing is implemented. 
    
\end{enumerate}

\subsection{Branching Ratios}
Another fundamental observable is the branching ratios to the ground and excited states of the daughter nuclei in $\beta$ decays. Branching ratios along with the half-life and $Q$-value,
in pure Fermi transitions, are the three ingredients needed to extract the $ft$ value of a transition which has special relevance for
the extraction of $V_{ud}$ and the test of
CKM unitarity.  Obtaining the precision necessary with large arrays such as 
the Gamma Ray Energy Tracking Array (GRETA) or DEGAi (Sec.~\ref{sec:betadelayedneutron}) is difficult due to the variation in efficiency of the component detectors.  Many of the branching ratios measured for $V_{ud}$ have been performed using a single, well-characterized HPGe detector \cite{TAMUHPGe,BordeauxHPGe}.
These experiments are continuing, for example with a recent effort to improve the $0^+\rightarrow0^+$ branching ratio of $^{10}$C at Texas A\&M University (TAMU), where the efficiency of the HPGe is known to $\pm0.1\%$, uncertainty, nearly 2 times smaller than the most-precise measurement to date \cite{savard-10C}. Although achieving enough events to keep the statistical uncertainty below 0.1\% will be challenging, the first improved $^{10}$C branching ratio in 30 years should finally be realized soon.
While not a trivial endeavor, the possibility exists for bringing the TAMU HPGe and fast-tape transport system to FRIB to take advantage of the unique beams, which can be produced.
\subsection{$\beta$-Delayed Neutron Emission\label{sec:betadelayedneutron}}

Beta-delayed neutron emission is a prevalent decay mode for neutron-rich nuclei \cite{DIMITRIOU2021144}. These exotic nuclei have large $Q_\beta$ values, and their decay daughters have low neutron separation energies, leading to strong $\beta$-delayed emission probabilities. For sufficiently neutron-rich nuclei, multi-neutron emission channels are open, which has direct consequences for astrophysics \cite{Mumpower_rep_2012a}, as demonstrated in previous experiments \cite{Mier13, Yoko2019,Phon2022,Dysz2025}. This two-step process is an important element of the landscape of $\beta$ decay studies because neutron emission uniquely illuminates the properties of $\beta$ decay \cite{Xu2023Rosetta} and is often the main, or even the only, experimentally accessible decay channel. 

This process requires population of unbound excited states in $\beta$-transitions. Nuclear structure effects provide natural explanations for the high neutron emission probabilities in heavier asymmetric nuclei. Their decays are dominated by Gamow-Teller transitions, which connect spin-orbit partner proton and neutron orbitals. In these transformations, highly excited configurations with neutron-hole and proton-particle states are created \cite{madurga2016evidence}. The residual nucleus, if created in a neutron-unbound excited state, will often, though not always, cool via the emission of one or several neutrons, sometimes followed by $\gamma$-ray emission \cite{Dysz2025}.

Beta-delayed neutron-emission branching ratios can be measured using large neutron counters with high detection efficiency \cite{TOLOSADELGADO2019133}. The FDSi design Super3Hen counter features a flat neutron detection efficiency curve, which is important for measuring neutron emission branching ratios, which are relatively insensitive to neutron energy. Neutron counters are also essential for measuring delayed multi-neutron branching ratios \cite{Mier13, Yoko2019, Phon2022}.

Neutron time-of-flight detectors are crucial for determining neutron energy spectra, which are used to establish the decay strength distribution. Because the neutron emission can populate excited states in the daughter nucleus, the neutron detectors are typically combined with $\gamma$-ray detection systems. Measuring the entire cascade is needed to establish the correct $\beta$ decay strength distribution \cite{madurga2016evidence, peltier2025evidence}. 

Neutron energy alone does not always provide sufficient information to extract the feeding pattern. However, detailed measurements of neutron-emission branching ratios provide an additional observable that is sensitive to the orbital angular momentum of the states involved. This approach was recently used to test the validity of the compound nucleus hypothesis \cite{Heid2019,Xu2024}. Neutron emission is not necessarily the only decay path available to excited neutron-unbound states. Recent observations of $\gamma$-ray emission from highly excited neutron-unbound states \cite{Pier2019_ngammacomp,xu2023prc} are significant because they indicate a strong mechanism that hinders neutron emission. Exotic neutron-rich nuclei will provide a key playground to study the nature of this phenomenon, which is likely linked to the structure of states populated in $\beta$ decay. Finally, recent observation of the $\beta$-delayed two-neutron emission from $^{134}$In opens up a new research avenue to study the nature of this process \cite{Dysz2025}.


 \begin{figure}[t!]
        \center\includegraphics[width=1.0\textwidth]{Measurment_methods/chart2d_qbn.pdf}
        \caption{\label{section:exp:figure:bd}
        $\beta$-delayed neutron emission across the chart of nuclei. Colors indicate nuclei with $Q_\beta-S_n >0$. All neutron-rich nuclei in yet unknown part of the chart of nuclei (yellow) will be $\beta$-delayed neutron emitters.
        Shown is an example of the level scheme of $^{134}$In from \cite{Dysz2025}, which is a $\beta$-delayed neutron and two-neutron precursor. $\beta$-delayed neutron emission in exotic nuclei is enabled by the Gamow-Teller transformations. Example of the such transformation in $^{132}$Sn region is shown. Limits of neutron-rich nuclei are taken from Ref.~\cite{Erler2012Limits}.}
    \end{figure}

\subsection{Discrete $\gamma$-Ray Spectroscopy}

Discrete $\gamma$-ray spectroscopy is a vital technique for decay studies of nuclei, including those far from $\beta$ stability and at the limits of nuclear existence, which are a focus for exploration at FRIB.  High- or medium-resolution $\gamma$-ray spectroscopy enables the characterization of bound (and unbound) states in daughter nuclei populated following $\beta$ decay of unstable isotopes, providing information on energies, $\beta$-feeding intensities, and level placement (through $\gamma$-$\gamma$ coincidences and intensity arguments).  With sufficient statistics and appropriate detector geometry, spin and potentially parity information can be extracted based on $\gamma$-$\gamma$ angular correlations and linear polarization. Lifetimes of excited states, in the range of picoseconds, can also be measured using techniques such as the centroid shift method \cite{smith1973centroid} with appropriately fast timing detectors.

Discrete spectroscopy setups are generally centered around high-resolution ($\sim$0.2\% FWHM) arrays based on high-purity Germanium (HPGe), or medium-resolution (2-3\% FWHM) detectors which offer superior timing characteristics, such as LaBr$_{3}$.  

At FRIB, while other $\gamma$-ray detector arrays such as GRETA may support niche decay experimental efforts, the DEGAi \cite{FDSi} represents the primary array for discrete $\gamma$-ray spectroscopy.  The configuration of DEGAi can provide up to 4$\pi$ solid angle coverage, enabling $\gamma$-$\gamma$ angular correlation measurements that are vital for constraining the spins of states populated in $\beta$ decay, and depending on the setup, DEGAi can support a maximum of 24  HPGe clover detectors across two hemispheres, offering increased sensitivity for high-multiplicity $\gamma$-ray cascades.  DEGAi has most frequently run with one hemisphere of HPGe, using 11 clovers.  For fast timing measurements, DEGAi can also include CeBr or LaBr$_{3}$ detectors within the support structure, providing a highly versatile capability.

In most $\beta$ decays of neutron rich nuclei, the daughter nucleus produced in the decay is left in a short-lived excited state. The ground state to ground state 
decay is strongly inhibited with large neutron excess as the parent and daughter 
nuclei have opposite parities. The total energy released by the decay
($Q_\beta$-value), which can be large for exotic nuclei, is shared as kinetic energy between the $\beta$ particle, the neutrino, and the recoiling nucleus, 
and as excitation energy of the daughter nucleus. 
The excited daughter nucleus quickly transitions 
to a lower, more stable energy state by emitting distinct characteristic $\gamma$ rays. 
Measuring the energy and efficiency corrected yields of the $\gamma$ transitions allows us to generate level schemes of daughter nuclei. The intensities of $\gamma$ rays feeding 
in and out of a level are used to decipher the $\beta$ feeding to discrete levels and eventually the $\log ft$ values. 
The time distribution of the emitted $\gamma$-rays in the daughter nucleus can be used to estimate the half-life of the parent.
Following the $\gamma$ transitions in the decay chain also allows for the determination of delayed neutron emission probabilities. 



\subsection{Total Absorption Spectroscopy \label{TAS}}
\label{sec:tas}

The Total Absorption Spectroscopy (TAS) technique was developed as a solution to the so-called pandemonium effect \cite{hardy1977essential}. 
This is the phenomenon where the $\beta$ decay feeding intensity distribution ($I_\beta$), when measured using low-efficiency $\gamma$-ray detectors, is artificially enhanced at lower excitation energies and reduced at higher excitation energies.
The TAS technique makes use of large-size $\gamma$-ray calorimeters with inherent high efficiency, which can extract the $I_\beta$ in a more accurate way. 
While TAS detectors were developed to measure $\beta$ decay feeding intensity distributions and to identify high-energy level reaction $\gamma$ rays \cite{simon2013sun}, they have proved effective impacting a wide array of physics research areas, such as reactor decay heat \cite{algora2010re, rasco2017co, wolinska2023co}, $\gamma$-strength functions \cite{spyrou2016strong}, reactor antineutrino predictions \cite{fijalkowska2017im, guadilla2019la}, nuclear structure \cite{dembski2025extreme}, and isolating small but important decay branches \cite{stukel2023ra}.
In terms of practical experimental applications, due to the inherent high efficiency of TAS detectors, they can be excellent active-veto detectors and allow for quick particle identification during online analysis at FRIB fast-beamline experiments. 

At FRIB, there are two Total Absorption Spectrometers available for measurements: MTAS and SuN.  The Modular Total Absorption Spectrometer (MTAS) from Oak Ridge National Laboratory (ORNL) is 1 ton of NaI composed of 24 NaI hexagonal and tetragonal-shaped modules organized in a unique radial pattern \cite{karny2016mo, karny2019de}. MTAS measures complete $\beta$ decay spectroscopy, including $\beta$-feeding intensities, dominant $\gamma$ decay paths \cite{fijalkowska2017im,shuai2022de}, and is a low-resolution neutron spectrometer \cite{rasco2017co,stepaniuk2025de}. MTAS is an integral part of the FRIB Decay Station initiator \cite{FDSi} (Fig.~\ref{section:exp:figure:bn}), and also performs experiments at Argonne National Laboratory (ANL).

The Summing NaI (SuN) detector \cite{simon2013sun} was developed at the NSCL and is now used at FRIB, as well as at Argonne National Laboratory. SuN is a 16 by 16 inch cylindrical detector, divided in eight optically isolated segments. The SuN group has measured $I_\beta$ distributions for a range of neutron-rich isotopes, e.g. Ref.~\cite{dombos2016total, dombos2021total, lyons2019decay, naqvi2022total, spyrou2016strong, dembski2025extreme}. SuN can be used at FRIB in conjunction with a mini DSSD for fast-beam correlation measurements, as well as in conjunction with the SuNTAN tape system \cite{harris2025suntan} for low-energy beam measurements.  

\subsection{Beta Energy Spectra\label{energySpectra}}

Precision measurements of $\beta$-energy spectra also provide a sensitive window to search for exotic scalar and tensor couplings contributing to the weak interaction. So far,
studies have been focused on pure Gamow-Teller transitions which are sensitive to exotic tensor couplings through the Fierz interference term (Sec.~\ref{sec:BSM}).
There are currently several ongoing efforts with nuclei, implementing different techniques to circumvent instrumental effects that would distort the shape of 
beta-energy spectra. To facilitate the interpretation, and provide an accurate theoretical comparison, all current projects use allowed transitions.

A low-mass multi-wire drift chamber, with a plastic scintillator used as a trigger
and detector, has been developed by a Leuven/Krakow/Caen collaboration,
and has been used for the measurement of the $\beta$-energy
spectrum with a radioactive source of $^{114m}$In \cite{Dek24}. The measurement uses
also a $^{207}$Bi source in parallel, for gain stability monitoring. Since the weak magnetism form
factor for the $^{114m}$In decay transition is not known, the authors aim at extracting both, the weak magnetism and the Fierz term. This strongly limits the sensitivity to exotic couplings
due to correlations.

Another measurement with $^{114m}$In at ISOLDE/CERN used two plastic scintillators located in a high magnetic field and readout with silicon photomultipliers (SiPM) \cite{Van23}. The magnetic field confines the $\beta$ particles such that those which are backscattered from the scintillators are recorded by the opposite detectors. The system is currently being upgraded to reduce nonlinearties induced by the plastic scintillators responses and the SiPM readout.

Cyclotron radiation emission spectroscopy (CRES), initially developed for tritium decay electrons
\cite{monreal2009relativistic}, is being extended to cover the full energy range of electrons from $^6$He decay.
The principle is to deduce the energy of decay electrons confined in a magnetic field from the cyclotron
frequency measurement as a function of time. The earliest possibly observed frequency gives the
electron energy at the time of decay. A project based on CRES is ongoing at CENPA, and the proof-of-principle of the
technique with $\beta$ particles from $^6$He and $^{19}$Ne has been demonstrated~\cite{Byr23}.

A calorimetry technique using CsI:Na inorganic scintillators has been explored and tested at NSCL with $^6$He and $^{20}$Fe beams \cite{Nav16,Hug18} and has been extended with 
YAP:Ce scintillators at GANIL with a low energy (25 keV) \cite{Kan23} and a high energy (312 MeV) $^6$He beam. The technique totally eliminates the backscattering of electrons on the detectors, which are still present in the setups using the $^{114m}$In source mentioned above. Although the technique is rather simple, the data analyses turn out to be
complicated: the great advantage of the calorimetry constitute also its main challenge since it is impossible to probe exactly the same scintillator volume with radioactive sources.

A novel technique using superconducting tunnel junctions (STJs) has enabled precision studies of nuclear $\beta$ decay and electron capture through direct detection of the low-energy nuclear recoil. This approach has been demonstrated by the BeEST experiment, where the electron-capture spectrum of $^7$Be was reconstructed directly from the recoiling $^7$Li daughter ion~\cite{beest2020direct,friedrich2021limits,smolsky2025wavepacket}. In this first implementation, the radioactive atoms were embedded directly into the STJ absorber, which requires isotopes with sufficiently long half-lives to allow for source preparation, detector integration, and cryogenic operation.

The Superconducting Array for Low-Energy Radiation (SALER) will extend this quantum-sensing technique to short-lived rare isotopes by coupling an STJ detector array directly to an on-line radioactive beam facility. At FRIB, radioisotopes produced in-flight can be slowed, implanted, and measured in the same experimental platform, removing the long half-life requirement imposed by off-line source preparation. This enables recoil-spectroscopy studies of isotopes that were previously inaccessible to embedded-source STJ measurements, including light and medium-mass nuclei with high discovery potential for precision tests of the weak interaction.

SALER is designed to measure the complete low-energy recoil response following nuclear decay with eV-scale energy resolution, high rate capability, and isotope-by-isotope selectivity. For $\beta$ decay, the daughter recoil spectrum provides direct sensitivity to the $\beta$--$\nu$ angular correlation, recoil-order effects, and possible scalar or tensor currents beyond the Standard Model. In electron capture, the same technique provides a direct calorimetric measurement of the daughter recoil and atomic de-excitation energy, enabling searches for sterile neutrinos, exotic weak couplings, and improved understanding of atomic and solid-state effects in precision nuclear decay measurements.  In the near term, precision studies of $^{11}$C and $^{19}$Ne $\beta$ decay will provide sensitive probes of non-Standard-Model scalar and tensor interactions in complementary nuclear systems. The broader SALER program will establish STJ-based recoil spectroscopy as a general platform for rare-isotope studies, connecting quantum sensor development with precision tests of fundamental symmetries at next-generation radioactive beam facilities.

The possibility to isolate individual $\beta$ decay energy components from the integral $\beta$ energy spectra would radically expand the number of individual $\beta$ decay energy spectra that can be studied.
These additional nuclei would allow $\beta$ energy spectral studies of nuclei further from stability with higher decay energies than the currently accessible simple low decay energy $\beta$ energy spectral studies. 
This technique allows study of individual $\beta$ energy spectra within the same $\beta$ decay with different forbiddenness (allowed, first-forbidden-unique, first-forbidden non-unique, etc), thereby probing the character of $\beta$ decay corrections and nuclear structure within a single nucleus.
The additional measurable $\beta$-energy spectra will allow for precision reactor antineutrino predictions and allow access to different parity states in many nuclei.
Using MTAS and the newly developed $\beta$-Spectrum Module ($\beta$SM), this isolation is progressing.
Individual $\beta$-energy separation will first target nuclei one or two further from stability than current measurements.
The nuclei to study can focus on nuclei which we believe can be calculated with current nuclear theory, such as low $Z$ nuclei and nuclei near shell closures.
Feedback from theorists identifying more easily calculable predictable $\beta$ spectra to guide future measurements is welcome.  
To fully evaluate the reliability of various theories for predicting $\beta$-energy spectra,
it would be an interesting exercise to establish theory predictions before revealing unmeasured spectra.
It should be possible to coordinate the experimental publication with the various theory publications directly connected to the experimental measurements.

\section{Nuclear Many-Body Methods\label{sec:many_body}}
As seen in the previous section, nuclear $\beta$ decay experiments cover a broad spectrum of physics problems and look to inform our knowledge of nuclear structure, astrophysics, and fundamental physics. To provide a theoretical description for such an extensive set of processes, an equally rich set of methods for solving the nuclear many-body problem is needed.

In this section, we cover approaches based on data-driven schemes, as well as {\it ab initio} approaches that start from interactions between nucleons.  The former, data-driven schemes allow one to flexibly study a number of transitions across the nuclear chart, making possible global analyses of $\beta$ decay which are particularly useful inputs for astrophysical simulations. The latter, {\it ab initio} approaches come with various levels of approximation schemes so that different approaches are better-suited than others to study particular problems. For example, experiments looking to understand the features of the nuclear interaction driving structure effects can be interpreted from {\it ab initio} approaches. Additionally, precision fundamental symmetries experiments that require understanding of how the nuclear Hamiltonian impacts predictions can also benefit from such calculations. Finally, we require not only a knowledge of discrete transitions between particular levels, but also quantities like strength functions. Therefore, approaches to study the dynamics of nuclear systems are also important to the study of $\beta$ decay and can be studied with both data-driven and {\it ab initio} approaches.

For the purposes of this white paper, we summarize approaches that are presently used to study nuclear $\beta$ decay while trying to highlight the strengths and drawbacks of each.  It is intended that this document can serve as a reference to better understand the literature based on these approaches.

\label{sec:manybody_methods}
\subsection{Density Functional Theory}\label{sec:DFT}

Nuclear density functional theory (DFT) provides a computationally efficient framework for describing medium-mass and heavy nuclei across the nuclear chart. It has achieved considerable success in predicting ground-state properties such as masses and radii~\cite{RocaMaza2018,Meng2016,Bender2003,Robledo2019,Stone2007a}. Beyond ground states, DFT can describe nuclear excitations within the time-dependent DFT (TDDFT) framework~\cite{Nakatsukasa2016}, and phenomena such as shape coexistence through multi-reference description of the generator coordinate method (GCM)~\cite{Reinhard1987}. Owing to its broad applicability and favorable computational scaling, DFT plays an important role in studies of nuclear structure and weak processes throughout the nuclear landscape.

DFT is based on the existence of a functional of the density $\rho(\boldsymbol{r})$, the so-called energy density functional (EDF), such that the energy of the system in an external field $v_{\rm ext}$ can be written as~\cite{Meng2016}
\begin{equation}
    E[\rho(\boldsymbol{r})] = F_{\rm HK}[\rho(\boldsymbol{r})] + \int d^3 \boldsymbol{r} v_{\rm ext}(\boldsymbol{r}) \rho(\boldsymbol{r}),
\end{equation}
where $F_{\rm HK}$ is the exact functional whose existence is guaranteed by the Hohenberg and Kohn theorems~\cite{Hohenberg1964}. For a system with $N$ particles, the many-body wave function depends on $3N$ spatial coordinates. In contrast, the one-body density depends only on the three spatial coordinates $\boldsymbol{r} = (x,y,z)$. Consequently, working with the density rather than the many-body wave function greatly simplifies the $N$-body problem, and leads to favorable scaling of the resulting equations with the system size. However, the Hohenberg-Kohn theorems do not provide a prescription for constructing the exact functional $F_{\rm HK}$.

A practical implementation of DFT was developed by Kohn and Sham, who introduced a set of auxiliary single-particle orbitals from which the density is constructed~\cite{Kohn1965}
\begin{equation}
    \rho(\boldsymbol{r}) = \sum \limits_i |\varphi_i(\boldsymbol{r})|^2.
\end{equation}
The exact functional is then decomposed as
\begin{equation}
    F_{\rm HK} = T + E_{H} + E_{xc},
\end{equation}
where $T$ is the kinetic energy contribution, $E_H$ is the Hartree energy, and $E_{xc}$ is the exchange-correlation term. 
Variation of the energy with respect to the density yields the Kohn-Sham (KS) equations for the auxiliary wave functions
\begin{equation}\label{eq:KS_equation}
    \left( - \frac{\nabla^2}{2m} + v_{\rm KS}[\rho(\boldsymbol{r})]\ \right)\varphi_i(\boldsymbol{r}) = \varepsilon_i \varphi_i(\boldsymbol{r}),
\end{equation}
where $v_{\rm KS} = \delta (E_H[\rho]+ E_{xc}[\rho])/\delta \rho$ is the KS potential and $m$ the particle mass. The equations are solved self-consistently because the potential depends on the underlying density.

Unlike the electron gas problem, where the exchange-correlation functionals can in principle be determined exactly, in nuclear physics the functionals are phenomenological, i.e., consisting of around a dozen parameters fitted to experimental data, often masses and radii.
As a result, many various functionals have been developed, each with its own strengths and limitations~\cite{RocaMaza2018,Bender2003}.

Depending on their formulation, nuclear EDFs can be categorized as non-relativistic and relativistic. In non-relativistic EDF theory, the functional is constructed from contact interactions involving powers and gradients of densities and spin currents, often supplemented by tensor terms, as in Skyrme functionals~\cite{Stone2007a}, or from finite-range interactions such as in Gogny functionals~\cite{Robledo2019}.
The resulting KS equations are Schr\"odinger-like, as in Eq. (\ref{eq:KS_equation}).

In contrast, relativistic EDFs start from an approach similar to field theory, by writing a Lagrangian density and deriving the KS equations of motion through the Euler-Lagrange variation principle. A detailed account of various types of relativistic EDFs can be found in Refs.~\cite{Meng2016,Niksic2011}. Furthermore, a natural extension to include the pairing correlations are discussed in Ref.~\cite{Vretenar2005a}. 

Although DFT is often used to obtain the ground state of a nucleus only, when coupled with other many-body methods, such as the Quasiparticle Random-Phase Approximation (discussed in Sec.~\ref{sec:QRPA}), it enables systematic calculations of $\beta$ decay properties throughout the nuclear chart.

\subsection{Quasiparticle Random-Phase Approximation}\label{sec:QRPA}

Quasiparticle random-phase approximation (QRPA) and its variants are commonly used in the $\beta$ decay studies in the literature. QRPA describes nuclear states as two-quasiparticle excitations and can hence handle large model spaces and wide range of nuclei with reasonable computational burden. Due to the limited number of many-particle configurations, QRPA often lacks accuracy in the description of the properties of individual states compared to for example nuclear shell model, however QRPA typically satisfyingly describes gross features of nuclei, such as giant resonances.
    
In the proton-neutron QRPA (pnQRPA), which is suited for computing charge-exchanging transitions such as $\beta$ decays, states in odd-odd nuclei coupled to angular momentum $J^\pi$ are obtained as 
    \begin{equation}
        \ket{J^{\pi}_k M}=\sum_{pn}\left(X_{pn}^{J^{\pi}k}[a_p^{\dagger}a_n^{\dagger}]_{JM}-Y^{J^{\pi}k}_{pn}[a_p^{\dagger}a_n^{\dagger}]^{\dagger}_{JM}\right)\ket{\rm QRPA}\;,
    \end{equation}
where $\ket{\rm QRPA}$ is the QRPA vacuum (the ground-state of the even-even reference nucleus, containing the QRPA correlations), and $a^{\dagger}_{p/n}(a_{p/n})$ are proton/neutron quasiparticle creation (annihilation) operators, with $[\cdot]_{JM}$ denoting the angular momentum coupling. The $X$ and $Y$ are the amplitudes solved from the QRPA equations \cite{suhonen2007nucleons}. In this formalism, the one-body transition densities needed in the computation of $\beta$ decay nuclear matrix elements can be conveniently written as
    \begin{align}
        \langle J^{\pi}_k\|[c^{\dagger}_p\tilde{c}_n]_J\|0^+_{\rm gs}\rangle&=\sqrt{2J+1}(u_pv_nX^{J^{\pi}_k}_{pn}+v_pu_nY^{J^{\pi}_k}_{pn})\\
        \langle 0^+_{\rm gs}\|[c^{\dagger}_p\tilde{c}_n]_J\|J^{\pi}_k\rangle&=\sqrt{2J+1}(v_pu_nX^{J^{\pi}_k}_{pn}+u_pv_nY^{J^{\pi}_k}_{pn})\;,
    \end{align}
where $|0^+_{\rm gs}\rangle$ is the ground-state of even-even nucleus, and $|J^{\pi}_k\rangle$ is the $k$-th excited state in daughter odd-odd nucleus with angular momentum and parity $J^\pi$. The occupation (vacancy) amplitudes of the proton/neutron orbitals $v_{p/n}(u_{p/n})$ are usually obtained within the Hartree-Fock-Bogoliubov (HFB), Bardeen-Cooper-Schrieffer (BCS), or HFBCS theory.

The nuclear interaction used in the QRPA equations is often based on a Bonn \cite{holinde1981two,machleidt1996nonlocal} or Argonne \cite{wiringa1995accurate} G-matrix
but can also be derived either from schematic interactions~\cite{Sarriguren1998}, \textit{ab initio} interactions~\cite{Paar2006,Beaujeault2023} (Sec.~\ref{sec:ab.initio}), or even self-consistently within the nuclear DFT~\cite{Paar2003,Paar2004} (Sec.~\ref{sec:DFT}). The latter is derived as a linear response approximation to the TDDFT. This particular derivation allows coupling with EDFs to obtain residual interaction directly from the underlying functional as a second derivative with respect to density. Assuming a weak external perturbation of the form $F(t) = F(\omega)e^{-i \omega t} + \text{h.c.}$, the density can be linearized as
\begin{equation}
    \rho(t) = \rho^0 + \delta\rho(\omega)e^{-i \omega t} + \text{h.c.},
\end{equation}
where $\rho^0$ is the static time-independent density and $\delta\rho$ is the induced density, $\omega$ being the excitation energy. As shown in Sec.~\ref{sec:DFT}, the mean field interaction is derived by varying the EDF with respect to the density $h = \delta E / \delta \rho$, such that the time-dependent mean-field can be expanded as
\begin{equation}
    h(t) = h^0 + \frac{\delta h}{\delta \rho} \, \delta \rho,
\end{equation}
where $h^0$ is the static mean-field Hamiltonian, and $\delta h / \delta \rho$ is the so-called residual interaction. Substituting this expansion into the TDDFT equation and retaining only terms linear in $\delta \rho$ yields the corresponding linear response equation.
By breaking the symmetries of the nuclear mean-field, the dimension of the QRPA matrix 
increases rapidly, making direct diagonalization infeasible. Advanced algorithms, such as the finite-amplitude method (FAM), exist that allow one to solve the QRPA problem for deformed systems~\cite{Nakatsukasa2007}.

Because the simple two-quasiparticle configurations included within the QRPA are, in general, not sufficient to capture the smearing widths of giant resonances and tend to overestimate their strength, additional corrections must be included. Such approaches are typically based either on the second RPA (SRPA), which considers more complex configurations~\cite{Gambacurta2020}, or on particle vibration coupling (PVC)~\cite{Litvinova2007,Litvinova2008}.

\subsection{Shell Model}
\label{section:theory:shellmodel}

The nuclear shell model is one of the oldest many-body methods providing very detailed information on nuclear states and transitions at low energies.
The method is based on the diagonalization of an $A$-body Hamiltonian containing nucleonic kinetic energies and inter-nucleon interactions in a spherically symmetric harmonic-oscillator basis.
Performing that in a sufficiently large model space is computationally doable for light nuclei, and this is the \textit{ab initio} approach known as No-Core Shell Model (NCSM) (Sec.~\ref{section:theory:ncsm}). For heavier nuclei, the required basis dimensions become computationally intractable, necessitating the use of truncation techniques. 

At low energies, a robust approximation consists of treating only the valence nucleons as active degrees of freedom in a model space spanned by one or two harmonic-oscillator shells beyond a closed-shell core, typically a doubly-magic nucleus. This severe reduction of the model space requires the construction of an effective Hamiltonian that accounts for the excluded degrees of freedom.
Written in Fock space, such a Hamiltonian is completely defined by a set of single-particle energies, $\varepsilon_i$, which determine the energies of the valence orbitals relative to the closed-shell core, and a set of two-body matrix elements, $V_{ij,kl}$, of the residual interaction (assuming the effective interaction is truncated at the two-body level):
\begin{equation}
H = \sum_i \varepsilon_i a^{\dagger }_i a_i+ \sum\limits_{ijkl}V_{ij,kl} a^{\dagger }_ia^{\dagger }_j a_l a_k \,.
\end{equation}
Either these Hamiltonian parameters can be derived microscopically from a realistic inter-nucleon potentials~\cite{stroberg2019nonempirical,coraggio2024role},
or they have to be determined phenomenologically from a fit of the excitation spectra of nuclei from a given model space to experimental data. Excellent description of low-energy nuclear structure can be achieved in the latter case~\cite{caurier2005shell}.

\subsection{\textit{Ab Initio} Methods\label{sec:ab.initio}}
Nuclear many-body methods have been used to provide critical information for various beta decay measurements. In particular, for the precision physics domain and the $V_{ud}$ extraction, this is the calculation of radiative corrections to super-allowed beta decays \cite{seng2021radiative,gennari2025ab}. For studies of Beyond the Standard Model (BSM) scalar and tensor currents in the weak interaction, nuclear recoil-order corrections have been computed. The calculation of {\it ab initio} nuclear matrix elements has also supported neutrino-less double beta decay measurements. Finally, calculation of theoretical beta decay and electron capture rates has been important for astrophysical simulations and measurements of beta-delayed neutron emission~\cite{li2025ab}.

Commonly used many-body methods for
solving the nuclear Schr\"{o}dinger equation mainly fall into two categories: coordinate-space methods that work directly with the wave
functions and optimize them in a certain way, or configuration space methods that expand the nuclear eigenstates in a basis of known many-body states \cite{hergert2020guided}. Generically, one would like to solve the $A$-body Schr\"{o}dinger equation with a translationally-invariant nuclear Hamiltonian that can be schematically represented as
\begin{equation}
    H = \sum_i T_i -T_{CM} + V_\mathrm{Coul} + \sum_{i<j} v_{ij} + \sum_{i<j<k} V_{ijk} + \ldots \, , \label{Eq:intH}
\end{equation}
where $T_i$ are the kinetic energies of the nucleons, $T_{CM}$ is the center-of-mass kinetic energy, $V_\mathrm{Coul}$ is the Coulomb interaction between the protons, and $v_{ij}$ and $V_{ijk}$ are two- and three-nucleon interactions, respectively. In principle, there could be four-nucleon and other many-nucleon forces; however, no models have implemented such interactions in many-body calculations so far.  

The interactions used in \textit{ab initio} calculations are now typically obtained using chiral effective field theory~\cite{hammer2019nuclear}.  Differences in choice of low energy constants and regulator result in a zoo of interactions, which gives rise to uncertainties in the many-body results.  In addition to these interaction uncertainties, there are additional uncertainties which arise in the numerical implementation of the interactions \cite{hergert2020guided}. 

    \subsubsection{Quantum Monte Carlo}

    Quantum Monte Carlo (QMC) methods are stochastic approaches to solve the nuclear many-body problem~\cite{carlson2014quantum}. In nuclear physics, QMC approaches typically follow a two step process. The first is to carry out a variational Monte Carlo (VMC)~\cite{wiringa1991variational} calculation, which begins with an ansatz for the trial wave function $\Psi_T$. The typical approach is to factorize the wave function in the following way,
    \begin{equation}
        \ket{\Psi_T} = \hat{F}\ket{\Phi}\, ,
    \end{equation}
    where $\hat{F}$ is an operator that encodes information about the short-range nature of the nuclear force, while $\ket{\Phi}$ is a model single particle wave function coupled to the right quantum numbers for the system under study. Embedded in the correlation operator are variational parameters that one optimizes by minimizing the energy expectation value, which is computed using Monte Carlo integration. The advantage of using such an approach is that both long-range and short-range physics can be treated in the nuclear wave function. As discussed in the review of Ref.~\cite{carlson2014quantum}, the model wave function $\ket{\Phi}$ can have a clustering encoded to capture the intrinsic structure of the nucleus, while the correlation operator allows one to embed the effect of short-range physics into the wave function via explicit two- and three-body correlations. While it is typical for structure calculations to take the dominant terms in the nuclear interaction, isospin breaking correlations can be included with a form inspired by perturbation theory~\cite{piarulli2026quantum}. 

    The next step is to use a diffusion Monte Carlo approach --- either Green's function Monte Carlo (GFMC)~\cite{carlson1987green} or Auxiliary Field Diffusion Monte Carlo (AFDMC)~\cite{lonardoni2018auxiliary} --- to remove excited state contamination from the wave function. This is done by performing a propagation in imaginary time $\tau$,
    \begin{equation}
        \ket{\Psi(\tau)} = \exp\left[-(H-E_0)\tau\right]\ket{\Psi_T}\, , 
    \end{equation}
    where $E_0$ is an energy offset used to control the normalization of the wave function. If $E_0$ is chosen to be sufficiently close to the true ground state energy of the system, we can project out the ground state energy, since we can imagine $\Psi_T$ as an expansion in the true, unknown eignenbasis of the Hamiltonian $\psi_i$; {\it i.e.}, 
    \begin{equation}
        \ket{\Psi_T} = \sum_{i=0}^{\infty} c_i \ket{\psi_i}\, .
    \end{equation}
    Then, we have the limit
    \begin{equation}
        \lim_{\tau \to \infty} \ket{\Psi(\tau)} \to c_0 \psi_0 \, ,
    \end{equation}
    which we achieve by dividing imaginary time into small slices $\Delta\tau$ and making $n$ short propagations. Thus, in practice, one computes
    \begin{equation}
        \ket{\Psi(\tau)} = \exp\left[ -(H-E_0)n\Delta\tau \right]\ket{\Psi_T}\, ,
    \end{equation}
    and averages energies and other observable quantities after convergence is reached. 
    
    While, in principle, the above would be true for any trial state, we encounter the fermion sign problem in practice. Briefly stated, the ground state of the Hamiltonian would be the nodeless, symmetric state; however, we have a system of interacting fermions which require an antisymmetric wave function. At sufficiently long imaginary-time propagations, this results in a signal-to-noise problem in the expectation values computed with Monte Carlo. To circumvent this, constrained path algorithms--- which essentially limit the propagation to regions where the propagated and trial states have a positive overlap--- were developed~\cite{carlson2014quantum}. Therefore, having more correlations in the trial wave function can help to improve the convergence and statistical uncertainty of final results in the Monte Carlo calculation. 

    The QMC approach is powerful for studying the lightest ($A \,\lesssim \, 20$) nuclei starting from the nuclear Hamiltonian; however, it is typical to use local interactions in these approaches to make the sampling efficient. In basis-expansion approaches, it is more common to take softer, non-local interactions that accelerate convergence. Thus, direct benchmarks between the approaches for quantities like nuclear structure radiative corrections (see Section~\ref{sec:ckm.c10}) and $\beta$ decay spectra (see Section~\ref{sec:bsm:struct.input}) are not exactly one-to-one because of differences in the underlying Hamiltonian. Recent efforts to include non-local terms in the nuclear Hamiltonian perturbatively~\cite{curry2024perturbative} could allow for benchmarks using consistent interactions across different approaches, and work along this line would be highly beneficial for $\beta$ decay and other nuclear structure applications. 


    \subsubsection{No-Core Shell Model}
    \label{section:theory:ncsm}
    \begin{figure}[t!]
        \center\includegraphics[width=0.6\textwidth]{{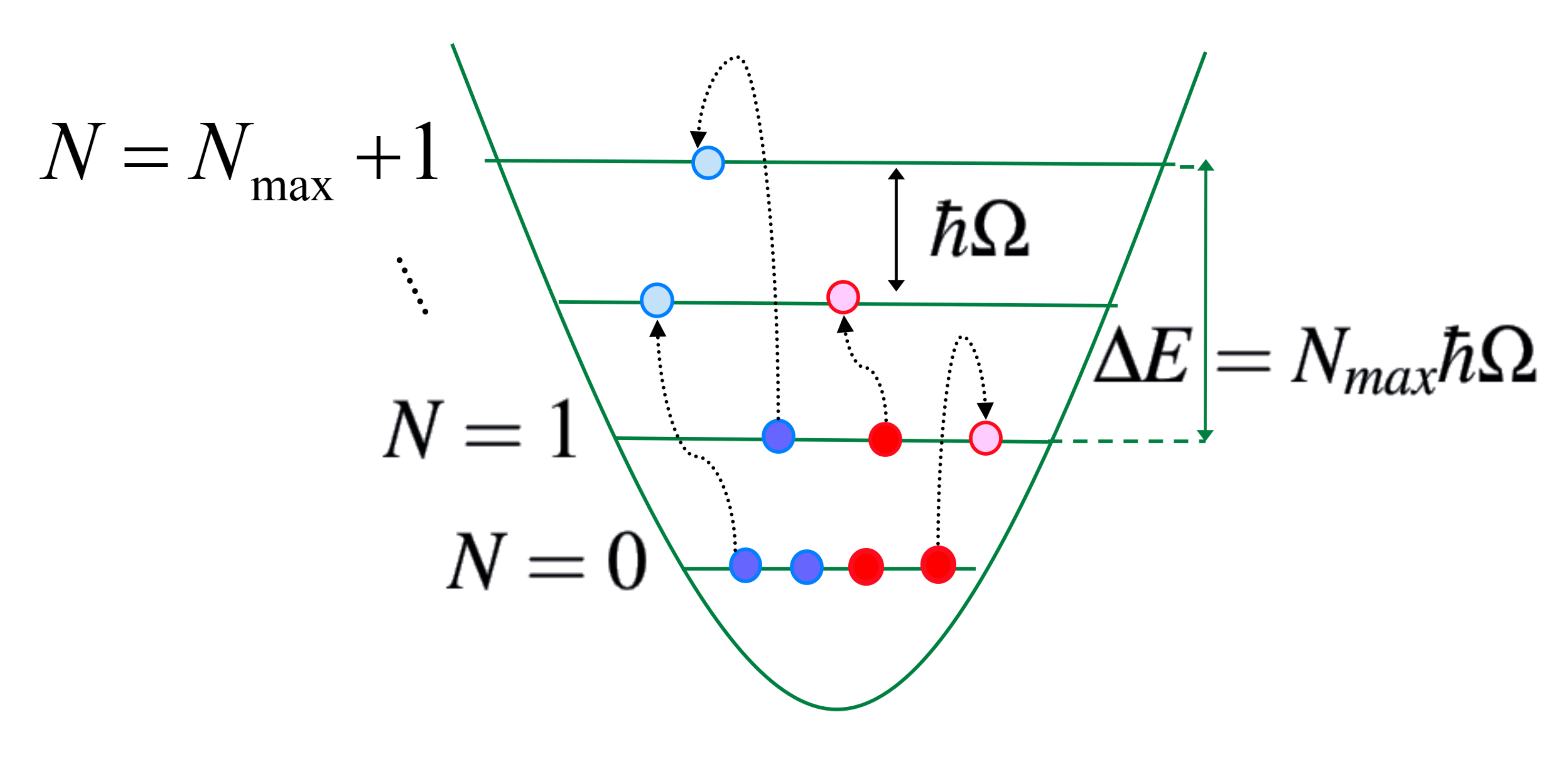}}
        \caption{\label{section:ncsm:figure:truncation}
        The truncation scheme for the many-body oscillator basis expansion with frequency $ \Omega $. For a given value of $ N_{ \mathrm{max} } $, all nucleonic excitations above the lowest Pauli configuration such that the sum of oscillator quanta $ N \le N_{ \mathrm{LPC} } + N_{ \mathrm{max} }$, are considered.}
    \end{figure}

    The \textit{ab initio} No-Core Shell Model (NCSM)~\cite{barrett2013ab, navratil2016unified} (see the original Refs.~\cite{zheng1994large, navratil2000large}) is a configuration space method for constructing the eigensolutions of the many-body Schrödinger equation with all $ A $ nucleons as active degrees of freedom. Working in terms of the fermionic Fock space $ \mathcal{F} $ provides a clean language for organizing the many-body basis and computing amplitudes relevant to particle theory. We define the fermionic creation and annihilation operators $ \{ c^{ \dagger }_{ \alpha }, c_{ \alpha } \} $, obeying the usual anti-commutation relations, which populate the vacuum with one-particle configurations $ \{ \phi_\alpha \} $. The one-particle configurations are frequently taken to be harmonic oscillator (HO) configurations with frequency $ \Omega $. The eigensolutions are then expanded in a set of antisymmetrized, many-body oscillator configurations, with truncation enforced on the total number of excitation quanta as
    \begin{gather}
        \overlap{ \vec{x}_{1} \cdots \vec{x}_{A} }{ \Psi^{A} }
        = \sum_{ N = 0 }^{ \nmax } \sum_{ \alpha } \ c_{ N \alpha }^{ \text{\tiny$(\mathrm{SD})$} } \
        \overlap{ \vec{x}_{1} \cdots \vec{x}_{A} }{ \psi_{ N \alpha }^{A} }_{ \mathrm{SD} }
        \quad , \label{section:ncsm:equation:cartesianexpansion}
    \end{gather}
    with the configurations naturally antisymmetrized via Slater determinants as
    \begin{align}\label{section:qmbp:equation:sdantisymmetrize}
        \overlap{ \vec{x}_{1} \cdots \vec{x}_{A} }{ \psi^{A}_{\alpha} }_{ \mathrm{SD} }
        = \overlap{ \vec{x}_{1} \cdots \vec{x}_{A} }{ \phi_{ \alpha_{1} } \, \cdots \, \phi_{ \alpha_{n} } }_{ \mathrm{SD} }
        = \frac{1}{ \sqrt{A!} } \
        \begin{vmatrix}
            \ \phi_{ \alpha_{1} } ( \vec{x}_{1} ) & \cdots & \phi_{ \alpha_{1} } ( \vec{x}_{A} ) \ \\
            \ \vdots & \ddots & \vdots \ \\
            \ \phi_{ \alpha_{A} } ( \vec{x}_{1} ) & \cdots & \phi_{ \alpha_{A} } ( \vec{x}_{A} ) \
        \end{vmatrix}
        \quad .
    \end{align}
    Valid configurations are generated by the repeated action of $ \{ c^{ \dagger }_{ \alpha }, c_{ \alpha } \} $ on the lowest Pauli configurations (LPCs) of the nucleus, as shown in Fig.~\ref{section:ncsm:figure:truncation}, while enforcing satisfaction of the truncation condition $ N \le N_{ \text{LPC} } + N_{\max} $. The basis is characterized solely by $ \Omega $ and $N_{\max} $, and results are independent of the choice of $ \Omega $ in the limit of $ N_{ \mathrm{max} } \rightarrow \infty $.
    
    The nuclear Hamiltonian discretized in this large-but-finite basis is then typically diagonalized via the Lanczos algorithm~\cite{lanczos1950iteration, komzsik2003lanczos}: given a pivot vector, an orthonormal basis for the Krylov subspace of $H$ is iteratively constructed via a vectorial three-point recursion relation. Extremal eigenvalues of $ H $ converge rapidly and there exist well-defined upper and lower bounds on the convergence of the approximate spectrum~\cite{paige1980accuracy, beresfordn.1995}. It was further recognized in the condensed matter theory community that such recursion methods may be used in the reconstruction of spectral quantities like the Green function~\cite{haydock1974inverse, haydock1980recursive, heine1980electronic, kelly1980applications, haydock1972electronic, pettifor2012recursion}. Referred to as the Lanczos Moments Method or Lanczos Strengths Method, targeted application of the Lanczos algorithm facilitates evaluation of objects like
    \begin{align}
        \mathcal{A}_{fi} = \int dz \ \sumintegral_{n} \ f ( z, \omega_{n} ) \ \bra{ \Psi_{f} } \, O_{2} \, \ket{ \Psi_{n} } \ \bra{ \Psi_{n} } \, O_{1} \, \ket{ \Psi_{i} }
        \quad ,
        \label{eq:Lanczos}
    \end{align}
    subject to the convergence properties of $ f $ folded with the amplitudes and the number of Lanczos iterations utilized.
    This approach has been exploited in a recent evaluation of the one-loop radiative correction $ \delta_{ \mathrm{NS} } $ to the $ \carbonfermidecay $ Fermi transition~\cite{gennari2025ab, gennari2025electroweak} with the main result discussed in Sec.~\ref{section:ckm:section:currentalgebra}.

     \subsubsection{Symmetry-Adapted No-Core Shell Model}
     \label{section:theory:sa-ncsm}

    The symmetry-adapted no-core shell model (SA-NCSM) \cite{launey2016symmetry, dytrych2020physics, launey2020emergent} reorganizes the conventional HO $N_{\max}$-truncated no-core shell model (NCSM) basis into subspaces classified by nuclear shapes and collective degrees of freedom. Concretely, many-body states are grouped into ${\rm SU}(3)$ irreducible representations (irreps) of given deformation and, more generally, into ${\rm Sp}(3,\mathbb{R})\supset{\rm SU}(3)$ subspaces, often termed ``symmetry-adapted'' (SA) bases, that track symplectic excitations associated with quadrupole collectivity. Each basis state is labeled by quantum numbers $(\lambda,\mu)$ and $S$ for ${{\rm SU}(3)}\times{\rm SU}(2)_S$, together with additional quantum numbers required for a complete specification.  Here, the ${{\rm SU}(3)}$ quantum numbers $(\lambda,\mu)$ are determined by the distribution of HO quanta $N=N_x+N_y+N_z$ via $\lambda=N_z-N_x$ and $\mu=N_x-N_y$, and $S$ is the total intrinsic spin. This organization motivates “SA model spaces,” in which the standard complete spaces at small $N$ (adequate for less-deformed shapes) are augmented at large $N$ only by those SA basis states that carry the physically dominant deformations and clustering patterns, thereby targeting the configurations most relevant for collective motion while drastically reducing dimensionality. SA-NCSM also uses the Lanczos algorithm to diagonalize the many-body Hamiltonian in the SA basis and can similarly use Eq. \ref{eq:Lanczos} for calculations of observables involving many intermediate states, such as certain beta decay corrections. 
    
    A practical advantage of the SA-NCSM construction is that the center-of-mass (c.m.) motion factors exactly. The c.m.~number operator $\hat N_{\rm c.m.}$ neither mixes HO excitations nor couples different ${\rm SU}(3)$ subspaces (it is an ${\rm SU}(3)$ scalar), so translational invariance is preserved without contamination from spurious c.m.~components. Beyond clean factorization, the SA-guided selection retains the \textit{ab initio} character of the calculation with realistic interactions while prioritizing low-$S$ and a small set of leading $({\lambda},{\mu})$ irreps and their associated ${\rm Sp}(3,\mathbb{R})$ ladder excitations—precisely those that encode large deformation, enhanced $E2$ strength, and spatially extended (cluster-like) configurations. In practice, this yields faster convergence of collective observables (e.g., radii and quadrupole moments) and enables calculations in spaces that would be prohibitive for the traditional NCSM. Hence, the SA-NCSM approach  can accommodate larger model spaces, and  reach heavier nuclei than traditional NCSM methods, including 
    $^{20}$Ne \cite{dytrych2020physics,DreyfussLESBDD20}, $^{21}$Mg \cite{Ruotsalainen19}, $^{28}$Mg \cite{william_PRC_2019}, $^{32}$Mg \cite{launey2025ab},  $^{32}$Ne \cite{LauneySOTANCP42018}, $^{40}$Ca \cite{Burrows_2025} and $^{48}$Ti \cite{LauneyMD_ARNPS21}.
    
    
    \subsubsection{Coupled-Cluster}
    \label{sec:theory_cc}

    Coupled-cluster (CC) theory~\cite{hagen2014coupled} relies on an exponential ansatz for the nuclear many-body wavefunction
    \begin{equation}
        \ket{\Psi_0} = e^{T} \ket{\Phi_0}\,.
        \label{expansatz}
    \end{equation}
    Here, $\ket{\Phi_0}$ is a reference state, typically a Hartree-Fock Slater determinant, and $T$ is the so-called cluster operator. The latter introduces many-body correlations on top of the reference, and can be written in terms of a sum of $n$-particle $n$-hole excitations as $T = T_1 + T_2 + T_3 + \dots + T_A$. Due to computational cost, truncations in the many-body expansion are introduced. The coupled-cluster singles and doubles (CCSD) scheme, including up to $2p$-$2h$ correlations, captures about 90\% of the full correlation energy in the Hartree-Fock basis~\cite{hagen2009ab,sun2022renormalize}. Almost 99\% of the correlation energy can be obtained when considering triples excitations in the CCSDT-1 approximation~\cite{watts1993coupled}. 
    
    Properties of medium mass and heavy nuclei up to $^{208}$Pb~\cite{hu2021ab} and beyond~\cite{bonaiti2025structure}, as ground
    state energies and radii, as well as bound excited spectra and transition strengths, can be described in this framework (see e.g.~\cite{hagen2012continuum,hagen2012evolution,hagen2016neutron,hagen2016structure,morris2018structure,gysbers2019discrepancy}). 
    Moreover, one can address the calculation of response functions within coupled-cluster theory by coupling it with the Lorentz Integral Transform method~\cite{bacca2013first,bacca2014giant}. The latter allows one to reduce the problem of computing continuum states to the solution of a bound-state like equation of motion. Up to now, these efforts have mostly focused on response properties of electric dipole transitions~\cite{simonis2019first,kaufmann2020charge,fearick2023electric} and lepton scattering~\cite{sobczyk2021ab,sobczyk202440,acharya202516}. 
    
    Such developments can be exploited also in the study of $\beta$ decay strength functions. Calculations of Gamow-Teller responses in closed-shell nuclei as $^{14}$O~\cite{giraud2023beta+} and $^{24}$O~\cite{neupane2024first} show good agreement with experiment when an \textit{ab initio} based estimate of the quenching factor is taken into account. Recently, the LIT-CC framework has been extended also to open-shell nuclei via particle-attached/removed techniques~\cite{bonaiti2024electromagnetic,marino2025structure}. With this approach, predictions for electric dipole polarizabilities in the oxygen and calcium isotopic chains suggest the need for higher order correlations to accurately resolve states with higher excitation energy. Work is in progress to perform LIT-CC calculations starting from axially-symmetric reference states~\cite{novario2020charge,hagen2022angular,sun2025multiscale}.

    \subsubsection{In-Medium Similarity Renormalization Group}
    \label{section:theory:imsrg}
    A cousin of CC is the in-medium similarity renormalization group (IMSRG) technique \cite{hergert2016Theinmedium}, which also decouples many-body excitations from the Hamiltonian and allows computation of medium mass nuclei due to favorable polynomial scaling with system size.
        
    IMSRG relies on a continuous unitary transformation found by integrating the operator flow equation:
    \begin{equation}
        \frac{d}{ds} H(s) = [ \eta(s), H(s) ],
        \label{SRG}
    \end{equation}
    where $H(s=0)$ denotes the starting Hamiltonian and $s$ is the flow parameter.
    This type of equation is typically used in pre-processing of the \textit{ab initio} internucleon interactions (i.e. ``free-space'' SRG \cite{bogner2007similarity}), which ``softens'' the interaction by decoupling short-range, high-momentum modes while maintaining low-energy observables \cite{schuster2014operator}.
    
    A suitable choice of ``in-medium'' generator can also partially- or block-diagonalize the many-body Hamiltonian. First, the Hamiltonian is formulated in terms of normal-ordered operators relative to a reference, i.e., creation and annihilation operators are contracted with the reference density matrix. In the original formulation, the reference was a single Slater determinant or Hartree-Fock state, but this has since been generalized to the ``multi-reference'' formulation \cite{hergert2016Inmedium}.
    Second, the generator of the transformation is chosen relative to off-diagonal matrix elements, defined as those which couple the reference to few-particle few-hole excitations.
    
\begin{figure}[tbp]
        \centering
        \includegraphics[width=0.99\linewidth]{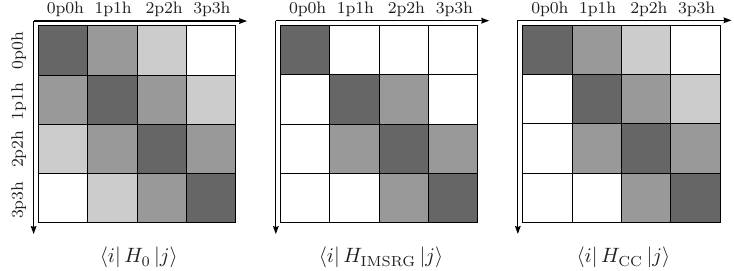}
        \caption{Decoupling of particle-hole excitations from a 0p0h reference state: the schematic
    matrix representation of the initial Hamiltonian $H_0$ (a) and the transformed Hamiltonians
    obtained from IMSRG (b) and CC (c), respectively. Adapted from Ref.~\cite{johnson2020white}.}
        \label{fig:decoupling}
    \end{figure}

    The many-nucleon Hamiltonian is never explicitly constructed, but $A$-body operators are included approximately by using their normal-ordered 0-, 1- and 2-body parts. The flow equation is integrated by evaluating commutators of those few-body operators and truncating induced higher-body terms. Schematically, this results in the transformation of Fig. \ref{fig:decoupling}(a) to Fig. \ref{fig:decoupling}(b) (with (c) showing the CC transformation for comparison).
    
    IMSRG evolution results in the energy of an eigenstate of the many-body Hamiltonian, the ground state if the reference has sufficient overlap. To obtain other observables, the resulting Hamiltonian must be further processed, e.g., with the equations of motion method \cite{parzuchowski2017abinitio} or the shell model.
    In the latter case, this may be the no-core shell model. The
    in-medium no-core shell model (IM-NCSM) \cite{gebrerufael2017ab} takes an eigenstate of an NCSM calculation with a small $N_{max}$ as the reference. Once decoupled via IMSRG flow, a similarly low-dimensional diagonalization is needed to extract wavefunctions and observables of the full Hamiltonian.
    This technique allows the use of Lanzcos Strengths Method, as discussed in Sec.~\ref{section:theory:ncsm}, to evaluate transition amplitudes. Thus, the analysis of Ref.~\cite{gennari2025ab} may be applied to $A=22$ and beyond with this technique. Effort is underway to consistently evolve the relevant operators.
    
    Another variant is the valence-space IMSRG (VS-IMSRG) \cite{stroberg2016ground} which chooses $\eta$ to decouple a particular subset of single-particle states. A two-stage IMSRG evolution is carried out, first decoupling the reference as a core, then decoupling a single major shell above the core from further excitations. The evolved Hamiltonian matrix elements in this ``valence space" may be used as effective shell model interactions, providing an \textit{ab initio} source for the standard shell model (Sec.~\ref{section:theory:shellmodel}). This technique has been used to investigate isospin breaking \cite{stroberg2021beta} and beta decay quenching \cite{gysbers2019discrepancy}.
    
    Another hybrid technique is the in-medium generator coordinate method, which constructs an explicitly deformed reference with collective correlations \cite{yao2018generator}.
    It has been applied to neutrinoless double beta decays \cite{yao2020ab,belley2024ab}.
    
\subsection{Perspectives and Outlook}
\label{sec:perspectives} 


The preceding sections have provided an overview of the many-body techniques that are employed in theoretical studies of nuclear beta decays. 
Broadly, they fall into two major paradigms --- data-driven approaches that use energy density functionals (EDFs) or interactions that are fitted to many-body data, and \emph{ab initio} approaches that start from nuclear two- and three-body interactions, usually constructed in chiral effective field theory (EFT). 

The data-driven interactions can be understood as attempts to parameterize the in-medium nuclear interactions at low resolution while incorporating relevant symmetries, not unlike an E(F)T. Indeed, there is work that seeks to apply more formal principles to the selection of operators for empirical parameterizations, e.g., through power counting ideas \cite{Furnstahl:2012fu,Dobaczewski:2012qw,Huth:2018si}. In parallel, applications of renormalization group techniques, in particular (IM)SRG, have shown how quantitatively correct (albeit not directly observable) single-particle shell structure and effective interactions and operators emerge from chiral forces, while avoiding some of the challenges of earlier approaches that were rooted in Many-Body Perturbation Theory \cite{stroberg2019nonempirical,Ding:2026fs}. 

Data-driven approaches can offer a highly accurate description of nuclear properties through high-quality global fit and an additional local fine-tuning of the interaction or effective field theory (EFT) parameters, if necessary, but it is not clear at all how empirical models should be combined with transition operators from modern Standard Model (SM) and Beyond Standard Model (BSM) EFTs. This is especially important for precision measurements, which require the consideration of higher-order operators. In the \emph{ab initio} framework, these operators appear by coupling the chiral Lagrangian to the electroweak sector, with consistency conditions that link their parameters --- i.e., low-energy constants (LECs) --- to those of the nuclear interactions \cite{Pastore:2009zr,Krebs:2020te,cirigliano2021toward}. Just like the interactions, the higher-order transition operators involve two-body (and more) contributions, hence it is unlikely that they are fully addressed by the usual empirical approach, which merely introduces effective charges or quenched couplings in the leading-order one-body transition operators. Multiple efforts are underway to implement this program.

As discussed in the preceding sections, there has been much progress in the many-body techniques that rely on the aforementioned interactions and operators. To develop efficient treatments of collective correlations in medium-mass and heavy nuclei, \emph{ab initio} approaches are adopting ideas like symmetry breaking and restoration~\cite{yao2020ab,belley2024ab,zhou2025ab,novario2020charge,hagen2022angular,sun2025multiscale,Bally:2021sq,Frosini:2022hj,Frosini:2022xg,Frosini:2022wd,Porro:2024gd,Porro:2024ag,Porro:2024co,Porro:2024ao,porro2025impact,bonaiti2026ab}.  (Such ideas have historically been used with phenomenological effective interactions, leveraging thee simplified structure of such interactions, e.g., contact or Gaussian radial dependencies.)  In \emph{ab initio} methods, this been made possible through the use of high-performance computing resources (and suitable access to such systems), as well as the use of model reduction techniques that are rooted in many-body theory, e.g., through expansions in the irreducible representations of dynamical groups, as in the SA-NCSM \cite{launey2016symmetry,McCoy:2020ol,Dytrych:2020db}, or rooted in data science (principal component decompositions, tensor factorization, etc.) \cite{Frosini:2024id}.


In the near- to mid-term, there are several important research thrusts in the nuclear many-body community which are directly relevant to beta-decay programs:

\begin{enumerate}
    \item Several groups have derived higher-order EFT corrections to the electroweak transition operators, both for single- and 
    double beta decay, and first applications in light- and medium-mass nuclei have been published (see, e.g., Refs.~\cite{cirigliano2024ab,cirigliano2024radiative,king2025quantum,cirigliano2018neutrinoless,cirigliano2018neutrinoless2,cirigliano2019renormalized,chambers-wall2025three-nucleon,jokiniemi2021impact,jokiniemi2023neutrinoless,castillo2025neutrinoless,fasano2025:diss-springer} and Secs.~\ref{sec:structure} and~\ref{sec:ckm}--\ref{sec:neutrino}). 

    While the ideal approach combines these transition operators with consistent EFT-based nuclear interactions, there have been several studies in which the nuclear wave functions are generated with data-driven EDFs or effective interactions. 
    More work is needed to assess theoretical uncertainties stemming from regulator dependence and variations of the LECs that appear in the transition operator, so that there are no hidden artifacts due to the inconsistent treatment. If issues arise, a near-term solution could be to constrain the EDFs or effective interactions with \emph{ab initio} results, e.g., by matching results for relevant observables \cite{Duguet:2023nk}. 
    
    Going further, the IMSRG is a powerful tool for constructing both effective interactions and operators from a consistent starting point, which can shed light on operator structures that are missing from the ans\"atze in data-driven models. Conversely, there have been successful efforts to learn empirical density functionals and interactions directly from data across many scientific fields \cite{Unke:2021lw,akashi2025machineslearndensityfunctionals}, including nuclear science \cite{Bakurov:2025to}, which may be extensible to transition operators.

    \item Data-driven EDFs and effective interactions are usually tailored to be used within a fixed many-body scheme, e.g., Hartree-Fock(-Bogoliubov)-like equations and (Quasiparticle) Random Phase Approximation for nuclear DFT, or exact diagonalization within a given valence space for the Shell model. In contrast, approximate \emph{ab initio} techniques like CC, Self-Consistent Green's Functions (SCGF) or IMSRG-based methods have inherent truncations that are (in principle) systematically improvable by explicitly accounting for more correlations in the many-body expansion. Major ongoing efforts seek to develop and implement improved truncations in these frameworks \cite{He:2024vl,Stroberg:2024cr}, while also boosting the efficiency of quasi-exact methods like QMC, (SA-)NCSM, or Lattice EFT that can be used to thoroughly benchmark calculations in nuclei that are accessible with both types of approaches.

    \item A related item is the expansion of the capabilities for treating collective correlations in empirical or \textit{ab initio} calculations through symmetry-breaking, restoration, and configuration mixing, as an efficient alternative to performing calculations involving many-particle many-hole configurations \cite{zhou2025ab,Zhou:2026hya}.

    \item Last but not least, there is growing demand for microscopic inputs for masses and electroweak rates in stellar environments, i.e., at finite temperature. Calculations using FT-HF(B), FT-(Q)RPA or Shell Model Monte Carlo (SMMC) have employed empirical EDFs or interactions \cite{Lang:1993dl,Alhassid:1994mv,Schunck:2016fk,Litvinova:2018df,Ravlic:2021wm,Ravlic:2023rd,Ravlic:2025sw}, while \emph{ab initio} calculations have largely been limited to SCGF and Lattice EFT calculations for infinite matter \cite{Rios:2020jz,Lu:2020zo,Ma:2024bj}. In the near term, Valence-Space IMSRG interactions and transition operators could be used as an alternative to empirical interactions in SMMC, and the newly developed FT-IMSRG opens many new pathways for exploring energies and transition rates in a no-core approach \cite{Smith:2025tf}.   
\end{enumerate}

These developments directly map onto experimental efforts discussed in this white paper. The capabilities for efficiently describing nuclear wave functions with complex intrinsic structures are essential for studying nuclear shape phenomena and isomerism (Sec.~\ref{sec:nuclear_structure}), possibly including forbidden beta decays. In \emph{ab intio} approaches, the implementation of  shape-based configurations complements the development of improved truncation schemes because of tradeoffs in precision between these configurations and particle-hole excitations. These theoretical developments are supplemented by major advances in the development and application of surrogate models in nuclear theory, which can accurately emulate the results of microscopic many-body calculations at a tiny fraction of the computational cost. These emulators make it feasible to perform ``proper'' statistical uncertainty analysis, by treating the parameters of the input interactions and transition operators, e.g., EDF parameters or chiral LECs, as distributions instead of fixed values. One can then sample from these distributions to generate probability distributions for the resulting observables, sample form them to generate error bars and explore correlations, or systematically explore the sensitivity of the observables to parameter variations. Together, these efforts will help direct both theoretical and experimental work to identify the most impactful measurements and calculations, and to achieve controlled and sufficiently small theoretical uncertainties in the nuclear structure inputs for astrophysics (Sec.~\ref{sec:astro}), high-precision tests of CKM unitarity or searches for BSM physis (Secs.~\ref{section:theory:ncsm}, \ref{sec:ckm} and \ref{sec:neutrino}).

\section{$\beta$ Decay for Nuclear Structure\label{sec:structure}}
\label{sec:nuclear_structure}
\begin{figure}[!h]
    \centering
    \includegraphics[width=0.75\linewidth]{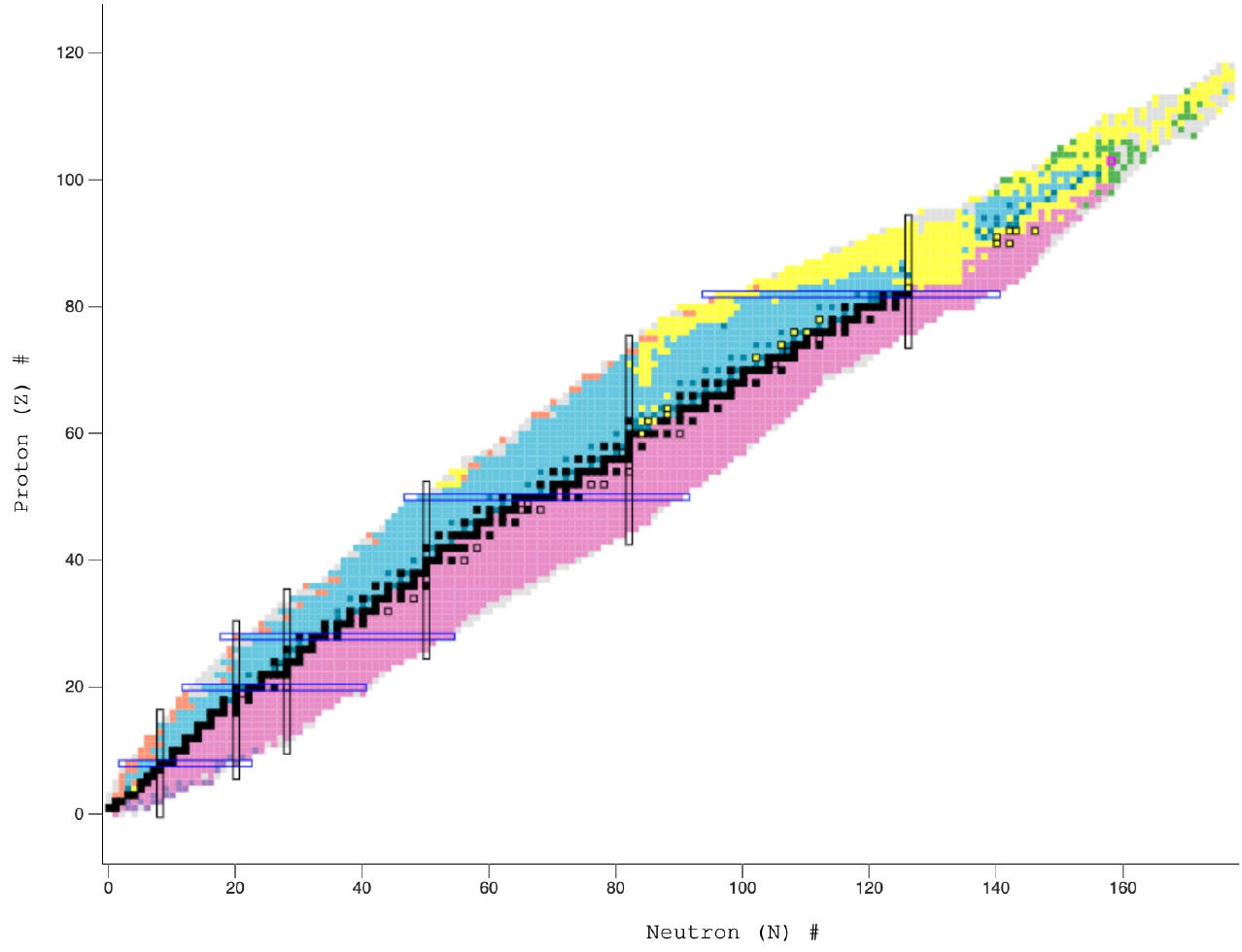}
    \caption{The Segrè chart with colors indicating the prominent decay mode. $\beta^-$-decay (pink), $\beta^+$ (blue), $2n$ (purple), $\textit{p}$-emission (orange), spontaneous fission (yellow), $\alpha$-decay (green). $\beta^{+/-}$ decays are shown to dominate the nuclear landscape, illustrating the reach of the $\beta$ decay mechanism as a tool for decay spectroscopy. From Ref.~\cite{nndc_chart}}
    \label{fig:bdecay_chart}
\end{figure}

\begin{figure}[!h]
    \centering
    \includegraphics[width=0.5\linewidth]{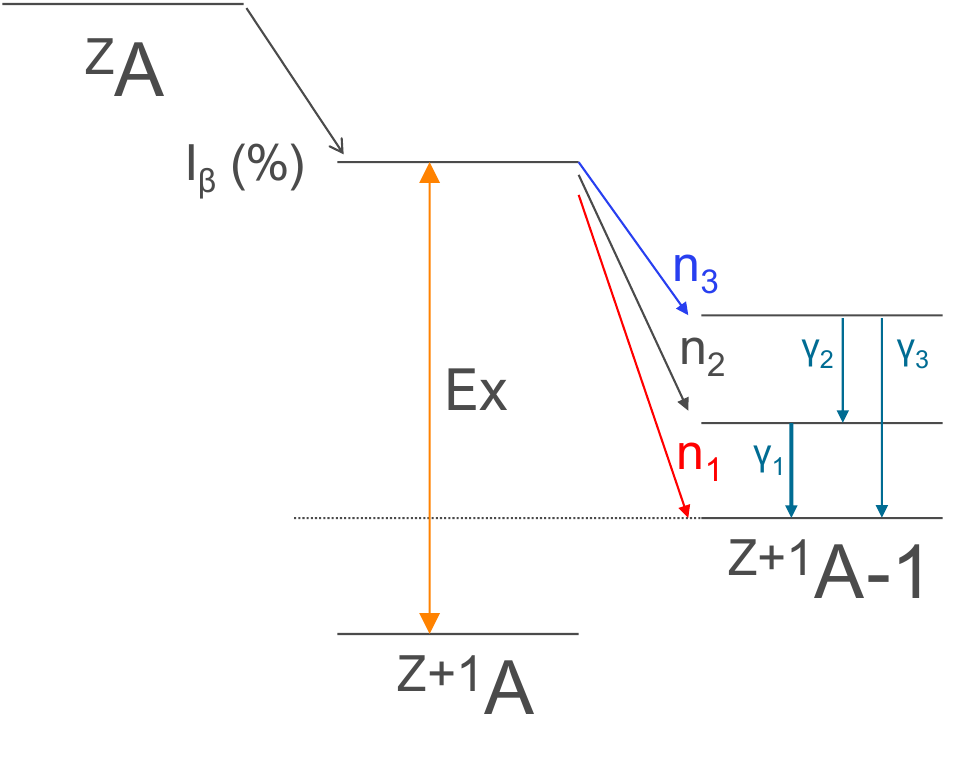}
    \caption{Diagrammatic scheme of $\beta$ decay populating excited states in a $\beta$-daughter and $\beta n$-daughter allowing for $\beta$-delayed $\gamma$ and neutron spectroscopy.}
    \label{fig:decay_scheme_simple}
\end{figure}

For many atomic nuclei on the Segrè chart, $\beta$ decay is a natural and spontaneous process (Fig.~\ref{fig:bdecay_chart}). This process occurs, on average, a time interval $\tau$ after the nucleus is created. With a sufficient $Q_\beta$, the daughter nuclei are often left in excited states following the $\beta$ decay. These excited states again decay by various decay modes such as electromagnetic transition or neutron emission. Measuring these decay products allows for spectroscopic measurements of the daughter nuclei, so-called ``decay spectroscopy". By producing radioactive isotopes at radioactive beam facilities, we are inherently capable of performing decay spectroscopy by using the natural $\beta$ decays to populate excited nuclear states. A schematic diagram of the decay spectroscopy mechanism is shown in Fig. \Ref{fig:decay_scheme_simple}.

From decay spectroscopy there are many fundamental properties that can be measured. This technique and methods therein give access to the nuclear level structure, decay modes and strengths, and nuclear shapes. These properties are important for the understanding of atomic nuclei and play an important role in many other areas of interest such as nuclear structure and reactions, nuclear astrophysics, fundamental symmetries, as well as other benefits to society. 

Many experimental systems exist at radioactive beam facilities world wide to perform decay. This include arrays of detectors capable of high resolution, fast timing, and high efficiency detection. These systems make it possible to systematically study atomic nuclei which can be compared to modern nuclear theory. 

\subsection{Open Quantum Systems}

Central to the experimental program at FRIB are the most neutron-rich nuclei, approaching, at, or even beyond the neutron dripline.  Nuclei at the very limits of binding present unique opportunities and challenges for both experimental studies and theoretical description.  

Experimentally, the most neutron rich species are generally among the lowest production cross sections, meaning that experiments are limited in the observables that are accessible.  Low statistics for measurements generally means that initial information is limited to integral quantities such as half life, $P_n/P_{2n}/.../P_{xn}$ probabilities, and minimal daughter excited-state spectroscopy.  However, despite these challenges the most weakly bound systems offer significant discovery potential.  These are the systems which offer the best path to understanding the importance of the continuum on nuclear structure. 

In particular, the emergence of halo nuclei, characterized by weakly-bound nucleons orbiting a compact core at large distance, is a unique phenomenon at the limits of stability. Given the limited amount of experimental information that can be obtained on such exotic systems, identifying halo signatures from the few observables at disposal, as half-lives and the excited daughter spectrum, becomes crucial. From the point of view of theory, determining the half-life of a nucleus involves computing the full $\beta$ decay strength, which instead requires much more statistics in an experimental setting. With the rapid progress experienced by nuclear theory in the last few years, also \textit{ab initio} calculations of $\beta$ decay strengths start to be available also for exotic neutron-rich nuclei as $^{24}$O \cite{neupane2024first} and the $N=50$ region around $^{78}$Ni \cite{li2025ab}. 

The experimental search for new halo nuclei has focused mostly around the $N=28$ ``island of inversion", where $^{40}$Mg is considered as a halo candidate \cite{crawford2019first}. In this region, nuclear deformation and coupling to the continuum are dominating effects that need to be properly taken into account. Recent coupled-cluster results, including continuum effects, indicate the coexistence of prolate and oblate shapes in the ground state of $^{40}$Mg \cite{sun2025multiscale}. Similar calculations are also possible within the IMSRG framework combined with the projected generator coordinate method \cite{zhou2025ab}. More work is needed to extend such progress to the computation of strength functions. 

While currently the effect of halo structure on half-lives is unclear, the possibility of comparing experimental and theoretical half-lives along isotopic chains can be helpful in following the evolution of shell structure towards the dripline. For instance, such an analysis led to understanding the erosion of the $Z=14$ sub-shell closure in the $N=28$ region \cite{crawford2022crossing}.

\subsection{Isomerism and Shape Coexistence}
Away from the valley of stability on the Segrè chart, the picture of nuclear structure from the spherical shell model breaks down. As these structures evolve, configuration mixing of nuclear states and the deformation of nuclei occur. The difference in initial and final state configuration hinders transitions between these states. Isomers come about when the life-time of a state is significantly increased by these differences \cite{Walker2022}. These isomers are states that have half-lives longer than $10^{-9}\ \textnormal{s}$ compared to typical nuclear states with half-lives of $10^{-12}\ \textnormal{s}$. 

Systematic experimental observations, several decades back, started to indicate these deviations in mass and in the nuclear shape in $N=20$ nuclei \cite{Thibault1975, Detraz1979, GUILLEMAUDMUELLER198437,Huber1978} from what was predicted. As the neutron and proton numbers increase away from stability the $NN$ force causes higher lying shells to traverse the nuclear shell structure \cite{Otsuka2020,Sorlin2008}. In some scenarios, intruder shells can become low relative to the the low-lying natural orbitals. Isomerism can come about when a mismatch in spin, shape, or collectivity and pairing occurs between excited states and the ground state. Through nuclear $\beta$ decay it is possible to populate these isomeric states, but also these states are sometimes $\beta$ decaying themselves. By performing decay spectroscopy it is possible to probe the nuclear shapes and spin-parity assignments of states decaying to and from these isomers. Isomerism has also been shown to play a role in the astrophysical $\textit{r}$-process causing delayed release of energy \cite{misch2020astromers, misch2024astromers}.

At FRIB, the nuclear landscape is rich, and rich with isomers. As FRIB ramps up in power and production, it is expected to reach the neutron and proton drip lines up to at least the $N/Z=82$ shell closures at the 1/week level \cite{frib400}. The $\beta$ decay mechanism can be used to experimentally populate isomeric states across the chart of nuclides. Additionally, in the last century it was shown that during fragmentation, when beam energies reach the range of GeV/nucleon and ion transport times are short compared to isomer half-lives, it is possible to produce and deliver isomeric states to experimental stations \cite{YOUNG1993}. With the power of existing facilities and the FRIB ramp up, one can expect many new isomers to be discovered. Since FRIB's first experiment in 2022, several new isomers have already been found at FRIB with more under analysis \cite{Gray2023, Lubna2023, Ogunbeku2025, FDSi}. A compilation of some previously discovered isomers are compiled in Ref. \cite{osti_2572219}.

\begin{figure}
    \centering
    \includegraphics[width=0.5\linewidth]{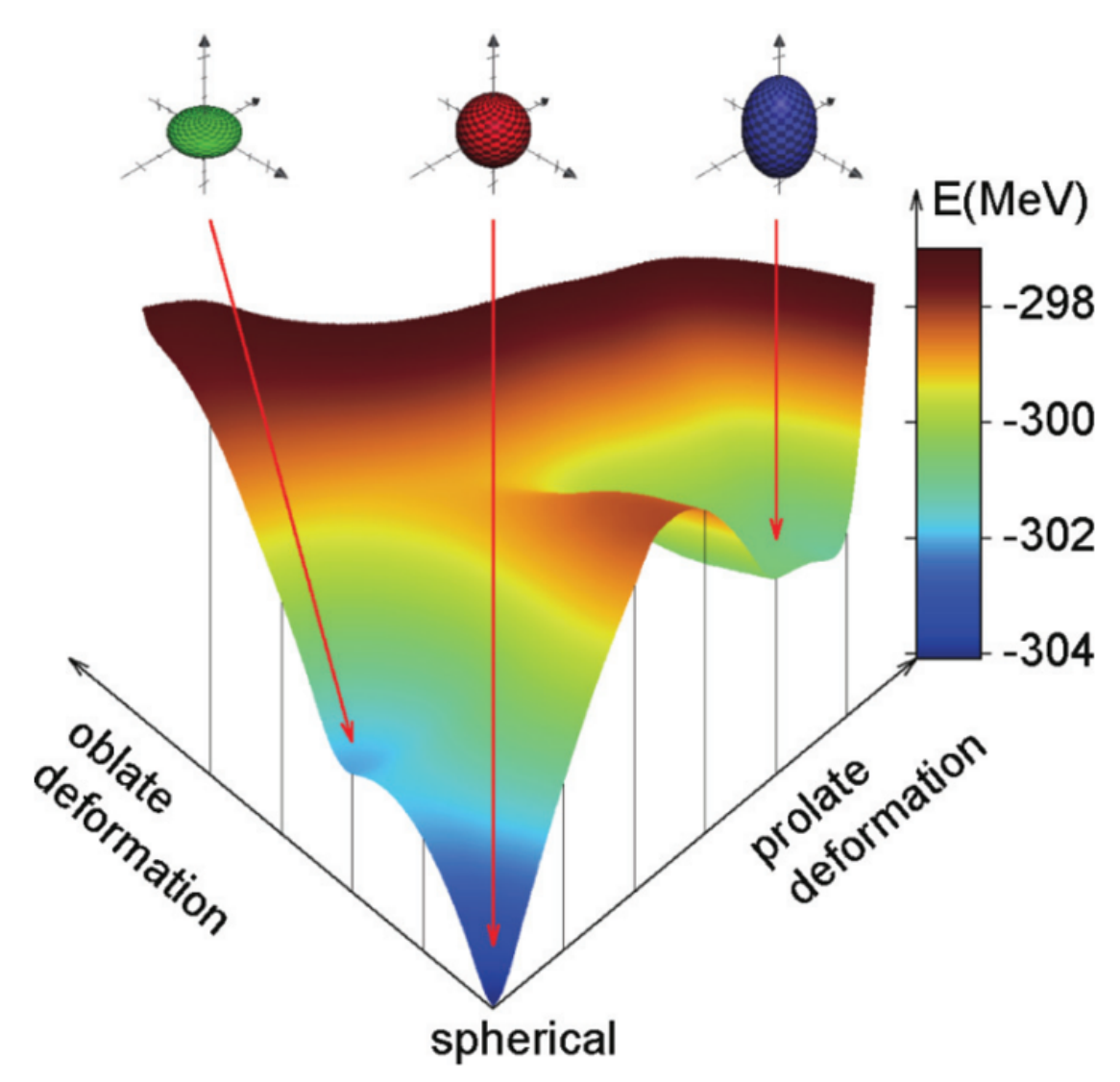}
    \caption{Energy surface of the shape coexisting nucleus $^{68}$Ni from Hartree-Fock calculations constrained by experimental results \cite{Suchyta2014}. The nucleus was found to exhibit a $0^+$ g.s. with low-lying $2^+$ and $0^+$ states. While the g.s. is spherical, these two excited states exhibit oblate and prolate deformation respectively.}
    \label{fig:deformed}
\end{figure}

The even-even nuclei have fully paired g.s. configurations. Some nuclei exhibit low-lying $0^+$ isomeric states. These states are indicative of collective behavior of the paired nucleons leading to shape changes. The wavefunctions of the g.s. and isomeric $0^+$ states have small overlaps that suppress the $E2$ transition strength. The $E0$ and $E2$ transitions give direct information about the wavefunction overlaps and collectivity. Measuring E0 transitions is a tool to understand shape coexistence. The transition probability depends on the initial and final state wavefunctions
\begin{equation}
    \rho^2(E0)=\frac{\langle f|M(E0)|i\rangle}{eR^2} \,.
\end{equation}
Shape coexistence has been studied previously using $E0$ transitions at the NSCL and continuing at FRIB \cite{Suchyta2014,Crider2016}. In Fig. \ref{fig:deformed}, the energy surface triple-shape coexisting nucleus $^{68}$Ni is shown. These experiments required a novel technique to measure the decay radiation of the $0^+\rightarrow0^+$ transitions. This technique has been extended to the FRIB era with one experiment already performed in 2025 using this technique at the FDSi.

\subsection{$\beta$-Strength Function}
The $\beta$-strength function $S_{\beta}(E)$ is defined as the distribution of $\beta$ decay strength --- {\it i.e.}, the magnitude squared of the $\beta$-transition matrix elements --- as a function of the daughter's excitation energy. It is one of the most important properties of atomic nuclei: together with nuclear masses, they determine all $\beta$ decay properties, including half lives, the shapes of the emitted leptons' spectra, and the probabilities of $\beta$-delayed particle emission; in addition, because the transition matrix is proportional to the overlap of wave functions between the initial and final states against the $\beta$ decay operators, the strength function unveils crucial information about the microscopic structures of parent and daughter nuclei. Thus, precise determination of experimental $S_{\beta}(E)$ is quite beneficial, providing the most stringent test of the nuclear theories that aim for modeling decay properties in various applications, and giving us a sensitive approach to probe the nuclear structure in parent and daughter nuclei far from the stability line.

Experimentally, the $\beta$-strength function could be obtained by measuring the $\beta$-feeding probability $I_{\beta}$ comprehensively within the  $\beta$ decay energy window $Q_{\beta}$. The Total Absorption Spectroscopy (TAS) introduced in Sec.\ \ref{sec:tas}, is a powerful approach to address those $\beta$-feeding strengths up to the neutron separation energy, $S_n$. Above $S_n$, neutron emission becomes the dominant radiation mode for those high-lying states, and it is more efficient to study the decay strength via $\beta$-delayed neutron spectroscopy. The Versatile Array of Neutron Detectors at Low Energy (VANDLE) is a neutron time-of-flight (TOF) spectrometer that has been installed at various decay stations \cite{madurga2016evidence, neupane2024first}. It consists of an array of $120\times6\times3$ cm$^3$ plastic scintillator modules mounted to an arch-shaped frame with a radius of about one meter. FDSi features the two-focal-plane configuration: at the first, or discrete, focal plane, 88 VANDLE modules are coupled with 11 HPGe clovers and 15 LaBr$_3$ detectors to measure neutron and $\gamma$ rays with high energy resolution; at the second focal plane, the Total Absorption Spectrometer, either with MTAS or SuN, provides extremely efficient $\gamma$ detection. This unique, hybrid detection system of FDSi is designed for efficiently measuring a variety of radiations following $\beta$ decay, perfectly suited to the studies of $\beta$-strength function. Many exotic nuclei have been investigated at the FDSi, and the measured $\beta$ decay strength revealed crucial information about shell evolution far from the stability line, see , e.g., Refs.\ \cite{cox2024proton, peltier2025evidence}. More importantly, as the beam power at FRIB is being ramped up, many more short-lived isotopes further away from stability will be produced, and their detailed decay properties, including the $\beta$-strength function, will be unraveled in the near future. It will provide a massive amount of experimental data to benchmark nuclear theories aiming to describe $\beta$ decay accurately across the chart of nuclei.

On the theoretical side, an accurate description of $\beta$-strength functions requires a proper treatment of the weak interaction operators themselves. The Gamow-Teller transition operator is given in terms of the $J = 1$ transverse electric multipole of the charge-changing axial-vector current. The latter can be expanded consistently with the nuclear interaction within the framework of chiral effective field theory. Up to next-to-next-to-leading order (N$^2$LO) in such expansion, the Gamow-Teller operator includes not only the standard one-body 
operator, but also two-body current terms: a pion-exchange and a contact contribution. The importance of such effects was suggested by the so-called ``quenching problem''. In fact, shell model calculations using only the one-body operator systematically overestimate GT strengths and require an empirical reduction of the axial coupling constant $g_A$ by approximately 20--30\% to match experimental transition rates. \textit{Ab initio} calculations \cite{gysbers2019discrepancy,king2020chiral} demonstrated that for GT transitions in nuclei up to $^{100}$Sn the explicit inclusion of two-body currents largely resolves this discrepancy without the need for quenching, validating the microscopic origin of the effect. More recently, the effect of two-body currents has been explored also in B(GT) strength distributions. Data from the $\beta$-delayed spectroscopy of $^{24}$O have been recently compared to \textit{ab initio} employing the CC and VS-IMSRG methods and incorporating two-body currents \cite{neupane2024first}. Agreement with experimental data was found at low excitation energies but theory struggled to reproduce the high-energy tail of the strength function. This may be due to continuum coupling effects, which still remain a challenge for \textit{ab initio} frameworks. More recently, VS-IMSRG calculations \cite{li2025ab} explored B(GT) strength distributions in the astrophysically relevant $N=50$ region, finding that two-body currents reduce the low-lying strength and consequently increase the predicted half-life in agreement with the trend seen for GT transitions in Ref. \cite{gysbers2019discrepancy}. While the study of Ref.~\cite{li2025ab} focuses only on two chiral interactions, a broader exploration was undertaken in Ref.~\cite{brase2025two}. The correlations between B(M1) and B(GT) in $^{48}$Ca were investigated using a set of non-implausible $\Delta$-full interactions developed in Ref.~\cite{hu2021ab}. This study pointed out that the inclusion of two-body currents can lead to either an increase or a reduction in the total B(GT) strength depending on the underlying interaction. Such a finding highlights the need for a more systematic order-by-order study of B(GT) strength distributions where the nuclear potential and currents are treated consistently within chiral EFT, enabling proper quantification of theoretical uncertainties.

\subsection{Open Questions and Current Challenges}
Measurements of half-lives and $\beta$ decay strength distributions provide a crucial test of state-of-the-art nuclear theory. As an example, the longstanding quenching problem in GT transitions has been solved from first principles through the inclusion of two-body currents in \textit{ab initio} calculations. Building on this success, it is desirable to systematically investigate the effects of two-body currents on the full strength distribution. This would be best accomplished through an order-by-order approach within chiral EFT, treating the nuclear potential and currents on an equal footing to properly quantify theoretical uncertainties. Such systematic studies are particularly important because experiments in exotic nuclei such as those possible at FRIB predominantly measure half-lives, which can only be predicted theoretically through calculations of the complete $\beta$-strength function. Another open avenue for future work is understanding continuum coupling effects, which become increasingly important in regions such as the island of inversion, at the center of experimental investigations at FRIB. 
On a different note, it would also be useful to investigate more systematically the assumptions entering the analysis of $\beta$ decay experiments. For instance, $\beta$-delayed neutron emission is thought of as a two step process where the $\beta$ decay daughter is produced as a compound nucleus, an hypothesis whose limitations should be more closely scrutinized.

\section{Weak Decays in Astrophysics\label{sec:astro}}

\subsection{$\beta$ Decay in the Rapid Neutron-Capture Process}
The rapid neutron capture process (\textit{r}-process) is considered to be responsible for roughly half of the abundances of elements heavier than iron in the Universe \cite{BBFH, Cameron}. 
It is expected to occur in explosive neutron-rich astrophysical environments, such as compact binary mergers with at least one neutron star \cite{lattimer1974black, lattimer1976tidal} and certain types of core-collapse events of massive stars, e.g., collapsars \cite{popham1999hyperaccreting} or magnetorotational supernovae \cite{winteler2012magnetorotationally}. In recent years, further exotic scenarios involving compact objects have also been proposed as hosting conditions conducive to the $r$-process, for example magnetar giant flares \cite{patel2025rprocess} or accreting white dwarfs \cite{cheong2025gamma}.
The slowly decaying, long-wavelength observations accompanying the gravitational wave event GW170817 confirmed that the ejecta from binary neutron star mergers harbor conditions capable of producing $r$-process elements up to and possibly beyond lanthanides \cite{barnes2013effect, tanvir2017emergence}.

$\beta$ decay plays an essential role in the $r$-process, by transferring the material from isotopic chain with proton number $Z$ to $Z+1$, thereby building up a population of heavier elements. Especially, during the $(n, \gamma) \leftrightarrow (\gamma, n)$ equilibrium, the elemental abundance pattern (i.e. as a function of proton number, $Z$) is effectively determined by the half-lives of the most abundant isotopes in a given isotopic chain. These isotopes are called the ``waiting-point'' nuclei. Isotopes with a magic neutron number are often among these waiting-point nuclei, and due to the longer half-lives closer to stability along the magic neutron number, accumulation of abundance occurs, which results in the second and third $r$-process peaks at $A\sim130$ and 196 \cite{Horowitz2019,Cowan2021}. After $(n,\gamma)\leftrightarrow(\gamma, n)$ breaks down and material decays back to stability (freeze-out), competition between $\beta$ decays and neutron captures may affect the abundance pattern in the rare-earth region ($A\sim160$) \cite{Surman_rep_1997, Mumpower_rep_2012a}. Furthermore, in the early phase of the $r$-process, when the temperature is close to 10~GK, the finite-temperature effect may have an impact on the outcome of nucleosynthesis \cite{Minato2009, Ravlic2024, Langanke2001, Litvinova2020a, saito2025effects}.

Global theoretical predictions of $\beta$ decay properties play an essential role in $r$-process nucleosynthesis studies. Currently, notable available tabulations include the phenomenological Finite-Range Droplet Model (FRDM) + QRPA~\cite{Moller1997, Moller2003, Moller2019}, spherical relativistic energy density functional (EDF) theory + QRPA~\cite{Marketin_RQRPA_2016}, and the axially deformed Skyrme EDF theory + charge-exchanging Finite-Amplitude Method (pnFAM) \cite{Ney2020}. New tabulations based on the axially deformed relativistic EDF theory are in preparation by FRIB scientists. These new tabulations and Ref.~\cite{Ney2020} suggest that the $\beta$ decay rates, especially for the region past $N=126$, are significantly slower than previously calculated in Refs.~\cite{Moller1997, Moller2003, Moller2019, Marketin_RQRPA_2016}. Especially for heavier nuclei beyond $N=126$, this could have a significant implication on the contribution of fission or $\alpha$-decay to the nuclear heating profile of material decaying back to stability. Given that this heating powers the electromagnetic kilonova signal, it is essential to be able to predict these decay paths. Competition between different decay methods can significantly impact the magnitude and shape of the kilonova heating and light curve, as shown in Refs.~\cite{lund2023influence, kullmann2023impact, saito2025effects}. 

Experimentally, there are two overarching goals with respect to $\beta$ decay studies for the $r$-process. The first is to measure as many of the relevant decays as possible focusing on half lives and $\beta$-delayed neutron emission probabilities. These measurements are compared to the theoretical models mentioned in the previous paragraph. In addition, a better probe for  constraining theoretical models is the measurement of the full $\beta$ decay strength distribution. This is done using the TAS technique (Sec.~\ref{sec:tas}). Previous experiments with the SuN detector focused on TAS measurements of neutron-rich nuclei and comparisons to the models used in astrophysical calculations \cite{dombos2016total, spyrou2016strong, lyons2019decay, naqvi2022total}. The results have shown that generally these global models are not able to reproduce the various features observed in the experimental $\beta$ decay strength distributions. 

With future experiments at FRIB utilizing FDSi/FDS and SuN, stringent tests of these global $\beta$ decay models will be possible for neutron-rich heavy nuclei. It will be essential to test, constrain, and calibrate the global $\beta$ decay models with new experimental data, including half-lives, $\beta$-delayed neutron emission properties, and the $\beta$ decay strength functions. Experimentally informed theoretical $\beta$ decay models will also benefit the predictions of $\beta$-delayed neutron emission and fission properties, which are also essential ingredients for $r$-process studies.

\subsection{Stellar Weak-Interaction Rates}
Late stages of massive stars are sensitive to reaction processes mediated by the weak nuclear force. The most prominent one is electron capture (EC)~\cite{Langanke2021,Langanke2003}. In this process, the electron is captured by the nucleus, turning a proton into a neutron and emitting an electron neutrino. Therefore, EC directly impacts the $Y_e$, \textit{i.e.} electron-to-baryon ratio, lowering the number of available electrons in the system and making the matter more neutron-rich. The dynamics of the collapsing star is, in turn, heavily determined by $Y_e$ as the mass of the inner core is given by $M_{\rm inner} \sim Y_e^2$~\cite{Bethe1990}. However, EC occurs in the environment with very high temperatures (above 10~GK) and densities ($\rho > 10^9$ g/cm${}^3$), which can significantly impact the rate. As these conditions are certainly beyond the experimental reach, astrophysical simulations have to rely on theoretical estimates. The first EC rate tabulations were formulated by Fuller, Fowler, and Newman in a series of seminal papers~\cite{Fuller1980,Fuller1982a,Fuller1982b,Fuller1985}, based on the independent particle model, which recognized the importance of Gamow-Teller (GT) transitions in determining the total rate. It was later recognized \cite{Langanke2001b,Langanke2003b}, based on large-scale shell-model (LSSM) calculations, that finite-temperature effects can lead to significant enhancement of the EC rates through temperature unblocking. Despite having great agreement with experimental strength functions at zero-temperature~\cite{Cole2012}, due to computational limitations, LSSM calculations have been mostly applied to $pf$-shell nuclei~\cite{Langanke2001}. EC rates can be very efficiently calculated within the finite-temperature QRPA (FT-QRPA), either based on schematic models~\cite{Juodagalvis2010} or EDF theory~\cite{Paar2009,Niu2011a,Ravlic2020,Dzhioev2020,Giraud2022a}. The latter approach is especially promising since it offers a great extrapolation ability throughout the nuclide chart. However, as demonstrated in Ref.~\cite{Sullivan2016}, current supernova models rely on patches of theoretical calculations, each in a different mass region, or utilize simple analytical approximations. Only recently, large-scale calculations covering all nuclei of relevance from $Z = 20$ to $Z = 52$, have been performed based on a self-consistent model within the FT-QRPA and relativistic EDF~\cite{Ravlic2025a}. On the other hand, in the presupernova stages of a dying star, $\beta^-$ decay may compete with EC, until $\rho \sim 10^{10}$g/cm$^{3}$, above which $\beta^-$ decay becomes Pauli blocked~\cite{Martinez-Pinedo_2000}. Furthermore, also within the EDF+FT-QRPA framework, a study \cite{dasher2025enhanced} demonstrates significant enhancement of anti-neutrino spectra due to $\beta^-$ decays in collapsing stars.

Sensitivity studies were also performed \cite{Sullivan2016,Ravlic2025a}, and found that nuclei undergoing EC and driving the deleptonization are located around $N = 50$ and $N = 82$ shell-closures. Similar conclusions were found for $\beta^-$-decay \cite{dasher2025enhanced}, further motivating experimental campaigns. These sensitivity studies directly inform the experimental programs at FRIB, highlighting the most critical nuclei for measurement. In particular, FRIB’s unique capabilities enable precision measurements of charge-exchange strength functions for nuclei near these shell closures, providing stringent benchmarks for theoretical EC and $\beta$ decay rate predictions under astrophysical conditions. Comparisons between theoretical calculations and $B(\rm GT)$ data from charge-exchange measurements at FRIB, such as $(d,{}^2\text{He})$ in inverse kinematics~\cite{giraud2023beta+}, are now possible for many previously inaccessible isotopes. This tight synergy between FRIB experiments and theory accelerates progress in constraining weak interaction rates relevant to core-collapse supernovae and neutron-rich stellar environments.

\section{Determining $V_{ud}$ from Nuclear $\beta$ Decay for Tests of CKM Unitarity\label{sec:V_ud}}
\label{sec:ckm}
According to the Standard Model (SM), the flavor eigenstates of the three generations of quarks mix to form mass eigenstates through the Cabibbo-Kobayashi-Maskawa (CKM) matrix~\cite{cabibbo1963unitary,kobayashi1973cp}:
\begin{equation}
V_{\text{CKM}}=\left(\begin{array}{ccc}
V_{ud} & V_{us} & V_{ub}\\
V_{cd} & V_{cs} & V_{cb}\\
V_{td} & V_{ts} & V_{tb}
\end{array}\right)~.
\end{equation}
Its matrix elements appear in the coupling between quarks and the W-boson, and thus can be measured from charged weak decays of hadrons or nuclei. 

The CKM matrix is unitary according to the SM, which imposes specific relations between its matrix elements. As an example, its first-row matrix elements satisfy the following relation:
\begin{equation}
    |V_{ud}|^2+|V_{us}|^2+|V_{ub}|^2=1~.~\label{eq:1strow}
\end{equation}
This is a definite prediction of the SM which can be tested experimentally, and such tests impose very strong constraints on new physics models. For instance, assuming new physics occur at a high energy scale $\Lambda\gg M_W$, then a simple dimensional analysis suggests that testing Eq.\eqref{eq:1strow} to sub-permille precision probes the new physics scale at $\Lambda\sim (10^0-10^1)~\text{TeV}$, which is competitive to experiments at high-energy colliders. 

At sub-permille level, the very small $|V_{ub}|^2\sim 10^{-5}$~\cite{navas2024review} in Eq.\eqref{eq:1strow} can be dropped and one needs only to determine $|V_{ud}|^2$ and $|V_{us}|^2$. $|V_{us}|^2$ is obtained primarily from kaon decays, and less precisely from tau decays. Meanwhile, $|V_{ud}|$ can be obtained from $\beta$ decays of pion, free neutron and nuclei. In what follows, we will discuss the experimental and theoretical efforts required to precisely measure $|V_{ud}|$ from nuclear $\beta$ decays. 

\subsection{Experimental Efforts}
\label{sec:experimental.efforts}

The $|V_{ud}|$ matrix element can be extracted from two types of superallowed $\beta$ decay transitions: pure Fermi between $J^{\pi}=0^+$ states or mixed between mirror nuclei. The principal quantity from both types of transitions is the so-called corrected $\mathcal{F}t$ value:
\begin{equation}
\mathcal{F}t \equiv f_{V}t(1+\delta_{R}')(1+\delta_\mathrm{NS}-\delta_{C})\left(1 +\frac{f_A}{f_V}\rho^2\right) 
= \frac{K}{g_V^2 G_F^2 |V_{ud}|^2 |M_F^0|^2 (1+\Delta_R^V)},
\label{eq:master.formula}
\end{equation}
where the left-hand side includes $f_{V}t$ values determined from transition-dependent experimental quantities, such as $Q_{EC}$-values, that enters in the calculation of the statistical rate functions $f_V$ and $f_A$, half-lives, and branching ratios. The $\mathcal{F}t$ value also includes transition-dependent theoretical corrections such as the transition-dependent radiative correction $\delta_R’$, the nuclear structure correction $\delta_\mathrm{NS}$, and the isospin symmetry breaking correction $\delta_C$. While the Fermi-to-Gamow Teller mixing ratio $\rho$ is zero for pure Fermi transitions, it needs to be experimentally determined for mixed decays. The right-hand side of the equation includes fundamental quantities such as the $V_{ud}$ matrix element, the Fermi weak-interaction coupling constant $G_F$ (extracted from muon decay), a combinations of constants $K=2 \pi^3 \hbar^7 \ln(2)/(m_e^5 c^4)$, the Fermi matrix element between initial and final nuclear states in the isospin-symmetry limit $M_F^0 = \sqrt{2}$ for pure Fermi and 1 for mixed, and the transition-independent radiative correction $\Delta_R^V$. Additionally, $g_V=g_V(0)$ is the vector current form factor at zero momentum transfer and by setting $g_V = 1$, the conserved vector current hypothesis makes the right hand side of Eq.~\eqref{eq:master.formula} constant. Consequently, the left hand side must be transition-independent under the hypothesis of the Standard Model. If that is the case, one can take a weighted average of all $\mathcal{F}t$ values of a certain type of transition and extract $V_{ud}$ from Eq.~\eqref{eq:master.formula}.

While nuclear superallowed transitions suffer from the need for transition-dependent theoretical corrections not needed for neutron and pion decays, they have a statistical advantage as many nuclear superallowed transitions exist. Indeed, the fact that $V_{ud}$ can be extracted from a consistent ensemble of 15 transitions \cite{hardy2020superallowed}, currently makes the determination from superallowed pure Fermi transitions the most precise one.

\subsubsection{Superallowed Pure Fermi}
\label{sec:ckm:expt:superallowed.pure}

The latest evaluation \cite{hardy2020superallowed} of pure Fermi transitions includes 222 individual measurements resulting in 21 transitions with complete data sets. Of these, 15 transitions have a precision better than 0.3$\%$ and are used in the evaluation of the $V_{ud}$ matrix element. Since that evaluation, two more measurement have been published. The branching ratio of $^{62}$Ga \cite{MacLean2020BR62Ga} and the half-life of $^{14}$O \cite{Sharma2022halflife14O}. This small number of measurements and the comparatively larger number of publications on theoretical corrections (see Sections~\ref{section:ckm:section:isb_correction} and \ref{Sec:rad.corr}) is a clear indication that, for the most part, the existing experimental data is consistent and sufficiently precise. This at no point means that the case is closed. It only signifies that the future focus of experimental campaigns has shifted towards helping improve the reliability of the theoretical corrections (by testing them) rather than collecting data to improve the statistical precision of the data set. This is best illustrated by the recent charge radius measurement of $^{26m}$Al \cite{Plattner2023radius26mAl}, which is an important quantity that enters in the determination of the isospin symmetry breaking correction. The precise collinear laser spectroscopy measurement presented in \cite{Plattner2023radius26mAl} unveiled a 4.5$\sigma$ deviation in the charge radius of $^{26m}$Al resulting in a substantial shift by one standard deviation in the corrected $\mathcal{F}t$-value. In the future, more of such measurement, aimed at testing and improving calculation methods for the corrections are to be anticipated. For instance, the $\beta$-delayed proton decays between $T=2$ states, to be studied at TAMUTRAP \cite{Shidling2021Tamutrap}, will provide an alternate set of $0^+\rightarrow0^+$ data.  These near-proton-dripline cases will have vastly different experimental systematics and provide a demanding test of isospin-symmetry-breaking corrections ($\delta_C$). Finally, performing high-precision measurements of the branching ratio of $^{10}$C are of critical importance to assess the possible presence of BSM scalar currents once the various theoretical corrections are on a more solid footing.

\subsubsection{Superallowed Mixed}
\label{sec:superallowed.mixed}

The latest evaluation of mixed mirror decays \cite{Severijns2023} comprise a larger number of older independent measurements than the pure Fermi transitions. This recently triggered measurement of half-lives \cite{Long2022halflife13N,Shidling2018halflife21Na}, branching ratios \cite{Rebeiro2019BR19Ne}, and $Q_{EC}$-values \cite{Karthein2019Qvalue21Na}. Furthermore, because the Gamow Teller matrix element is non-zero in mixed decays, we also need to experimentally determine the mixing ratio $\rho$ in order to extract $V_{ud}$. This mixing ratio can be obtained from the measurement of one of three parameters: the $\beta-\nu$ angular correlation parameter $a_{\beta \nu}$, the $\beta$ asymmetry parameter $A_{\beta}$, or the neutrino asymmetry parameter $B_{\nu}$ \cite{naviliat2009test}. Because the determination of any of these parameter is challenging, there are currently only five mixed mirror decay from which $V_{ud}$ can be extracted at a reasonable level of precision \cite{naviliat2009test}. However, substantial enhancements are available through near-cancellation of the observable, exceeding that of the neutron (e.g.$^{17}$F) to up to a factor 13 ($^{19}$Ne) \cite{Hayen2020}. Several efforts to measure correlation parameters in mirror transitions are underway. These includes more precise angular correlation measurements ($a_{\beta\nu}$, $A_\beta$, recoil-asymmetry) of K and Rb isotopes with TRINAT \cite{Fenker2018TRINAT}, and the St.\ Benedict ion trapping system \cite{Brodeur2023StBenedict} at the University of Notre Dame that will be devoted to measuring $\beta$-$\nu$ angular correlations in multiple mirror transitions including the very sensitive $^{17}$F. Finally, the Superconducting Array for Low Energy Radiation (SALER) will aim at determining the $a_{\beta\nu}$ using superconducting tunnel junctions technology in various mixed mirror decays including $^{11}$C and $^{19}$Ne.




\subsection{Conserved Vector Current Verification}

Although the Conserved-Vector Current (CVC) hypothesis was introduced by Feynman, Gell-Mann, Sudarshan and Marshak~\cite{feynman1958theory,sudarshan1958chirality} in 1958, already five years before Sherr and Gerhardt noticed \cite{sherr1953experimental} referring to $^{10}$C and $^{14}$O cases that allowed favored $\beta $ decays, as they called them, offer a possibility for an experimental test of the Fermi theory. 
The framework was refined in the next decade when Cabibbo could explain the difference of $G_V$ obtained from the $\beta $ decay (a semileptonic process) and of $G_{\mu }$ obtained from a purely leptonic muon decay as due to the mixing of the first generation of quarks introducing the so-called Cabibbo angle $\theta $ with  $G_V = G_{\mu }\cos{\theta}$~\cite{cabibbo1963unitary}. After the formulation of the full CKM matrix formalism and replacement of $\cos{\theta}$ by $V_{ud}$~\cite{kobayashi1973cp}, the modern picture of 3-generation of quarks and their mixing matrix firmly set up in the basis of the Standard Model.
Since then, numerous experimental studies have been pursued to get high-precision data on nuclear superallowed $0^+ \to 0^+$ $\beta $-decay which provided the best testing grounds for the CVC hypothesis and, if the hypothesis is valid, the best grounds for an extraction of the $V_{ud}$ matrix element.

Through about five decades, a seminal work has been led by Hardy and Towner to evaluate the existing data on $0^+ \to 0^+$ $\beta$ decay halflives, branching ratios and $Q$-values and to analyze and apply theoretical corrections to the process \cite{hardy2020superallowed}. The master formula defining the so-called {\em absolute}, or corrected, $\mathcal{F}t$ value from experimentally extracted $ft$ values for these pure Fermi decays
is given by Eq.~\eqref{eq:master.formula} by setting $\rho = 0$.
The constancy of the $\mathcal{F}t$ values extracted from various transitions would confirm the CVC hypothesis. If confirmed, one can use the above relation to extract the $V_{ud}$ matrix element for the tests of the CKM matrix unitarity. 

As already stated,
in addition to the $0^+ \to 0^+$ $\beta$ decay, another possibility to test CVC
and extract $V_{ud}$ arises from mirror $T=1/2$ $\beta$ decays~\cite{naviliat2009test},
which are mixed transitions. To extract of the absolute $\mathcal{F}t$ value,
one has to determine the Fermi to Gamow-Teller mixing ratio 
($\rho \approx \lambda |M_{GT}^0|/|M_F^0|$ with $\lambda = g_A/g_V$) and use the expression in Eq.~\eqref{eq:master.formula}. Besides theoretical radiative and isospin-symmetry breaking corrections, an experimental challenge is to determine $\rho $ from a correlation measurement providing either the $\beta $-$\nu $ angular correlation parameter $a_{\beta \nu }$, the $\beta $-asymmetry parameter $A_{\beta }$, or the $\nu $-asymmetry parameter $B_{\nu }$~\cite{naviliat2009test}. The current status is summarized in Ref.~\cite{Severijns2023}.

Radiative and isospin-symmetry breaking corrections entered in the expressions of $\mathcal{F}t$  for mirror decays or $0^+ \to 0^+$ $\beta$ decays have to be evaluated theoretically with very high precision. The achievements up to date are summarized in the following sections.

\subsection{Isospin-Symmetry-Breaking Corrections\label{section:ckm:section:isb_correction}}

The isospin-symmetry breaking correction $\delta_C$ quantifies the deviation of the realistic Fermi matrix element squared form its model-independent (or isospin-symmetry limit) value:
\begin{equation}
|M_F|^2 = |M_F^0|^2 \left(1-\delta_C \right), 
\end{equation}
with $|M_F^0|^2=T(T+1) - T_{iz}T_{fz}$, where $T,T_{iz}$ and  $T,T_{fz}$ are the isospin and its third component of initial and final states.
The correction has to be evaluated within a many-body approach capable of describing isospin-symmetry breaking effects in nuclear states. There has been lots of work within different many-body approaches including shell-model, angular momentum and isospin projected density-functional theory, relativistic RPA and others (for a review, see Ref.~\cite{hardy2020superallowed} and references therein).

\begin{figure}[!h]
    \centering
    \includegraphics[width=0.75\linewidth]{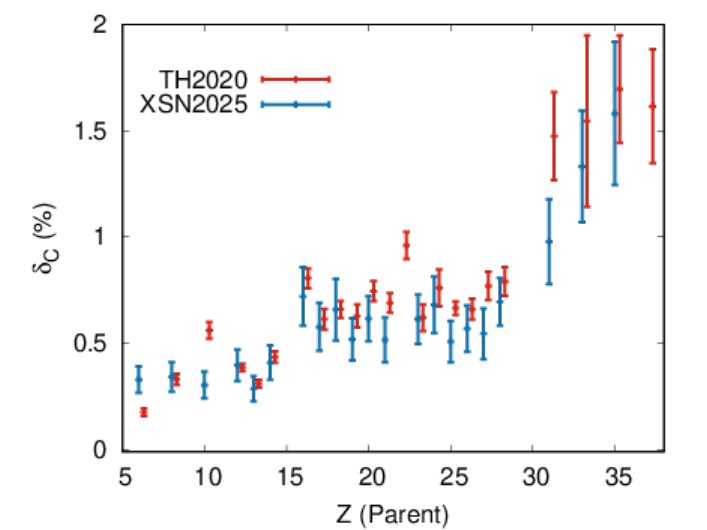}
    \caption{Shell-model plus WS calculations of $\delta_C$:
   ``TH2020'' is from Ref.~\cite{hardy2020superallowed},
    ``XSN2025'' is from Ref.~\cite{xayavong2025refined}.}
    \label{fig:delta_C}
\end{figure}

The phenomenological shell model proposes an experimentally constrained approach to the problem. Namely, the realistic Fermi  matrix element is calculated using an isospin-nonconserving shell-model interaction and realistic Woods-Saxon (WS) or Hartree-Fock (HF) wave functions. The isospin-symmetry breaking correction is typically evaluated in the lowest order approximation as a combination of two terms (see Ref.~\cite{xayavong2024higher} for justification):
\begin{equation}
\delta_C \approx \delta_{C1} +\delta_{C2}\label{eq:deltaCsplit}
\end{equation}
where the first term accounts for the deviation of one-body transition densities from their isospin-symmetry limit, while the second term appears due to the deviation of the realistic single-particle radial wave functions from unity.
To provide high accuracy, the isospin-nonconserving interactions are tuned to reproduce the splittings of the isobaric multiplets through the model space~\cite{smirnova2023isospin} or at least adjusted locally to isobaric multiplets involved in the transition~\cite{towner2008improved}. 
To get reliable radial wave functions, the WS potential parameters are thoroughly adjusted to reproduce proton and neutron separation energies within the parent and daughter nucleus, as well as a known charge radius. This largely removes ambiguity known from various WS parametrizations~\cite{hardy2020superallowed,xayavong2018radial,xayavong2025refined}.  The current status of $\delta_C$ obtained within this approach is shown in Fig.~\Ref{fig:delta_C}.

New experimental data on charge radii of parent and daughter nuclei, on the spectroscopic factors to cross-check the single-particle strength distribution in the intermediate states summation, as well as rates of competing non-analogue transitions are highly desirable to constrain theoretical models.

The implication of HF radial wave functions is still an open question due to the role of post-HF effects~\cite{xayavong2022radial}. More work in this direction is required.


Further efforts should be pursued to get reliable corrections for nuclei beyond $^{74}$Rb, since they are in the focus of experimental programs. From a theoretical point of view, the robust description of deformation in the region of $^{80}$Zr is vital.
Another future direction to pursue is to refine the calculation of isospin-symmetry breaking corrections for mirror transitions.

One should also be alerted that the aforementioned theory framework for $\delta_C$ is not without ambiguity. It was pointed out \cite{miller2008isospin,miller2009isospin} that the splitting in Eq.\eqref{eq:deltaCsplit} was not using the correct isospin operator, and correcting for this might lead to a
substantial reduction of the $\delta_C$ values. Instead, a perturbative expansion of $\delta_C$ with respect to the isospin-symmetry breaking interaction was proposed (the leading contribution is just the proton-proton Coulomb interaction). As we discuss below, this perturbative approach can be related to certain terms in the radiative corrections which appear at  $O(\alpha^2)$ \cite{plestid2025vertex}.
Further developments along this direction \cite{seng2023electroweak,seng2024toward} pointed out a more direct relation between $\delta_C$ and nuclear observables such as charge radii.   This is beneficial since experimental and theory improvements in the precise extraction of the latter will also benefit the former. In this sense, precision measurements of charge radii in superallowed candidate nuclei at FRIB can help constraining $\delta_C$. One particularly interesting nucleus is $^{26}\text{Si}$, because a measurement of its charge radius will directly probe the isospin breaking effect in the $A=26$ isotriplet, which is anticipated to be unusually large~\cite{ohayon2025critical}. 

Recently, a renewed interest in {\it ab initio} calculations of this quantity has emerged. Using GFMC, it is possible to compute the isospin raising and lowering operator directly for the lightest superallowed transition in $^{10}$C~\cite{piarulli2026quantum}. By evaluating the operator for wave functions whose imaginary time propagation was initialized with an explicitly isospin-breaking VMC wave function, it was possible to compute the result and compare with the value of $\sqrt{2}$ obtained in the isospin symmetric limit. The values that were obtained fell in the range $\delta_C \approx 0.15\% ~\text{--}~ 0.25 \% $, but with relative uncertainties ranging from $34\%$ to $65\%$. In this context, the different models were consistent with one another; however, when propagating to $V_{ud}$, this uncertainty on $\delta_C$ becomes the dominant theoretical uncertainty. Investigating potential improvements to the trial wave function to include more correlations could help to mitigate statistical uncertainty in the propagation in the future.

\subsection{Radiative Corrections in Superallowed $\beta$ Decay}
\label{Sec:rad.corr}

The precision needs of $\beta$ decay experiments, and in particular the superallowed transitions, place stringent demands on quantum electrodynamic (QED) corrections. A detailed accounting of one-loop effects, including precise calculations that incorporate nuclear structure, present a demanding task for the nuclear many-body community. Furthermore, numerically enhanced contributions at higher loop order require a detailed understanding of the general structure of electromagnetic corrections to $\beta$ decay.

Due to their stringent theory needs, we will focus on superallowed decays and the renormalization of the vector current. Conventionally the leading electromagnetic effects related to the static field of the nucleus are factorized from other electromagnetic effects and included in the ``$ft$'' value. The remaining radiative corrections then define $\mathcal{F}t\equiv (1+{\rm RC}) ft$, with the radiative corrections given as a product of terms, 
\begin{equation}
    (1+{\rm RC}) \equiv (1+\Delta_R^V) (1+\delta_{\rm NS} -\delta_C) (1+\delta_R')~. \label{eq:rc}
\end{equation}
We have included the isospin breaking correction $\delta_C$ since it arises (in part) from the Coulomb field and is related to electromagnetic corrections that appear at two-loop order. The term $(1+\Delta_R^V)$ accounts for short-distance corrections (momentum transfers above the hadronic scale). 
The term $(1+\delta_{\rm NS})$ encodes corrections that arise from nuclear structure. 
Finally, $(1+\delta_R')$ or the ``outer corrections'' contains long-distance effects. 
The precise definition of each of these terms varies in the literature, and so one must be careful to use consistent schema to avoid double counting. We discuss explicitly the conventions of the current algebra and EFT approaches below. 

Historically, these corrections were first developed at one-loop order in the Fermi theory \cite{sirlin1967general} and subsequently in the formalism of current algebra \cite{sirlin1978current}. We discuss the current algebra formalism below which allows for a one-loop definition of $\delta_{\rm NS}$ and $\Delta_R^V$. To facilitate progress at higher loop orders, it is helpful to use machinery from EFT to simplify computations, which we discuss next \cite{plestid2025manifest,plestid2025vertex,hill2024field,cirigliano2024ab}. The current algebra results can {\it always} be used to inform the one-loop analysis, even when using the EFT formalism. 

\subsubsection{Current Algebra Approach}\label{section:ckm:section:currentalgebra}

Sirlin developed the current algebra approach for the radiative corrections (RC) to $\beta$ decays at order $\mathcal{O}(\alpha/\pi)$ \cite{sirlin2013radiative}.
It was originally designed for superallowed $\beta$ decays, but later generalized to deal with RC to generic semileptonic decays \cite{seng2021radiative}.
A detailed account of the theory can be found in Ref.~\cite{seng2021radiative}.
Here we summarize a few important points.
First, loop diagrams involving two or more heavy gauge boson propagators are only sensitive to physics at the scale $q\sim m_W$ and thus can be computed perturbatively. Consequently, the only two one-loop diagrams at $\mathcal{O}(\alpha/\pi)$ that depend on details of the strong interaction in the non-perturbative regime are: (1) the photon-loop correction to the hadronic/nuclear form factor, which includes both the wavefunction renormalization and the vertex correction, and (2) the $\gamma W$-box diagram. These diagrams contain infrared divergences that have to be canceled out by tree-level diagrams with the emission of an extra photon. The latter, however, is not sensitive to the details of hadronic and nuclear structures of the decaying nucleus due to the small decaying phase space.


%

For superallowed $0^+\rightarrow 0^+$ nuclear $\beta$ decays, the structure-dependent part from the form factor correction cancels out with that from the vector $\gamma W$-box diagram, and gives rise to analytically-calculable pieces that are identified as the Fermi function~\cite{fermi1934attempt} and the Sirlin function~\cite{sirlin1967general}. The remaining piece in the radiative corrections that is sensitive to hadron and nuclear structure resides in the axial $\gamma W$-box diagram, which we denote as $\Box_{\gamma W}$. The dependence on strong-interaction physics is entirely contained in the following tensor:
\begin{equation}\label{2-point-function}
    T^{\mu\nu}=\int d^4 x e^{iq\cdot x}\langle f|T\{J_\text{em}^\mu(x)J_{W,A}^\nu(0)\}|i\rangle~,
\end{equation}
where $J_\text{em}^\mu$ and $J_{W,A}^\nu$ are the electromagnetic and axial charged weak current, and $\{i,f\}$ are the initial and final nuclear states.  $\Box_{\gamma W}$ is then given by the $q$ integral of the above tensor, weighted by $q$-dependent factors arising from the electron, photon, and $W$ propagators.
Applying this formula to the free neutron decay provides a one-loop definition of $\Delta_R^V$: 
\begin{equation}
    \Delta_R^V=2\Box_{\gamma W}^n+\dots~,
\end{equation}
where ``$+\dots$'' denotes other contributions to $\Delta_R^V$ that depend only on physics at the scale $q\sim m_W$ and are perturbatively calculable. Meanwhile, the nuclear-structure dependent correction $\delta_\text{NS}$ simply arises from the difference between the nuclear and nucleon axial box diagrams:
\begin{equation}
\delta_\text{NS}=2(\Box_{\gamma W}^\text{nucl}-\Box_{\gamma W}^n)~.
\end{equation}
Thus, in the current-algebra formalism the theoretical prediction of $\delta_\text{NS}$ requires a reliable modeling of all the ingredients that enter the nuclear tensor $T^{\mu\nu}$, including the intermediate nuclear states and the electromagnetic and weak currents used in the calculation.
A brief discussion and comprehensive list of references regarding the nuclear
modeling performed in extraction of $ \delta_{\mathrm{NS}} $ for the $ \carbonfermidecay $ are given in Ref.~\cite{gennari2025ab} (Sec.~\ref{section:theory:ncsm}).


Notice that the $\mathcal{O}(\alpha/\pi)$ analysis above does not include $\delta_C$, because the latter arises starting from the second order in the isospin-symmetry breaking interaction in accord with the (generalized) Behrends-Sirlin-Ademollo-Gatto theorem~\cite{behrends1960effect,ademollo1964nonrenormalization}.

We also note that Eq.~\eqref{2-point-function} involves only two current insertions. This is surprising since, on general grounds, the one-loop radiative corrections will involve vertex corrections to the nuclear vertex, which necessarily involve one weak current and two electromagnetic currents. These ``three-point'' contributions can be related, in the limit of isospin conservation and zero-recoil, to the two-point function with vector charged weak current \cite{sirlin1978current}, which results in the aforementioned cancellation. This is why all $O(\alpha)$ radiative corrections can be expressed using Eq.~\eqref{2-point-function}. As we discuss below, the vertex corrections at two-loop order, specifically those at $O(Z^2\alpha^2)$, can be identified with $\delta_C$ \cite{plestid2025vertex}.  

\subsubsection{Effective Field Theory Approaches}
The current algebra formalism is complete through $O(\alpha)$; however, the precision needs of $\beta$ decays demand two- and three-loop input \cite{hardy2020superallowed}. It becomes difficult, if not impossible, to track such high-loop-order effects without separating scales for a hadronic process like $\beta$ decay that receives contributions from scales of order the $W$-mass all the way down to the electron mass. 

This has motivated a re-framing of the problem using effective field theory (EFT) formalism. At $O(\alpha)$ this must reproduce the result of the current algebra formalism, as well as other $O(\alpha)$ effects such as the Fermi function. At higher loop orders, however, the EFT formalism allows one to work with one scale at a time in terms of a tower of EFTs each of which is tailored to describe the physics of a particular length scale \cite{cirigliano2023effective,hill2024all,hill2024field,cirigliano2024radiative}. 

At the shortest distances one can work in terms of Standard Model degrees of freedom: propagating $W$-bosons, quarks, and gluons. One then matches between this theory and a 4-Fermi effective theory with $n_f=5$ dynamical quarks, sometimes called low-energy effective theory (LEFT) in the literature~\cite{Jenkins:2017jig}, where the $W$, $Z$, Higgs, and top-quark are all integrated out. This matching can be computed through higher loop order using both $\alpha$ and the strong-coupling $\alpha_s$ as an expansion parameter \cite{Sirlin:1981ie,hill2020effective,Dekens:2019ept,moretti2025beyond}. Renormalization group equations are then solved to ``run'' the couplings down to lower scales. Heavy quarks are integrated out of the theory at their mass thresholds and a similar matching and running is performed in LEFT with $n_f=4$ and $n_f=3$ dynamical quarks. 

After this matching and running has been performed, one is left with a 4-Fermi effective theory involving only $u,d,s$ quarks, leptons, and photons \cite{czarnecki2004precision, czarnecki2019radiative,hill2020effective}. It is convenient to work in $\overline{\rm MS}$ and work at $\mu\simeq 2~{\rm GeV}$. The Wilson coefficients multiplying the 4-Fermi contact operators contain all of the physics from quantum fluctuations with virtuality larger than $\sim 2~{\rm GeV}$. 

Next, one must face the realities of hadronic physics and relate the above Lagrangian to the necessary matrix elements between nucleons. Currently, this is done using a data-driven dispersive formalism \cite{seng2018reduced,seng2019dispersive,seng2020joint,shiells2021electroweak}. However, there has been rapid progress towards a first-principles lattice QCD calculation \cite{feng2020first,yoo2023electroweak,ma2024lattice}. One then defines a new effective Lagrangian, known as heavy baryon chiral perturbation theory (HB$\chi$PT), involving neutrons, protons, and light mesons as the relevant degrees of freedom.

The form of the interactions in this theory are constrained by the symmetries of QCD, while their Wilson coefficients (often called low-energy constants in the $\chi$PT literature) are fixed by
demanding that the matrix elements computed in the theory with nucleon fields match those computed using the theory with quark and gluon fields \cite{cirigliano2023effective,Cirigliano:2024nfi}. At this stage, the single-nucleon short-distance radiative correction is obtained and the theory can be used to make predictions for neutron decay \cite{Cirigliano:2022hob,Ando:2004rk}.

To move beyond one-nucleon processes, one must relate the nucleon-level Lagrangian to nuclear matrix elements. The formalism for this matching has been initiated in Ref.~\cite{cirigliano2024ab}, and is already making contact with many-body methods \cite{king2025quantum}. 
The radiative corrections within HB$\chi$PT can again be organized by the photon momentum. In the language of Refs.~\cite{cirigliano2024ab,cirigliano2024radiative}, there are several momentum regions that contribute: {\em i)} photons with `hard' momenta, $q\sim$ GeV, are responsible for the short-distance corrections to neutron decay and lead to contact operators involving two nucleons whose low-energy constants are currently unknown; {\em ii)} `potential' photons with $|\vec q|\sim m_\pi\gg q_0$ contribute through diagrams in which the electron exchanges a photon with a second nucleon leading to so-called potentials or two-body currents;
{\em iii)} `ultrasoft' photons with $|\vec q|\sim q_0$ of the order of the $Q$ value, which induce the outer corrections and the  Fermi function, discussed in more detail below. Within this framework, the nuclear-structure dependent correction, $\delta_{NS}$, is then given by the nuclear matrix elements of the two-body currents generated by the hard and potential photons, which scale as either $O(\alpha {m_\pi}/{m_N})$ or $O(\alpha {Q}/{m_\pi})$. The first calculations of these have been performed for $^{10}$C and $^{14}$O \cite{cirigliano2024ab,king2025quantum}, where the largest uncertainty arises from the coupling constants of the contact operators that are currently poorly known.


Before descending to lower energies, let us comment on the isospin breaking correction, $\delta_C$. Recent work has explored how to construct the LSZ reduction formula for low-energy processes involving bound states for which superallowed $\beta$ decay is a prime example. It can be shown in this formalism that the isospin breaking correction (as formalized in Refs.~\cite{miller2008isospin,miller2009isospin}) arises at $O(Z^2\alpha^2)$ from vertex correction diagrams. Thus, $(1+\delta_{\rm NS}-\delta_C )$  should be thought of as parameterizing $O(\alpha)$, $O(Z^2\alpha^2)$, and other higher order radiative corrections that are sensitive to nuclear scales. As we will now discuss, it can be interpreted as corrections to a Wilson coefficient in a further low-energy effective theory. 

Finally, let us descend to the energy scales that characterize the on-shell electron emitted in $\beta$ decay $\sim 5 ~{\rm MeV}$. In this theory, we must explicitly incorporate the electron's kinematics because the photon virtuality is comparable to the electron energy and mass, and so loop corrections depend on the electron momentum, $p_e$, with no nuclear suppression (i.e., they are not suppressed by $(p_e R)^n$ for $n\geq1$ and $R$ the nuclear radius). 

Having defined higher order currents and matrix elements at nuclear-scale we may match onto the final low-energy theory discussed above. This theory treats the single-particle nuclear states as point-like infinitely heavy particles whose weak and electromagnetic interactions are encoded in a set of effective operators with their own Wilson coefficients. At leading power the theory for superallowed decays is very simple, depending only on a single operator \cite{hill2024field}, 
\begin{equation}
    \mathcal{L} \supset -C_V(\mu) h_v^\dagger v_\mu \tau^+ h_v \bar{e} \gamma^\mu (1-\gamma_5) \nu. 
\end{equation}
Within this final effective theory long-distance corrections (traditionally denoted as $1+\delta_R'$) may now be computed as low-energy matrix elements $\mathcal{M}$. The decay rate is then written as $\Gamma \sim |C_V|^2 |\mathcal{M}|^2$ where $|C_V|^2 \propto [1+\tilde{\Delta}_{V}(\mu)]\times[1+ \tilde{\delta}_{\rm NS}(\mu)]$ and $|\mathcal{M}|^2 \propto [1+ \tilde{\delta}_R'(\mu)]$. 

Putting everything together, we arrive at a factorization theorem \cite{hill2024field}, 
\begin{equation}
    \label{factorization-theorem}
    \mathcal{F}t = ft \bigg[1+\tilde{\Delta}_{V}(\mu) \bigg]\times \bigg[1+ \tilde{\delta}_{\rm NS}(\mu) \bigg] \times \bigg[ 1+ \tilde{\delta}_R'(\mu) \bigg] + O(\alpha \times  QR) ~,
\end{equation}
where $\tilde{\delta}_{\rm NS}$ is the analog of $\delta_{\rm NS}-\delta_C$. Note that all ladder graphs of order $(Z\alpha)^n$ are resummed in the $ft$ value; $O(Z^n\alpha^n)$ remain in $\delta_C$. We have allowed for $O(\alpha QR)$ power corrections, which can also be computed if needed at higher orders; at $O(Z^n\alpha^n)$ they are contained in the $ft$ value while at $O(\alpha)$ they can be obtained from the current algebra formalism and correspond to matrix elements of two-body operators within chiral EFT discussed above. We focus on the form of the factorization theorem at leading-power for simplicity, and because these are the contributions which need higher order perturbative input to match the $10^{-4}$ precision goals of superallowed $\beta$ decays. 

In this formula we have assumed the nuclei to be fully ionized, such that no ``atomic scale'' exists in the problem. In nature atomic orbitals introduce a further infrared scale which induces corrections at $O(10^{-3})$. Current theory treats these corrections phenomenologically with a screened Coulomb potential~\cite{rose1936note,garrett1967potential,salvat1987analytical}, the effect of which is included in $f$. The effective field theory techniques of the lowest-energy theory can be lifted to analyze this portion of the corrections, but this piece of formalism remains to be completed. 

Equation \eqref{factorization-theorem} looks very similar to the traditional factorization formula, Eq.~\eqref{eq:rc}, but the quantities with tildes are defined each in their own respective EFTs and satisfy well defined renormalization group equations (see Ref.~\cite{cirigliano2024ab,cirigliano2024radiative} for a related discussion). Notably, the Fermi function is a renormalization scale-dependent (i.e., ``running'') object that emerges naturally in the point-like effective theory \cite{hill2024all,hill2024field}. They are furthermore separately gauge invariant. For this reason one can work at different orders in perturbation theory, which is pragmatic when there are well known numerical enhancements from certain logarithms, and/or coherent enhancements proportional to $Z^n$ with $Z$ the charge of the nucleus. We now turn to a discussion of some of these higher order corrections. 

\subsubsection{Higher-Order Corrections}
The $O(\alpha)$ radiative corrections demand nuclear matrix elements at $\sim 10\%$ precision (since, using the scaling mentioned above, $(\alpha {m_\pi}/{m_N})\times (0.10) \sim 10^{-4}$ matches the precision target of superallowed decays). By way of contrast, higher order radiative corrections do not demand the same level of scrutiny for nuclear matrix elements. Instead, the challenge is a proper organization of perturbation theory in $\alpha$ and $Z\alpha$ that avoids any double counting, and allows for the systematic identification of numerical enhancements. In what follows we sketch recent progress on each of the three terms in the factorization theorem of Eq.~\eqref{factorization-theorem}. 

First, let us break $1+\tilde{\Delta}_R$ into two pieces: {\it i)} a finite matching at the hadronic scale $\mu \sim 2~{\rm GeV}$; and {\it ii)} logarithmic enhancements from renormalization group evolution. We will attack these pieces in turn, requiring higher-order terms in the renormalization group equations than for the finite matching computation. The renormalization group evolution down to $\mu\sim 2~{\rm GeV}$, up to $O(\alpha^2)$ and $O(\alpha_s\alpha)$ is discussed in Refs.~\cite{czarnecki2004precision,czarnecki2019radiative,cirigliano2023effective} with new state of the art input at $O(\alpha_s^2\alpha)$ having been recently computed in \cite{moretti2025beyond}. A matching calculation using the four-Fermi theory with quarks and gluons onto the hadronic theory can be performed with lattice methods \cite{ma2024lattice}, or (as is currently done) with a semi-phenomenological approach using dispersive methods \cite{cirigliano2023scrutinizing}.  

Next let us work at the nuclear scale and discuss $1+\tilde{\delta}_{NS}$. Most of the effort here is focused on the nuclear matrix elements required for the $O(\alpha)$ correction; these are discussed in Sec.~\ref{sec:ckm.c10}. There must also be $O(Z\alpha^2)$ corrections to $\tilde{\delta}_{NS}$ (this can be seen by demanding consistency with matrix elements computed in the point-like theory for $\tilde{\delta}_R'$ \cite{cao2025z,crosas2026radiative}), and these have not yet been fully formulated in the literature. One explicit contribution that has been identified is contained in the $O(Z\alpha^2)$ contributions to the potentials in Ref.~\cite{cirigliano2024ab,cirigliano2024radiative}.
Further work is needed to fully characterize the $O(Z\alpha^2)$ piece of $\tilde{\delta}_{NS}$. 

Finally, at low energies there has been substantial progress on the formulation of $1+\tilde{\delta}_R'$ in terms of a point-like effective theory. First, by defining all objects (including the Fermi function) in a point-like theory, double counting issues are completely removed and the formalism has been made systematic \cite{plestid2024generalized,hill2024field,hill2024all}. Second, since the corrections are now formalized as effective field theory matrix elements, renormalization group improvements can be made to the perturbation theory catching higher-order logarithmically enhanced terms \cite{borah2024renormalization}. Finally, a number of new coefficients that had never been computed before, have recently been obtained in the literature \cite{borah2024renormalization,cao2025z,crosas2026radiative}, substantially improving our control over long-distance radiative corrections. 

To briefly summarize recent developments: {\it 1)} The Fermi function and its relation to an all-orders resummation of Feynman diagrams was developed in Refs.~\cite{hill2024all,hill2024field}. {\it 2)} The anomalous dimension of $C_V(\mu)$ was computed at $O(Z^2\alpha^3)$ in Ref.~\cite{borah2024renormalization} and a symmetry identified which fixes all coefficients in terms of results in the literature up to (and including) $O(Z^3\alpha^4)$. {\it 3)} The $Z\alpha^2$ correction (both for $0\gamma$-emitted and $1\gamma$-emitted contributions) was recently obtained by two independent groups. The authors of  Ref.~\cite{cao2025z} extracted the result using factorization techniques \cite{griend2025fermi,cao2025factorization}, while the authors of Ref.~\cite{crosas2026radiative} performed an independent calculation; both results agree. Finally, {\it 4)} A renormalization group improved perturbation theory, including constraints from the Kinoshita-Lee-Nauenberg theorem, has been constructed with theoretical error estimates at the level of $\sim 10^{-4}$ \cite{cao2025z}. These more recent calculations supplant previous work (by Jause and Rasche \cite{Jaus:1970tah} and Sirlin and Zucchini \cite{Sirlin:1986cc}) and can be seamlessly combined with minimal subtraction nuclear matrix elements defined in the formalism of Ref.~\cite{cirigliano2024ab}. 

So far the discussion has been on the corrections that appear at leading-power which require high orders in perturbation theory. Let us briefly comment on further refinements that remain outstanding at sub-leading power. First, the point-like effective theory should be extended to include power corrections; some of the necessary terms can already be read off from the analysis of Refs.~\cite{cirigliano2024ab,cirigliano2024radiative}. Second, an explicit matching calculation between the ``chiral-potentials'' formalism and the long-distance point-like EFT remains to be performed. Finally, a merging of the effective field theory approaches described herein with the traditional formalism for $ft$ values should be fully pursued.

\subsubsection{Nuclear Structure Correction ($\delta_{\rm NS}$) for $^{10}$C}\label{sec:ckm.c10}

   \begin{figure}[t!]
        \center\includegraphics[width=0.9\textwidth]{{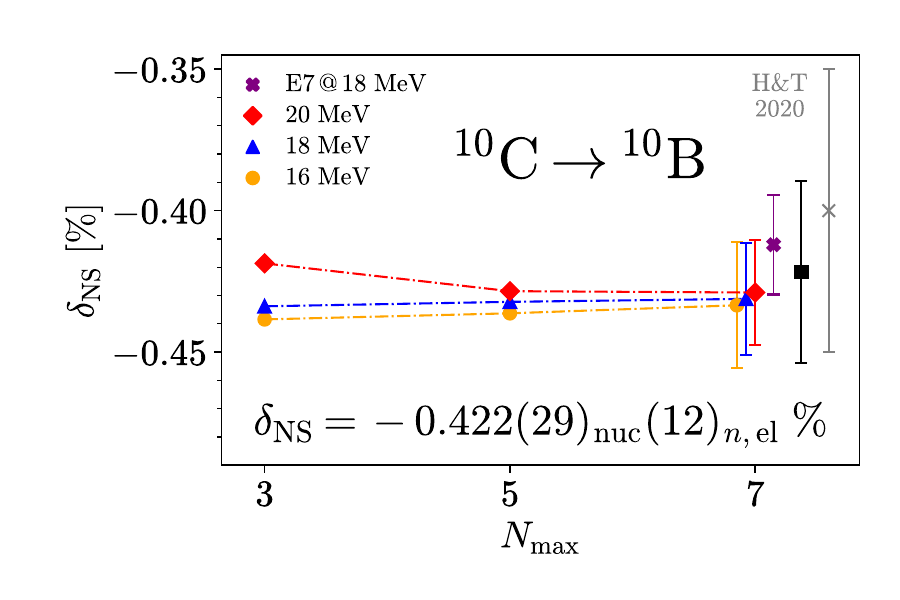}}
        \caption{\label{section:theory:figure:NCSM_deltaNS_convergence_C10_to_B10}
        Evaluations of $\delta_{\text{NS}}$ in the NCSM for the $ \carbonfermidecay $ with $ N_{ \text{max} } = 3,\, 5,\, 7 $ configuration space truncations and, for the $ \emninteraction $, oscillator frequencies in the range of $ \hbar \Omega = 16 - 20 \ \mathrm{MeV} $. For the $E_{7}$ point generated with the $ \emninteractionstar $ interaction, we use a sole frequency of $ \hbar \Omega = 18 \ \mathrm{MeV} $. The black square and corresponding error bar indicate the value of $ \delta_{ \text{NS} } $ as extracted from the set of NCSM evaluations as described in Ref.~\cite{gennari2025ab}.}
    \end{figure}
    
A first {\it ab initio} calculation of the nuclear structure radiative correction in the dispersive approach~\cite{seng2021radiative} was performed by Gennari et al. \cite{gennari2025ab} using the NCSM (Fig.~\ref{section:theory:figure:NCSM_deltaNS_convergence_C10_to_B10}). The NCSM evaluations were performed over a sequence of $ N_{ \text{max} } = 3,\, 5,\, 7 $ configuration space truncations and oscillator frequencies $ \hbar \Omega = 16 - 20 \ \mathrm{MeV} $, for two different chiral interactions $ \emninteraction $ and $ \emninteractionstar $. The important feature to note is that, with increasing configuration space dimension, the dispersion over the oscillator frequency of the NCSM evaluations for a given interaction tightens; the evaluations approach the frequency-independent, infinite-basis result, producing a value of 
\begin{equation}
    \delta_{\rm NS} = -4.22(31) \times 10^{-3} \, ,
\end{equation}
where the uncertainty represents estimates of truncation error in the model space and multipole decomposition truncations used in the NCSM calculations, the specific choice of harmonic oscillator frequency in the many-body calculation, the difference resulting from two choices of the nuclear interaction, and the modeling of higher-energy single nucleon contributions to the correction. 

The nuclear-structure dependent corrections have recently also been evaluated using QMC, following
 the Chiral EFT approach to $\delta_{\rm NS}$ of Refs.~\cite{cirigliano2024ab,cirigliano2024radiative}. The extracted value for $\delta_{\rm NS}$ in this approach is~\cite{king2025quantum},
\begin{equation}
    \delta_{\rm NS}^{(0)} = -[ 4.46 (48) - 4.64(77) ] \times 10^{-3} \, ,\qquad
    \overline{\delta_{\rm NS}^{E}} = (0.97-1.17 ) \times 10^{-3} \, ,
\end{equation}
where $\delta_{\rm NS}=  \delta_{\rm NS}^{(0)} + \overline{\delta_{\rm NS}^{E}}$, and $\delta_{\rm NS}^{(0)}$ is $O(\alpha m_\pi/m_N)$ and arises from electron-energy independent interactions, while  $\overline{\delta_{\rm NS}^{E}}$ is energy-dependent 
and appears at $O(\alpha Q/m_\pi)$ and would be counted as a power correction in Eq.\ \eqref{factorization-theorem}
(the former is closer to $\delta_{NS}$ in the traditional approach, while the latter is traditionally collected in so-called finite-size and shape corrections).
The range of calculations represents the spread over four different models adopted for the nuclear interaction, and the uncertainty accounts for undetermined low-energy constants in the EFT.

Although these two approaches represent a major advance in {\it ab initio} computations of nuclear structure corrections to $\beta$ decay, nuclear interaction uncertainties are at present hard to quantify. As detailed above, the approaches take a subset of different chiral EFT models to perform the calculation; however, the correlations between these calculations are not well quantified, nor is it clear if the full parameter space for the interactions is being explored. To provide insight into whether there truly is a tension between the Standard Model and experimental data, it will be necessary to provide theoretical calculations with well-quantified uncertainties. In recent years, the nuclear physics community has made a number of advances in the quantification of nuclear uncertainties~\cite{bub2024bayesian,wesolowski2021rigorous,hu2021ab,somasundaram2023maximally}, as well as in constructing emulators of observable quantities~\cite{konig2019eigenvector,odell2023rose,becker2023ab,somasundaram2024emulators,armstrong2025emulators}. The radiative corrections to $^{10}$C superallowed $\beta$ decay represent an opportunity to apply this rigorous uncertainty quantification to a problem that calls for the best possible error estimates from theory. In the future, the fundamental symmetries community will benefit from a concerted effort to perform robust uncertainty quantification of nuclear structure corrections. Further, it would be useful to develop approaches to benchmark different many-body methods using the same nuclear interaction. Progress in developing soft local chiral interactions~\cite{somasundaram2023maximally} and ways to incorporate non-local terms perturbatively in QMC~\cite{curry2024perturbative} represent progress toward performing such a benchmark in the future.

\subsubsection{Nuclear Structure Correction ($\delta_{\rm NS}$) for $^{14}$O}

Similar to $^{10}$C, an EFT approach was followed to compute the nuclear structure radiative correction in $^{14}$O \cite{cirigliano2024ab}. The calculation of $\delta_{\rm NS}$ was performed with one model of the nuclear interaction~\cite{lynn2015chiral}, and the value extracted in this approach is~\footnote{The quoted number for $ \delta_{\rm NS}^{(0)} $ differs from that of \cite{cirigliano2024ab} due to a mistake in the spin-orbit potential of that reference, see \cite{king2025quantum} for more details.
},
\begin{equation}
   \delta_{\rm NS}^{(0)} = -2.84(88) \times 10^{-3} \, ,\qquad
    \overline{\delta_{\rm NS}^{E}} = 2.06(41)  \times 10^{-3} \, ,
\end{equation}
 where again the uncertainty in $\delta_{\rm NS}^{(0)} $ accounts for undetermined low-energy constants in the EFT and that in $\overline{\delta_{\rm NS}^{E}}$ is due to missing higher-order terms in the EFT. Determining the low-energy constant --- either through lattice QCD, theoretical modeling as done for similar constants in $0\nu\beta\beta$ \cite{cirigliano2021determining}, or by fitting these constants to the $0^+\to0^+$ transitions --- would bring down the error of $\delta_{\rm NS}$ within a given nuclear interaction model when using this approach. It would be useful to have calculations performed in both dispersive and EFT approaches for this nucleus with many different many-body methods to help estimate the current uncertainties arising from the nuclear theory input to extracting $V_{ud}$.

\subsubsection{A Strategy for $^{26m}$Al}
\label{Sec:rad.corr:26mal}

\begin{figure}[tbp]
    \centering
    \includegraphics[width=0.50\linewidth]{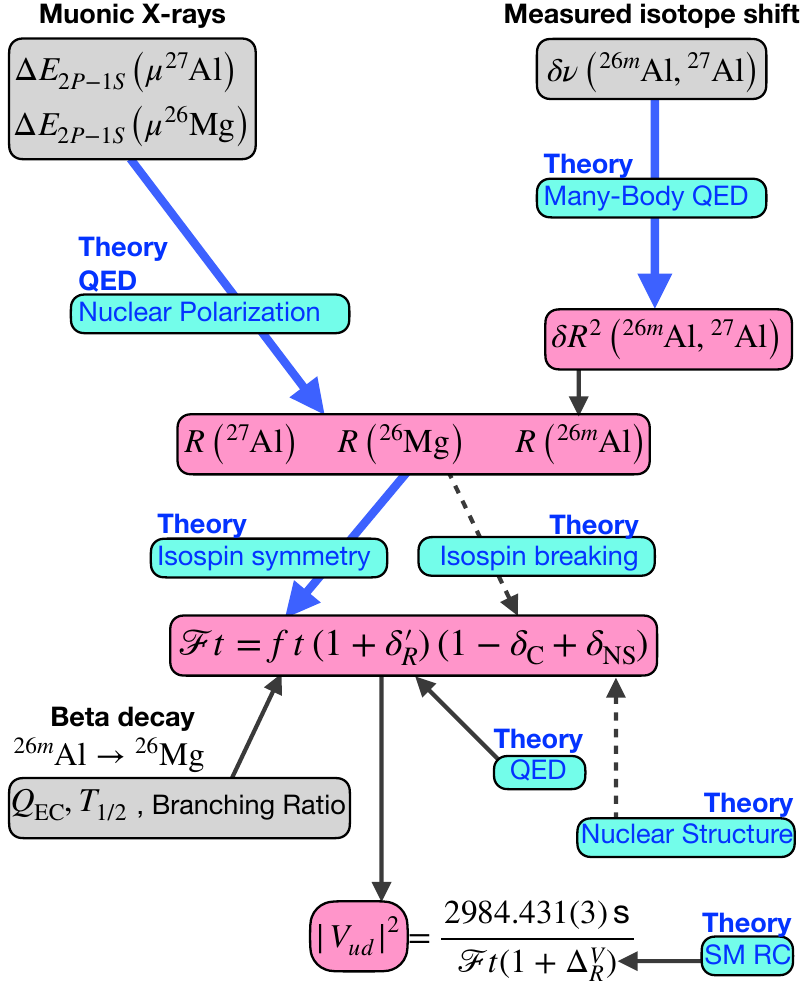}
    \caption{Schematic representation of the approach taken in this work to connect the measurements of the nuclear $^{26m}$Al$\,\rightarrow^{26}$Mg beta-transition to the atomic transitions used to determine the nuclear charge radii and to $V_{ud}$. A detailed explanation of all ingredients is given in the main text.
    }
    \label{fig:Scheme}
\end{figure}
$V_{ud}$ from superallowed $\beta$ decay is usually extracted by averaging
the 15 best measured transitions~\cite{hardy2020superallowed}. However, 
it was shown \cite{gorchtein2025robust} that the same precision can be achieved by a single transition $^{26m}\text{Al}\rightarrow{}^{26}\text{Mg}$, due to fact that its lifetime, branching ratio and $Q_\text{EC}$ value are most precisely measured. Therefore, comparing $V_{ud}$ extracted from this single transition versus the average of the rest provides a test of CVC. For that purpose, we need all the theory inputs to this transition to be well under control (Fig.\ref{fig:Scheme}). This includes the following: 
\begin{enumerate}
    \item[(i)] The statistical rate function $f$, which encodes the finite-size effect of the nucleus at tree level. This requires the precise knowledge of the nuclear charge distributions and (charged) weak distribution, the latter can be obtained from the former through isospin symmetry~\cite{seng2023model,seng2024data}. For that we need the information of the nuclear charge radii of $^{26}\text{Mg}$ and $^{26m}\text{Al}$ to high precision. The former is obtained from energy levels of the muonic atom $\mu^{26}\text{Mg}$, account for corrections from nuclear polarizability. For the latter, one starts with the $^{27}\text{Al}$ charge radius obtained from muonic atom + nuclear polarizability, and with an extra measurement of the $({}^{26m}\text{Al},{}^{27}\text{Al})$ isotope shift, plus atomic calculation of the mass and field shift. 
    \item[(ii)] The ``outer'' radiative correction $\delta_{R}'$ that involves elementary QED calculation.
    \item[(iii)] The nuclear-structure dependent ``inner'' radiative correction $\delta_\text{NS}$ that requires ab-initio nuclear many-body calculation.
    \item[(iv)] The isospin-breaking correction $\delta_{C}$ to the Fermi matrix element that again requires ab-initio calculation.
\end{enumerate}

The following section details efforts toward one of the aforementioned points; namely, the evaluation of $\delta_{\rm NS}$ in heavier systems.

\subsubsection{Towards Heavy Systems and Systematic $V_{ud}$ Extraction}\label{sec:ckm_heavy}

The developments outlined in the preceding subsections establish a consistent construction of the radiative and structure-dependent electroweak operators required for precision superallowed $\beta$ decay calculations. The next natural step is to embed these operators in many-body frameworks capable of treating the full range of candidate emitters with controlled uncertainties. To date, applications have largely focused either on phenomenological shell-model calculations for medium-mass superallowed transitions~\cite{barker1992determination, towner1992nuclear, towner1994quenching, towner2002calculated, towner2008improved} or on ab initio treatments of very light systems, where quasi-exact results can be obtained~\cite{cirigliano2024ab, king2025quantum, gennari2025ab}. Extending these efforts to the medium-mass region with systematically improvable ab initio methods, such as coupled-cluster theory (Sec.~\ref{sec:theory_cc}) and related approaches, is essential for producing a coherent and internally consistent set of the ``outer'' nuclear-structure–dependent corrections required for global $V_{ud}$ extraction~\cite{hardy2020superallowed}, particularly as FRIB enables access to superallowed decays beyond the traditional $sd$-shell cases~\cite{frib400}.

Our current effort centers on implementing two-body electroweak currents derived from chiral EFT within ab initio many-body calculations of selected light- and medium-mass superallowed emitters~\cite{cirigliano2024radiative, cirigliano2024ab}. These two-body contributions represent the leading nuclear-structure–dependent terms beyond the impulse approximation and are expected to show nontrivial variations across the chart of nuclides. As an initial step, we focus on nuclei with the relatively small experimental uncertainties or tractable many-body structure, such as the previously mentioned $^{10}$C, $^{14}$O, and $^{26m}$Al, as well as $^{34}$Cl, $^{34}$Ar, $^{38}$K, $^{38}$Ca, $^{42}$Sc, $^{46}$V, $^{50}$Mn, and $^{54}$Co. The current experimental values and uncertainties relevant to the $V_{ud}$ extraction are taken from Ref.~\cite{hardy2020superallowed} and summarized in Tab.~\ref{tab:superallowed_exp}. Additionally, we utilize fictitious benchmark transitions, such as $^{6}\mathrm{Be}\rightarrow ^{6}\mathrm{Li}$ and $^{6}\mathrm{Li}\rightarrow ^{6}\mathrm{He}$, to validate the implementation against quasi-exact methods~\cite{cirigliano2024ab, king2025quantum, gennari2025ab}. These calculations enable us to assess the impact of EFT two-body currents, explore regulator and interaction-dependent systematics, and establish uncertainty estimates that will carry forward as we move to heavier systems, ultimately allowing the radiative-correction operators developed above to be applied coherently across the medium-mass region.

The longer-term objective is to extend this program across the full set of superallowed emitters, ultimately combining two-body current corrections with consistent ab initio calculations of isospin-symmetry-breaking effects (Sec~\ref{section:ckm:section:isb_correction}), using the same chiral interactions and many-body techniques. Within this unified framework, we are also pursuing a global extraction of $V_{ud}$ in which all superallowed transitions are analyzed simultaneously using EFT-motivated contact terms as free parameters, allowing experimental data across many nuclei to directly constrain the short-distance contributions. In parallel, higher-order EFT operators, comprehensive uncertainty quantification, and benchmarking against complementary ab initio methods, such as QMC and the no-core shell model, will be essential for ensuring that theoretical systematics remain under control. Addressing issues such as spurious isospin breaking and operator evolution within many-body frameworks~\cite{farren2024spurious} will further strengthen the reliability of these results in the future. As FRIB delivers increasingly precise data on branching ratios, half-lives, and $Q$ values across an expanded set of superallowed emitters, a coherent ab initio treatment of the nuclear-structure–dependent radiative corrections will enable a more robust and systematically improvable global determination of $V_{ud}$.


\begin{table}[h]
    \centering
    \begin{tabular}{l@{\hspace{9mm}}c@{\hspace{8mm}}c@{\hspace{9mm}}c@{\hspace{7mm}}c}\hline\hline
    Transition  & $Q_{EC}$ (keV) & $t$ (ms)  &  $\sqrt{\langle r^2 \rangle}$ (fm)  & $\mu_\pi$ (MeV) \\ \hline
    $^{10}$C $\rightarrow$ $^{10}$B &  $1907.994 \pm 0.067$ & $1321800 \pm 1800$ & $2.4277(499)$ & 58.3 \\ 
    $^{14}$O $\rightarrow$ $^{14}$N & $2831.543 \pm 0.076$ & $71075 \pm 15$ & 2.5582(70) & 55.3 \\ 
    $^{26}$Al $\rightarrow$ $^{26}$Mg & $4232.72 \pm 0.15$ & $6351.24\pm 0.55$ & $3.0337(18)$ & 46.6 \\ 
    $^{34}$Cl $\rightarrow$ $^{34}$S & $5491.662 \pm 0.046$ & $1527.77^{+0.47}_{-0.44}$ & $3.2847(21)$ & 43.1 \\ 
    $^{34}$Ar $\rightarrow$ $^{34}$Cl  & $6061.83 \pm 0.08$ & $896.55 \pm 0.81 $ & $3.3654(191)^{\dagger}$ & 42.0 \\ 
    $^{38}$K $\rightarrow$ $^{38}$Ar & $6044.240 \pm 0.048$ & $925.42 \pm 0.28$ & $3.4028(19)$ & 41.6 \\ 
    $^{38}$Ca $\rightarrow$ $^{38}$K & $6612.12 \pm 0.07 $ & $574.8 \pm 1.1$ & $3.4264(51)$ & 41.3 \\ 
    $^{42}$Sc $\rightarrow$ $^{42}$Ca & $6426.34 \pm 0.12$ & $681.44 \pm 0.26$ &$3.5081(21)$ & 40.3 \\ 
    $^{46}$V $\rightarrow$ $^{46}$Ti & $7052.45 \pm 0.10$ & $423.114^{+0.068}_{-0.053 }$ & $3.6070(22)$ & 39.2 \\ 
    $^{50}$Mn $ \rightarrow$ $^{50}$Cr & $7634.453 \pm 0.066$ & $283.68\pm 0.11$ & $3.6588(65)$ & 38.8 \\ 
    $^{54}$Co $ \rightarrow$ $^{54}$Fe & $8244.38 \pm 0.26$ & $193.495^{+0.086}_{-0.063}$  & $3.6933(19)$ & 38.3 \\ \hline\hline
    \end{tabular}
\caption{The superallowed candidates up to $A=54$ and the experimental inputs used to extract the fully-corrected $\mathcal{F}t$ values. $^\dagger$ radius of $^{35}$Cl. Values from Ref.~\cite{hardy2020superallowed}. }
\label{tab:superallowed_exp}
\end{table}

\subsubsection{Possible Test of $\delta_{\rm NS}$ Using Photon Spectrum}
A precision measurement of the real photon spectrum from $0^+ \rightarrow 0^+$ decays would have potential to test calculations of $\delta_{NS}$ at low momentum transfer. The dominant nucleus-independent contribution from $\beta$ bremsstralung scales like $1/E_\gamma$, so it might be possible to isolate the small
contribution from photons coupling to nuclei by its scaling with $E_{\gamma}(Q-E_\gamma)^3$, along with its less-focussed angular distribution. Although this separation was calculated in detail for neutron $\beta$ decay~\cite{bernard2004radiative} the $\gamma$ coupling to the nucleon could not be distinguished in NIST's precision measurements of the radiative $\beta$ decay branch~\cite{bales2016precision}.
Although the $\gamma$ energies, limited by the $\beta$ decay energy release,
would not reach the most interesting continuum region, a precision measurement would be a helpful test of $\delta_{NS}$ at low momentum transfer. 
Measurements in $0^+ \rightarrow 0^+$ decays with Compton-suppressed Ge arrays fronted by segmented plastic scintillator could be done in conjunction with ongoing searches for small decay branches. 
A full calculation with known uncertainty likely needs to be revisited. A recent review points out several disagreements between experiments and theory~\cite{italiano2024on}, e.g., early measurements in agreement with theory in the allowed decay of $^{35}$S~\cite{boehm1954internal} were thought to be in disagreement by a later experiment~\cite{powar1976internal}.

\section{Precision $\beta$ Decay as a Probe of BSM Tensor and Scalar Interactions\label{sec:BSM}}
\label{sec:bsm}
Assuming that interactions do not involve field derivatives, beta decay can generally be described in terms of parity-even and parity-odd couplings that must be determined experimentally for the vector (V), scalar (S), axial vector (A), tensor (T), and pseudoscalar (P) Lorentz-invariants,
as described in the Lee and Yang Hamiltonian~\cite{lee1956question}:
\begin{equation}
\label{eq:LY_Hamiltonian}
\mathcal{H}_{\beta} \;=\;
\sum_{i=S,V,T,A,P}
\left[
C_i\, (\bar{p}\,\Gamma_i\, n) \; (\bar{e}\,\Gamma_i\,\nu)
\;+\;
C_i'\, (\bar{p}\,\Gamma_i\, n) \; (\bar{e}\,\Gamma_i\,\gamma_5\,\nu)
\right]
\;+\;\text{h.c.}\,,
\end{equation}
where the operators $\Gamma_i$ span the five independent Lorentz-invariant bilinears, 
$\Gamma_S = 1$, $\Gamma_V=\gamma^\mu$, $\Gamma_T=\sigma^{\mu\nu}/\sqrt{2}$, $\Gamma_A=\gamma^\mu\gamma_5$, and $\Gamma_P=\gamma_5$, 
and the complex couplings $C_i$ and $C_i'$ encode, respectively, interactions with left- and right-chiral leptons.

In allowed nuclear $\beta$ decay, V and S interactions can contribute to Fermi transitions and A and T interactions to Gamow-Teller ones. A convenient language for correlation measurements was provided by the work of Jackson, Treiman, and Wyld~\cite{jackson1957possible}, who derived the leading-order differential rate for allowed $\beta$ decay for a general mixture of Lorentz structures. For a non-oriented initial nucleus, that differential distribution may be written as:
\begin{equation}
\label{eq:JTW}
d\Gamma
\;\propto\;
p_e E_e (E_0-E_e)^2
\left[
1
+ a\,\frac{\vec{p}_e \cdot \vec{p}_{\nu}}{E_e \, E_{\nu}}
+ b\,\frac{m_e}{E_e}
+ ...\right]dE_e \,,
\end{equation}
where $a$ is the $\beta$--$\nu$ angular correlation coefficient and $b$ is the Fierz interference term. The ellipsis denotes terms that enter when polarization is present (e.g., the $\beta$ asymmetry $A$).

After a number of wrong experiments in the 1950s, which brought much confusion in the study of beta decay, pioneering measurements in the 1960s have deduced
the $\beta$--$\nu$ angular correlation coefficient $a_{\beta\nu}$ from
recoil momentum distributions. This confirmed the theoretically postulated $V-A$ structure of the weak interaction.

However, beyond the Standard Model (BSM) extensions can lead to exotic S and T currents which would result in deviations from the expected values for any correlation coefficient. In $a_{\beta\nu}$, this expected deviation would be squared in the exotic S or T coupling. A linear deviation enters the Fierz interference term $b$, which is zero in the SM, but is not sensitive to right-handed couplings.

Unique forbidden beta decays are also sensitive to the A and T currents, with the advantage that the angular correlation does not vanish in the beta spectrum, as it does in allowed decays, thus preserving the sensitivity to right-handed couplings~\cite{glick2017beta}. It has recently been shown that they may also be sensitive to BSM light degrees of freedom, which are unobservable in allowed beta decays~\cite{seng2024intruding}.

\subsection{Experimental Efforts}
\label{sec:bsm:expt}

\subsubsection{Mirror Transitions}
\label{sec:bsm:expt:mirror}

    Mirror transitions in nuclei occur between isospin doublets.
    As discussed in the introduction of Sec.~\ref{sec:experimental.efforts} and
    in Sec.~\ref{sec:superallowed.mixed},
    they offer an
    additional set of transitions to extract $V_{ud}$. However,
    in contrast to pure Fermi transitions, they require the determination of the 
    Gamow-Teller-to-Fermi mixing ratio, $\rho$.
    Measurements of the electron-asymmetry parameter, $A$, and of the $\beta-\nu$ angular correlation coefficient, $a$, in mirror decays enable the extraction of $\rho$ which
    is currently more accurate than theoretical determinations.

    A recent global fit of nuclear and neutron beta-decay data shows a hint of a possible tensor contribution, at the 3$\sigma$-level, when assuming
    right-handed neutrinos \cite{falkowski2021comprehensive}.
    This effect 
    is only due to neutron-decay data, and especially to the single measurement of the $\beta-\nu$ angular correlation coefficient by the aSPECT collaboration \cite{Bec20}, which also adds tension to constraints on
    both, scalar and tensor couplings to the fit with left-handed neutrinos.
    As shown in Fig.~5 of Ref.~\cite{falkowski2021comprehensive}, and commented in that paper,
    the addition of data from mirror transitions does not change much the tension produced by the neutron data. Similar level of tensions on the tensor coupling involving right-handed neutrinos have been observed in previous global fits \cite{Sev06} with a totally different data set from neutron decay. Indeed, the beta-asymmetry parameter, $A_n$, and
    the neutron lifetime, $\tau_n$, significantly changed in the past two decades.
    This illustrates
    that the importance of tensions by 2-3$\sigma$ should not be exaggerated, in particular when 
    quickly moving data sets are at play. Nevertheless, it is important to improve the precision on
    correlation measurement in mirror decays, with devices such as the St.\ Benedict at Notre Dame,
    because they have shown to provide improvements by a factor of up to 2.8, on
    constraints on exotic couplings involving right-handed neutrinos \cite{falkowski2021comprehensive}.

\subsubsection{$\beta$ decay Paul Trap}
The $\beta$ decay Paul Trap (BPT) \cite{scielzo2012the} program at the ATLAS facility at Argonne National Laboratory searches for exotic T currents using the $\beta^-$ decay of $^8$Li \cite{li2013tensor,sternberg2015limit,burkey2022improved} and the $\beta^+$ decay of $^8$B \cite{gallant2023angular,longfellow2024improved}. In addition, the neutrino energy spectrum from $^8$B $\beta$ decay is an important input for the solar neutrino astrophysics community \cite{longfellow2023determination}. Both the $^8$Li and $^8$B $\beta$ decays predominantly proceed via a nearly pure Gamow-Teller transition from the $J^{\pi}=2^+$, isospin $T=1$ $^8$B ground state to the 3-MeV $J^{\pi}=2^+$, $T=0$ resonance in $^8$Be, which nearly immediately breaks apart into two $\alpha$ particles. The large $Q$ values for the decays and the relatively low masses of the parents lead to MeV scale $\alpha$ and $\beta$ decay products and the energies of these charged decay products are measured by four $32\times32$ double-sided silicon strip detectors surrounding the trapping region of the BPT. The decay rate for $\beta$-delayed $\alpha$ emission from an unpolarized nucleus includes an additional ``triple" correlation term \cite{holstein1974recoil,holstein1976erratum}. For the spin sequences in the $^8$Li and $^8$B decays, this enhances the effective $a_{\beta\nu}$ from $-1/3$ to $-1$ for the case of the $\beta$ emitted in the same direction as the $\alpha$. To take advantage of this enhancement, the $\beta$-neutrino angular correlation coefficient $a_{\beta\nu}$ is extracted by fitting the energy difference spectrum for the two $\alpha$ particles for the case where the $\beta$ hits the same detector as one of the $\alpha$ particles with a linear combination of detailed simulations assuming a pure T and a pure A interaction and assuming $b_{Fierz}=0$. The most recent iteration of the $^8$Li experiment resulted in $|C_T/C_A|^2 = 0.0012 \pm 0.0019_{stat} \pm 0.0028_{sys}$ \cite{burkey2022improved}. Under the assumption $C_T=-C_T'$, this converted to $a_{\beta\nu} = -0.3325 \pm 0.0013_{\text{stat}} \pm 0.0019_{\text{sys}}$ \cite{burkey2022improved} and provided the first improvement by 25\% of the relative uncertainty on a single measurement of $a_{\beta\nu}$ from a Gamow-Teller transition in over 60 years \cite{johnson1963precision}.
However, note that $a_{\beta\nu}$ has a quadratic dependence on exotic couplings. See, e.g.,~Ref.~\cite{Sev06} for global constraints from beta-decay data for tensor couplings involving right-handed neutrinos.
The improvement in systematic uncertainties for $^8$Li was enabled by new symmetry-adapted no-core shell model (SA-NCSM) calculations of the recoil-order terms \cite{sargsyan2022impact}. The most recent $^8$B measurement employed the same SA-NCSM technique and was of similar precision to the $^8$Li value \cite{longfellow2024improved}.
It should be noted that the procedure for the extraction of couplings using the so-called $\tilde{a}$ prescription in \cite{burkey2022improved,gallant2023angular,longfellow2024improved} is inconsistent with the assumption made during the data analysis that the Fierz term is zero and in
future analyses the effect of the Fierz interference term should be included explicitly when simulating the $\alpha$ energy difference. The next generation of BPT experiments has recently begun and utilizes a new trap design with optimizations to decrease $\beta$ scattering, an important experimental systematic \cite{varriano2024the}. On the theory side, one complication from the SA-NCSM calculations is the prediction of an extremely broad ``intruder" $2^+$ level in $^8$Be \cite{sargsyan2022impact}, the existence of which is currently unclear experimentally. Furthermore, there are a number of typos, ambiguous terms, and missing induced electromagnetic corrections in Holstein's review article \cite{holstein1974recoil,holstein1976erratum} detailing the recoil-order terms used in the simulation and a modern version with controlled order of truncation would be greatly appreciated. Finally, a calculation of the radiative correction to the ``triple" correlation term for $\beta$-delayed particle emission is needed.

\subsubsection{Calorimetry with implanted beams produced by fragmentation}
As introduced in Sec.~\ref{energySpectra}, a calorimetry technique has been explored
at NSCL using beams of suitable $\beta$-unstable candidates to measure
$\beta$-energy spectra over the widest possible energy range and with the
smallest possible distortions. In a standard configuration, i.e. with the $\beta$
source external to the detector, the dominant instrumental effect distorting
the spectra is the back-scattering or out-scattering of $\beta$ particles from
detectors or surrounding materials.
The accuracy of models describing such effects in Monte-Carlo simulations \cite{Kan26}
is still too poor to reach competitive levels of precision. Such effects are totally
eliminated when the
source is embedded inside the detector and fully surrounded by active material
without any dead layer in between. The implantation of beams produced by fragmentation
enables achieving such conditions in a rather simple way.
The beams are implanted at
a depth which is larger than the range of the emitted $\beta$ particles. Due to the
well defined range and small range-straggling of high-energy ion beams,
the source is well localized and the implantation depth can accurately be controlled.
The technique enables the selection of the most suitable candidates for searches
of exotic interactions in terms of their
SM background. This is in contrast to partial calorimetry techniques in
which a source is sandwiched between two detectors which, in addition to the
presence of dead layers, often require isotopes
with longer half-lives for which the accurate theoretical description of
$\beta$-energy spectra can be problematic. For Gamow-Teller transitions, the largest SM
background is due to the weak magnetism form factor and its most precise extraction
is from experimental data through the strong form of CVC within an isospin triplet.
This and other
beam production considerations, has conducted to select $^6$He and $^{20}$F as
sensitive candidates \cite{Nav16}. Even if produced by fragmentation, the beam
purities achieved at NSCL and GANIL for such light elements have been excellent.

With an implantation/decay time sequence, the full $\beta$-energy spectrum
is efficiently measured at once, unlike sequential measurements with
classical magnetic spectrometers.
This avoids a precise external normalization which is often difficult to
implement for
g.s. to g.s. transitions. Measurements carried out at NSCL and GANIL showed that
with $^6$He beams implanted in CsI(Na), NaI(Tl), and YAP(Ce) and with a $^{20}$F
beam implanted in CsI(Na), the beam-induced backgrounds in the detectors
is sufficiently small and can properly be described. The resolution
functions of the inorganic scintillators have been obtained with standard
$\gamma$ sources and do not constitute a limitation.

The analyses of energy spectra require also Monte-Carlo simulations
to account for distortion effects which are typically two orders of magnitude
smaller than 
back-scattering and can be described more accurately. These are energy losses
due to ``external'' Bremsstrahlung photons produced
during the slowing down of $\beta$ particles and which could escape the
detectors \cite{Huy18}, as well as ``internal'' Bremsstrahlung
photons due to radiative higher-order processes along with regular $\beta$ decay.
Dedicated studies, using radioactive sources and MTAS, are in progress to benchmark
simulations which model the production and absorption of Bremsstrahlung photons in
YAP(Ce) scintillators.

The calorimetry technique is particularly well adapted to FRIB. For example,
with a 132 MeV/nucleon $^{20}$F beam at NSCL, the implantation depth in CsI(Na) was
11.6(2)~mm whereas the range for 5.5~MeV electrons is 8.6~mm, which is
rather at the edge. The higher energy
beams from FRIB enable to implant isotopes with larger masses and atomic numbers.

\subsubsection{He6-CRES}
\label{sec:bsm:expt:he6cres}

The He6-CRES experiment at the University of Washington is looking to measure the Fierz term $b$ through beta energy spectra from the decays of several isotopes - currently $^6$He and $^{19}$Ne. The energies are being measured with Cyclotron Radiation Emission Spectroscopy (CRES), a technique initially developed by Project 8 for a neutrino mass measurement \cite{monreal2009relativistic}. The essence of CRES is to non-destructively measure the frequency of an emitted beta at its birth as it undergoes cyclotron motion in a magnetic field. The beta energy can be extracted from the relation,
\begin{equation}
   E = \frac{q B c^2}{\omega} \,,
\end{equation}
where $\omega$ is the cyclotron frequency, $q$ is the charge of the $\beta$
particle, and $B$ is the strength of the magnetic field. The decay electron or positron will produce $\sim$fW power microwaves from its cyclotron motion. The fundamental of the cyclotron radiation has the same frequency as the cyclotron motion, and measuring the initial frequency produced thus constitutes the basis of measuring the initial energy. Generally, this radiation could be picked up in several ways, but the only technique demonstrated thus far is to have the decay occur within a waveguide. In this configuration, the cyclotron radiation resulting from the motion couples to propagating modes in the waveguide, which are amplified and digitized. 

This technique offers an advantage over traditional calorimetry because it allows determination of the energy immediately when the beta is created, rather than getting a signal after the particle has already been through many potential processes. It also offers the potential for excellent energy resolution (\textless 2eV \cite{ashtari2023tritium}). CRES has so far only been done with neutral molecular gas sources ($^{83}$Kr, $^3$H, $^6$He, and $^{19}$Ne), but atomic and ionized sources are also being developed. CRES is impervious to any source of background radiation that is not contained in the gas. Thus radio-pure sources are desired, but CRES may be done in a high-radiation environment.

To search for BSM physics it is necessary to calculate the $\beta$-energy spectra with accuracies smaller than the required sensitivity. The dominant uncertainties on the SM description of the spectra are radiative and recoil-order corrections \cite{hayen2018high}. Although these contributions are only of order of a few \%, the experiment is designed to go beyond the $10^{-3}$ level. A scheme for characterizing the effects of uncertainties can be found in Ref. \cite{glick2022formalism}. The case of $^6$He has been studied recently \cite{glick2022nuclear, king2022ab} with different approaches, concluding that uncertainties are at the level of $2\times10^{-4}$. The cases of $^{14}$O, $^{19}$Ne are in a less solid base. Fortunately, the dominant part of the recoil-order correction for $^{19}$Ne (so-called weak magnetism) can be related to the well-known magnetic moments of the initial and final states, while for $^{14}$O there is no such contribution to first order. Furthermore, for both of these superallowed decays, a symmetry implies that a matrix element which is a typical source of uncertainties, the so-called pseudo-induced tensor, is highly suppressed. Radiative corrections are dominated by the model-independent Sirlin term \cite{sirlin2013radiative}. Recent calculations~\cite{hill2024field, sheng2023dispersive} imply effects on the spectra at $10^{-4}$ level. Detecting these would be interesting for helping with determining possible effects on $V_{ud}$ and may be achieved in the next phase of the experiment. In summary, rough estimates indicate that the uncertainties in our estimations of the beta spectra $^{19}$Ne and $^{14}$O are at the $10^{-3}$ level. This is below the experimental sensitivity at the present phase of the experiment, but needs to be improved for reaching the ultimate sensitivity. This is a call to our theory colleagues to help us in moving beyond this limiting uncertainty.

\subsubsection{$\beta$ Spectroscopy with Levitated Nanoparticles}
\label{sec:bsm:expt:quips}

The Quantum Invisible Particle Sensor (QuIPS) located at Lawrence Berkeley National Lab is an experiment building upon a recent proposal to use levitated, radioactively-doped nanoparticles in order to fully reconstruct the final state kinematics of beta decay~\cite{carney2023searches}. The momentum of the child particle is measured via optomechanical techniques with a sensitivity set by the standard quantum limit, corresponding to a momentum uncertainty $\Delta p_\mathrm{SQL} \approx 15$~keV for a femtogram nanoparticle. By integrating the optomechanical set up with a beta calorimeter, the energy and momentum of the $\beta$ particle are also measured, fully determining the final-state phase space. The QuIPS experiment will aim to measure the differential decay rate as a function of both the beta energy $E_\beta$ and the angle $\theta_{\beta \nu}$ between the neutrino and the $\beta$ particle, which may be written at tree-level as 
\begin{align}
    \frac{d^2\Gamma}{dE_\beta \,d \cos \theta_{\beta\nu}} = A(E_\beta) \, \cos\theta_{\beta\nu} + B(E_\beta),
    \label{eqn:fullyDifferentialDecayRate}
\end{align}
where the functions $A(E_\beta)$ and $B(E_\beta)$ depend on the microscopic parameters of the electroweak interaction~\cite{jackson1957possible}. The ratio $A/B$ may be extracted from precise measurement of $d^2\Gamma/dE_\beta \,d \cos \theta_{\beta\nu}$, from which one may directly measure $b$ parameter which is linearly dependent on BSM scalar (tensor) currents in Fermi (Gamow-Teller) decays through the Fierz interference term. 

The above experimental program has many complementary aspects to the Paul trap program. For one, it suffers from very different systematic uncertainties: the
daughter nucleus momentum is directly accessible, and the initial position of the
$\beta$ particle is well-defined to within the micron-sized nanoparticle, which enables accurate reconstruction of the direction of the $\beta$ momentum by measuring its final position. Furthermore, any radioisotope may be implanted into the nanosphere as long as it satisfies a time-constraint on its half-life $T_{1/2} \gtrsim $ 1 day so that the experiment at Berkeley Lab may be set up before all the decays have occurred. On the other hand, the scattering of the $\beta$ particle
as it exits the nanoparticle introduces new uncertainties that must be mitigated. Ideally, a $\beta$ emitter with a large $Q$-value is used to minimize the effect of this scattering. The momentum recoil also scales with the $Q$-value, which implies that the effect of the nuclear recoil corrections must be balanced against scattering of the $\beta$ particle. Currently, the isotopes identified that satisfy the lifetime constraint with pure Gamow-Teller decay modes are $^{64}$Cu and $^{66}$Ni. It is interesting to note that creating a similar experiment at FRIB would allow \textit{in situ} creation of the radioisotopes, which would loosen lifetime constraints and thus enable access to more radioisotopes. 

Accurate evaluation of the recoil and nuclear structure corrections requires the input of the theory community to evaluate the matrix elements that may mimic the effects of a non-zero $b$ parameter in any measurement of Eq.~\eqref{eqn:fullyDifferentialDecayRate}. While precise quantification of the required uncertainty on the theory side is still ongoing, NLO corrections to the Fierz interference term arise at order $m_\beta/m_N$ and $m_\beta R P_\mathrm{Fermi}/m_N$, where $R$ is the nuclear radius, $m_N$ the nucleon mass, and $P_\mathrm{Fermi}$ the Fermi momentum~\cite{glick2022formalism}. For $^{66}$Ni, for instance, $m_\beta R P_\mathrm{Fermi}/m_N \approx 3\times 10^{-3}$, and so a $10^{-3}$ determination of any new physics contributions to $b$ requires the relevant matrix elements be evaluated approximately at the 30\% level, motivating further \textit{ab initio} study of relevant heavy nuclei.

\subsection{Nuclear Physics Input from Theoretical Calculations}\label{sec:bsm:struct.input}
For the theory to match the high precision aimed at by experiments, there is a need for higher-order theory corrections to the nuclear $\beta$ decay. In particular, nuclear recoil-order terms would have to be included in the analyses of
precision $\beta$ decay measurements. The expressions for recoil-order contributions to allowed $\beta$ decays have been commonly taken from work by Holstein, published over 5 decades ago \cite{holstein1974recoil,holstein1976erratum}. However, numerous typographical errors have been identified in this work over the years \cite{holstein1976erratum, melconian2011}. In addition, it lacks a method to quantify the uncertainties caused by omitted higher orders, and some recoil-order corrections are missing, as argued by a recent work in Ref.~\cite{glick2022formalism}. More reliable expressions for recoil-order corrections are presented in Ref.~\cite{glick2022formalism}, conveniently divided into separate corrections for each desired observable ($\delta_1$ for the spectrum shape, $\delta_a$ for correcting angular correlation measurements, and $\delta_b$ for $b$ Fierz term) and offering the advantage of uncertainty quantification, including a way to estimate to what order one should aim before going through the actual calculations. Another advantage of the corrections in Refs.~\cite{glick2022formalism, benatar2026leading} is that they are based on the well-established Donnelly and Walecka multipole expansion notion~\cite{walecka1995theoretical}, with conveniently calculated tables listed in Ref.~\cite{donnelly1979multipole}, and a Mathematica code in Ref.~\cite{haxton2008sevenoperators}, which is widely used in the \textit{ab initio} community, therefore ready for implementation in more accurate calculations.

The first \textit{ab initio} study of recoil-order terms was carried out for $^6$He $\beta$ decay, using the NCSM~\cite{glick2022nuclear}, based on the recoil-order formalism of Ref.~\cite{glick2022formalism}. A key ingredient was the Coulomb displacement energy $\Delta E_c$, which modifies the coefficient of the $\sigma \cdot r$ term in the Coulomb multipole operator $C_1^A$ in Walecka notation. This effect was identified in Ref.~\cite{behrens1982electron} and implemented in Walecka notation in Ref.~\cite{glick2022formalism}, but is missing in the Holstein framework, where it would appear as a correction to the induced tensor $d^I$.

The calculations showed that measurements of the Fierz term, assumed to be zero in the SM, will actually have a recoil-sourced value $\delta_b=-1.52\left(18\right)\cdot{10}^{-3}$, comparable to the ${10}^{-3}$-level BSM effect sought experimentally. In addition, unlike previous recoil-order corrections calculations, this correction value comes with a quantified uncertainty, of the order of ${10}^{-4}$, that cleanly separates nuclear-structure effects from genuine BSM signals.

Likewise, ${10}^{-3}$-order corrections for the spectrum shape ($\delta_1$) and for the angular correlation ($\delta_a$) were found, both with electron-kinetic-energy dependence and a quantified uncertainty of the order ${10}^{-4}$. Following that, it was shown that ${10}^{-3}$-level BSM signatures are accessible with ${10}^{-4}$-precision measurements of $a_{\beta \nu}$~\cite{glick2023multipole}, revisiting earlier expectations that ${10}^{-6}$-precision measurements required due to $a_{\beta \nu}$'s quadratic dependence on the BSM couplings.

A similar study of the $\beta$ decay spectrum of $^6$He was performed using QMC approaches~\cite{king2022ab}. This approach took inspiration from EFT by noting that the momentum transfer, $q$, was limited by the $Q$-value of the $\beta$ decay. Noting that the multipole expansion of the transition matrix element~\cite{walecka1995theoretical} is written in terms of multipoles, which are either purely even or odd polynomials in $q$, then one can perform an expansion of these quantities in $q/m_{\pi}$. Combined with the chiral expansion of the operators entering into the multipole, it is possible to derive a $\beta$ decay spectrum up a certain order in $q/\Lambda_{\chi}$ to reach the desired level of precision to compare with experiment. This expansion does not, however, account for the uncertainty coming from the underlying nuclear Hamiltonian. In Ref.~\cite{king2022ab}, an estimate of the errors induced from the nuclear interaction was made by taking four different models of the nuclear Hamiltonian to perform the computation~\cite{piarulli2016local,Piarulli2017light,baroni2018local}.  The spectrum received distortions of up to $\approx 1\%$, with the dominant correction coming from the weak magnetism term, which was shown to be equal to the value extracted from electromagnetic decay of $^6{\rm Li}(1^+;0)$ to its ground state within the experimental uncertainty. Thus, the experimental value was adopted to further constrain any potential model dependence. The dominant source of uncertainty on the spectrum comes from higher order terms in the low $q$ expansion of the multipoles, and the overall deviation in the distortion of the spectrum with respect to the SM value was less than the $0.1\%$ goal needed to compare with experiment. Finally, it was found that the SM Fierz term induced by recoil is $\delta_b=-1.47(3)\times10^{-3}$ in the QMC calculation. 

 Furthermore, {\it ab initio} calculations of weak magnetism, induced tensor and second forbidden axial recoil-order terms have been recently calculated for $^8$Li and $^8$B beta decays using SA-NCSM \cite{sargsyan2022impact, longfellow2024improved, sargsyan2026role}. These studies used several nucleon-nucleon (NN) interactions for the many-body calculations and showed a strong correlation between some of the recoil-order terms and the quadrupole moments of the parent nuclei across all the interactions and the model spaces used. Using this correlation and the experimentally measured quadrupole moments, enabled a precise prediction of these recoil-order terms with uncertainties stemming from the underlying $NN$ interactions and the many-body truncations.  Tables of predicted $^8$Li and $^8$B beta decay recoil-order terms are available in Ref. \cite{longfellow2024improved}. Using SA-NCSM, such \emph{ab initio} calculations of recoil-order terms are possible for up to Ca mass region. 

 {\it Ab initio} computations of recoil-order corrections to $\beta$ decay, have made great strides. However, as with other observables, because computations are costly, it is typical to perform calculations with only a few different models of the nuclear interaction to obtain a naive estimate of the uncertainty. To better aid in the interpretation of these experiments, a rigorously quantified uncertainty on recoil-order terms will be useful. Great progress has been made to quantify interaction uncertainties~\cite{bub2024bayesian,wesolowski2021rigorous,hu2021ab,somasundaram2023maximally} and emulate many-body calculations~\cite{konig2019eigenvector,odell2023rose,becker2023ab,somasundaram2024emulators,armstrong2025emulators}. Applying these tools to recoil-order corrections will help in providing uncertainty-quantified inputs for experimental analyses. 

 The {\it ab initio} calculations of recoil-order corrections in searches for tensor currents have been performed in light nuclei~\cite{sargsyan2026role}. Several candidate nuclei for precision measurements to investigate new physics have been identified which range up to mass $A \leq 144$~\cite{cirigliano2019precision}. Thus, also having calculations from both phenomenological and {\it ab initio} many-body approaches with methods that can address these heavy systems will be important for the interpretation of such measurements. In order to validate these approaches, benchmark studies comparing to NCSM and QMC calculations of spectra in light nuclei would be valuable. Further, benchmarks would provide more theoretical input to interpret the data already measured in $A \leq 8$ systems.

 \subsection{Revisiting Induced Second-Class Currents}

  The physics scale reach of the $V_{ud}$ CKM unitarity constraint is unique.
A number of experiments using new techniques
seek order-of-magnitude improvement in sensitivity
to Lee-Yang Lagrangian currents transforming as Lorentz scalars and tensors.
These can couple to SM left-handed neutrinos if they produce wrong-handed
$\beta$ particles, producing $m_\beta/E_{\beta}$ dependence
of the rate from a spinor
helicity factor. Thus low-$E_\beta$ experiments have naturally increased
sensitivity, which is the fundamental reason that results of such experiments can
achieve complementary sensitivity to experiments at the higher scales of
$\pi \rightarrow e \nu \gamma$ and  LHC $p+p \rightarrow e + m_T$ missing mass experiments.

Note the LHC constraints assume good isospin rotations. It is nearly a tautology
that the best constraints on first-generation second-class currents (SCC), defined as contributions to the nucleon's hadronic current changing sign as neutrons change to protons, come from precision nuclear $\beta$ decay. The well-explored phenomenology of a second-class term in the electron-nucleon axial vector current consistently parameterizing SCC nucleon and nuclear couplings, is summarized in Ref.~\cite{wilkinson2000limits} with more recent null results in Ref.~\cite{minamisono2011low-energy}.

Similarly, the hadronic vector current of the nucleon acquires a CVC-breaking and isospin-breaking term with, for other physics reasons, the same $E_\beta$ dependence as the Fierz term, so can be directly compared to existing limits on $C_S$. Termed the induced scalar contribution to the hadronic vector current, it can
be parameterized as $[({e}/{A})({m_\beta}/{m_N})({m_\beta}/{E_\beta})]$, with $A$ the nuclear mass number and $m_N$ is the nucleon mass.
(One should restrict how one regards CVC as a particle symmetry to the first generation, as 
the SM produces a substantial CVC-breaking term in strange baryon $\beta$ decay~\cite{holstein1982induced}.)
Although $e$ itself scales with $A$ in the impulse approximation~\cite{holstein1974recoil,holstein1976erratum}, so does $d_{II}$, which
despite Lipkin's conjecture based on $\rho-\omega$ mixing that it could scale with $A$~\cite{lipkin1971second},
was found not to scale with $A$ in detailed nuclear structure calculations using nucleon-nucleon and meson exchange models of the axial vector SCC~\cite{kubodera1973the}.
If one instead naively takes $e$ as a constant, one can explain the 3$\sigma$ claim of a Fierz term in the neutron $\beta$ decay with $e\approx -30$, yet maintain consistency with nuclear $\beta$ decay constraints suppressed by $1/A$. E.g., a fit to the 2020 (pre-$\Delta_{NS}$ and $f$ re-evaluations) data finds $e$=$-4\pm 32$.
This will remain a curiosity without an explicit model for $e$.

\section{$\beta$ Decay for Neutrino Physics\label{sec:neutrino}}

Beta decay remains central to neutrino physics.
At the most basic level, $\beta$ decay is the only laboratory process that directly couples charged leptons to electron-flavor neutrinos at low energies, making it a unique probe of neutrino properties. Decades of $\beta$ decay measurements have shaped our knowledge of neutrino masses, mixing, and the structure of the weak interaction, while continuing to provide sensitive tests of new physics beyond the SM.

Kinematic endpoint studies provide the only direct laboratory access to the absolute neutrino mass scale. Spectrum-shape studies probe sterile neutrinos and constrain exotic couplings. Correlation measurements test the fundamental chirality of neutrino interactions. Together with $0\nu\beta\beta$ searches, these programs form a comprehensive experimental strategy for addressing open questions about the mass, nature, and interactions of neutrinos in and beyond the SM.

\subsection{Neutrino Mass from Kinematic $\beta$ Decay}
The most direct laboratory method to determine the absolute neutrino mass scale is through precise spectroscopy of the electron energy spectrum in $\beta$ (or electron capture) decay~\cite{formaggio2021direct}. In the Standard Model, the electron spectrum near the endpoint is modified by the incoherent sum of neutrino mass eigenstates:
\begin{equation}
\frac{d\Gamma}{dE_e} \propto F(Z,E_e) p_e (E_0 - E_e) \sum_i |U_{ei}|^2 \sqrt{(E_0 - E_e)^2 - m_i^2} \, ,
\end{equation}
where $F(Z,E_e)$ is the Fermi function, $E_0$ is the endpoint energy, and $U_{ei}$ are PMNS matrix elements. A finite $m_i$ shifts and distorts the endpoint.

The decay isotopes being experimentally pursued are $^3$H (tritium) and $^{163}$Ho.
The statistical neutrino mass sensitivity scales like $\sqrt[4]{N}$, which drives experiments to isotopes with a low $Q$ value to optimize the statistics in the region of the endpoint. 
Tritium $\beta$ decays ($^3\mathrm{H} \rightarrow ^3\mathrm{He} + e^- + \bar{\nu_e}$) with $Q = 18.6$ keV and $T_{1/2} = 12.3$ yr.
$^{163}$Ho EC decays ($^{163}\mathrm{Ho} + e^- \rightarrow ^{163}\mathrm{Dy} + \nu_e$) with $Q = 2.86$ keV and $T_{1/2} = 4750$ yr.
$^{187}$Re $\beta$ decay ($^{187}\mathrm{Re} \rightarrow ^{187}\mathrm{Os} + e^- + \bar{\nu_e}$)  with $Q = 2.47$ keV and $T_{1/2} = 4.3 \times 10^{10}$ yr had been formerly pursued~\cite{sisti2004new}, but was abandoned due to practical difficulties with detector performance and its exceptionally long half-life~\cite{nucciotti2016use}.
An active research program is investigating potential ultra-low $Q $ value decays particularly to excited states~\cite{keblbeck2022updated,ruotsalainen2024ultralow}, but as yet no candidate isotope has viable properties (especially branching ratio) to motivate an experimental program.

The KATRIN experiment (MAC-E spectrometer)~\cite{katrin2022katrin}, using molecular tritium $\beta$ decay, has now reached sub-eV sensitivity. In 2019–2021 runs, KATRIN set the world-leading limit $m_\beta < 0.45$ eV (90\% C.L.)~\cite{katrin2025direct}, with the ultimate sensitivity goal of 0.3 eV with the full exposure.  Future kinematic experiments such as Project~8
\cite{monreal2009relativistic}) will use atomic tritium~\cite{ashtari2025dynamics} $\beta$ decay aiming to reach sensitivities down to $40$ meV, sufficient to test the inverted mass ordering~\cite{ashtari2022project}.  A demonstrator CRES measurement of the tritium endpoint placed the first RF-based neutrino mass limit of $m_\beta < 155$ eV~\cite{ashtari2023tritium,ashtari2024cyclotron}.  An alternate R\&D approach to a tritium CRES neutrino mass measurement is pursued by QTNM~\cite{amad2025determining}.

Holmium experiments utilize microcalorimeters to measure the EC decay spectrum.  Recent results from the HOLMES and ECHo-1k experiments place limits on the neutrino mass of $m_\beta < 27$ eV~\cite{alpert2025most} and $m_\beta < 15$ eV~\cite{adam2025improved}, respectively.
Scaling these experiments may yield next-generation sensitivity comparable to or exceeding that of KATRIN.

\subsection{Searches for Sterile Neutrinos}
\label{sec:neutrino:sterile}

$\beta$-decay spectra are also a sensitive tool to search for sterile neutrinos with eV–MeV masses~\cite{acero2023white}. A sterile neutrino mixed with $\nu_e$ would manifest as a kink in the spectrum at $E_0 - m_s$, with amplitude proportional to $|U_{es}|^2$.  Mixing strengths have been excluded down to $|U_{es}|^2 \sim 10^{-2}$–$10^{-4}$ for $m_s$ between a few eV and tens of keV.  This parameter space is highly relevant for sterile-neutrino dark matter models and cosmology and complementary to traditional oscillation searches for sterile neutrinos that primarily probe sub-eV mass scales.

Owing to the intensity of its tritium source, KATRIN places the most stringent limits on eV-scale sterile neutrinos in its accessible mass range using its neutrino mass dataset~\cite{acharya2025sterile}.
Previous experiments Mainz~\cite{kraus2013limit} and Troitsk~\cite{belesev2013upper} provide coverage of a somewhat expanded mass range also from their neutrino mass datasets.
These same tritium sources have been used with dedicated runs to extend to the keV-scale mass range~\cite{aker2023search, abdurashitov2017first}, with the TRISTAN detector upgrade to KATRIN offering a definitive measurement~\cite{mertens2019novel}.
Dedicated decay studies with other isotopes (e.g.\ $^{63}$Ni, $^{35}$S) provide additional windows of sensitivity.

From $0.1 \sim 1.0$ MeV, the most stringent limits come from nuclear recoil measurements on EC decays.  The BeEST experiment uses $^7$Be implanted in superconducting tunnel junction (STJ) sensors~\cite{friedrich2021limits}.
This quantum technology is allowing for rapid scaling with projected limits surpassing $|U_{es}|^2 \sim 10^{-7}$.

QuIPS (see Sec.~\ref{sec:bsm:expt:quips}), an experimental collaboration joint between Lawrence Berkeley National Lab and Yale University, will search for sterile neutrinos in the MeV mass range~\cite{carney2023searches}. It is based on a set-up involving a radioactive, levitated nanoparticle, in which the momentum of the neutrino is inferred by simultaneous optomechanical measurement of the daughter nucleus momentum and measurement of the electron momentum with a number of charged current devices in the $\beta^{-}$ decay of $^{90}$Y. Conservation of momentum converts measurement of the aforementioned quantities into a measurement of the neutrino momentum. This in turn enables one to constrain the presence of decay events with an anomalously large neutrino mass, which will lead to stringent constraints for $m_s$ in the 500 keV to 2 MeV range.

\subsection{Neutrinoless Double $\beta$ Decay}
Although technically distinct from single $\beta$ decay, neutrinoless double $\beta$ decay ($0\nu\beta\beta$) is another key process linking neutrino physics to nuclear $\beta$ transitions. Observation of $0\nu\beta\beta$ would establish the Majorana nature of neutrinos and violate lepton number, providing a direct window to BSM physics~\cite{agostini2023toward}. In the commonly studied scenario, $\beta\beta$-decay is mediated by the exchange of a light neutrino, which does not require introducing any additional particles. However, there are many other possible decay mechanisms including BSM particles that could lead into $0\nu\beta\beta$ decay \cite{cirigliano2018neutrinoless2,bolton2020neutrinoless,graf2022unraveling}.

In the light-neutrino-exchange scenario the $0\nu\beta\beta$-decay half-life of a given isotope can be written as
\begin{equation}
    \frac{1}{T_{1/2}^{0\nu}}=g_{\rm A}^4G^{0\nu}|M^{0\nu}|^2\left(\frac{m_{\beta\beta}}{m_e}\right)^2\;,
\end{equation}
where $g_{\rm A}$ is the axial-vector coupling constant, $G^{0\nu}$ the phase-space factor \cite{kotila2012phase}, $M^{0\nu}$ the nuclear matrix element (NME), and $m_{\beta\beta}=\sum_i(U_{ei})^2m_i$ the so-called effective Majorana mass. Hence, measurement of the decay would inform us on the unknown mass-scale of neutrinos. However, the interpretation of $0\nu\beta\beta$ rates in terms of an effective Majorana mass requires detailed nuclear matrix element calculations \cite{agostini2023toward} and is strongly complementary to single $\beta$ decay kinematic experiments. There are 35 candidate isotopes that can undergo $\beta^-\beta^-$-decay and 34 that can decay via $\beta^+\beta^+$, $\epsilon\beta^+$ or $\epsilon\epsilon$ processes \cite{tretyak2002tables}. It is important to note that in order to distinguish between different decay mechanisms and make conclusions about the neutrino properties, one would need to measure $0\nu\beta\beta$ decay in several isotopes \cite{graf2022unraveling}.

Current and future experiments \cite{agostini2023toward} focus mostly on the $\beta^-\beta^-$-decaying isotopes $^{76}$Ge, $^{100}$Mo, $^{130}$Te, and $^{136}$Xe.
Notable current-generation experiments have surpassed $T_{1/2}^{0\nu}$ sensitivity of $10^{25}$ years, with the most stringent limits now exceeding $10^{26}$ years~\cite{abe2025search,agostini2020final,acharya2025first} corresponding to a mass limit of $m_{\beta\beta} > 28-122$ meV (the range owing to discrepancy between $M^{0\nu}$ calculations).
Recently concluded experiments include those using $^{136}$Xe (EXO-200~\cite{albert2018search} and KamLAND-Zen800~\cite{abe2025search}) and $^{76}$Ge (GERDA~\cite{agostini2020final} and the \textsc{Majorana Demonstrator}~\cite{arnquist2023final}).
Currently operating experiments include CUORE ($^{130}$Te)~\cite{adams2025constraints} and LEGEND-200 ($^{76}$Ge)~\cite{acharya2025first}, with the KamLAND2-Zen upgrade underway ($^{136}$Xe); both LEGEND-200 and KamLAND2-Zen are expected to exceed $10^{27}$ year sensitivity.
Preparation for ``tonne-scale'' experiments LEGEND-1000 ($^{76}$Ge)~\cite{abgrall2021large} and CUPID ($^{100}$Mo)~\cite{alfonso2025cupid} is underway, with sensitivity largely covering the inverted mass ordering parameter space; other experiments are proposed~\cite{agostini2023toward} but constrained by funding availability.

$\beta\beta$-decaying nuclei with mass numbers $A\geq 48$ pose a challenge to nuclear theory, and hence most nuclear-theory predictions for $0\nu\beta\beta$ decay are based on phenomenological many-body methods such as pnQRPA, nuclear shell model, DFT or interacting boson model (IBM). The NMEs obtained from the different frameworks tend to disagree by a factor of a few for a given isotope \cite{agostini2023toward}, and since these methods are based on different approximations and adjustments to nuclear data, it is difficult to derive reliable uncertainties or compare the different methods with each other. Thanks to advances in {\it ab initio} nuclear theory, {\it ab initio} predictions for the nuclear matrix elements in the key isotopes have become available \cite{belley2021ab,belley2023ab,li2026ab} in recent years allowing robust uncertainty estimation \cite{belley2024ab}. Benchmarks in light nuclei show good agreement between quasi-exact and polynomially scaling {\it ab initio} methods \cite{yao2021ab}. The known $\beta\beta$-decay nuclei with mass numbers $A\geq 48$ are only accessible by the polynomially scaling methods such as VS-IMSRG, IM-GCM and CC.

Recently, an EFT framework for $0\nu\beta\beta$ decay has also been developed \cite{cirigliano2018neutrinoless,cirigliano2019renormalized}, which allows computing the nuclear matrix elements based on chiral EFT consistently with the nuclear wave functions in an {\it ab initio} framework. In particular, these studies have shown that a previously unrecognized contact term with an unknown coupling constant has to be introduced at the leading order to renormalize the theory \cite{cirigliano2018new}. By fitting this coupling to synthetic data \cite{cirigliano2021determining,cirigliano2021toward}, a recent {\it ab initio} study \cite{wirth2021ab} estimated that this new term enhances the value of NME of $^{48}$Ca by some 40\%, and VS-IMSRG studies in heavier nuclei suggest similar or even more significant effects \cite{belley2023ab,belley2024ab}. NSM and pnQRPA studies see similar effects by estimating the unknown coupling by the charge-independence-breaking (CIB) term of different nuclear Hamiltonians \cite{jokiniemi2021impact}. Lattice QCD is yet to confirm the size of the unknown coupling \cite{davoudi2021path}. The EFT analysis also introduces new next-to-next-to-leading-order (N$^2$LO) contributions to the operators, which have been shown to be consistent with the power counting \cite{pastore2018neutrinoless,castillo2025neutrinoless}, however these contributions may become dominant in more exotic decay mechanisms \cite{dekens2024neutrinoless}.

Given the difficulty of NME calculations, it is also important to probe the nuclear many-body calculations with experimental observables \cite{agostini2023toward}. For example, double-charge-exchange reactions \cite{cappuzzello2021numen,sakaue2024candidate,shimizu2018double,jokiniemi2023correlations,wang2024correlation}, double-$\gamma$ decays \cite{romeo2022gammagamma} and two-neutrino $\beta\beta$ ($2\nu\beta\beta$) decays \cite{vsimkovic20180,jokiniemi2023neutrinoless,horoi2022statistical,horoi2023predicting,el20252nubetabeta,lian2026ab,horoi2026uncertainty} and muon capture on nuclei \cite{araujo2024monument,jokiniemi2024muon,king2022partial} have been shown to be good probes for $0\nu\beta\beta$ decay. On the other hand, the potential to probe the structure of $\beta\beta$-decay nuclei by heavy-ion collisions has recently attracted interest \cite{li2025benchmarking}. FRIB is one of the key facilities, where charge-exchange reactions can be measured, and these measurements are important benchmarks for future {\it ab initio} calculations on $\beta\beta$ decays.

From the theory side, while the {\it ab initio} methods have become capable of computing predictions for $0\nu\beta\beta$ decays, future improvements are still called for. Current calculations in the literature either use the impulse approximation, neglecting that the decaying nucleons may simultaneously strongly interact with other nucleons in the decaying nucleus, or include partial contributions from higher-body currents by normal-ordered two-body currents \cite{wang2018quenching,menendez2011chiral,engel2014chiral,jokiniemi2023neutrinoless}. Very recently, a chiral EFT based study evaluated full higher-body currents up to N$^3$LO and found that the effect in light nuclei was non-negligible \cite{chambers-wall2025three-nucleon}. These currents should be included in {\it ab initio} calculations of the isotopes of experimental interest to evaluate their
impact on the experimental goals. On the other hand, many of the $\beta\beta$-decaying nuclei are deformed, but currently most of the {\it ab initio} calculations assume spherically symmetric nuclei. The effect of deformation should be studied more carefully in an {\it ab initio} framework such as the IM-GCM. Finally, the pioneering study on the uncertainty quantification of the $0\nu\beta\beta$-decay of $^{76}$Ge \cite{belley2024ab} should be extended to the other experimentally relevant candidate nuclei to allow for more reliable interpretation of experimental signals.

\section{Summary\label{sec:summary}}
In recent years, the nuclear science community went through the exercise of planning the long-range goals of the field for the future~\cite{dodge2024us}. Among the goals identified in this process were: {\it i)} understanding the rich patterns that emerge in the structure and reactions of nuclei starting from the underlying nuclear force; {\it ii)} better comprehending the nuclear processes driving the life-cycles of stars; and {\it iii)} using atomic nuclear to better understand physics beyond the Standard Model. Over the course of this workshop, and throughout this white paper, we identified where nuclear $\beta$ decay currently is impacting these goals and what it will take to address challenges in the future. FRIB will provide information about nuclei at the extremes of stability and for key astrophysical reactions with a wealth of discrete and continuous data coming from leveraging the unique capabilities of the FSDi. To take advantage of the experimental capability, nuclear theory will need to best leverage the strengths of {\it ab initio} and data-driven approaches to nuclear structure, assess the assumptions that enter into various formalisms, and ensure uncertainties can be robustly quantified on key observables. 

In Section~\ref{sec:exp_methods}, we summarized experimental approaches to measure quantities of interested to $\beta$ decay. Through the detectors in the FDSi, FRIB will provide both discrete and total absorption spectroscopy information. From these data, we can understand level schemes, $\beta$-feeding intensities for several systems, $\beta$-delayed particle emission probabilities, and $\gamma$ strength functions for several systems, as well as having the potential to isolate the $\beta $ decay spectra of key decay branches. Experiments like SALER at FRIB are demonstrating the cross-cutting capabilities of the field, leveraging quantum sensing to make precision determinations of $\beta$ decay spectra. While not at FRIB, other techniques like $\beta$ Paul Traps, CRES, and spectroscopy based on radioactively-doped, levitated nano-particles are providing precision spectra for fundamental symmetries, and could provide interesting routes in the future if coupled with the unique capabilities of FRIB. While it would be a non-trivial pursuit, implementing set-ups like the TAMU HPGe detector at a facility like FRIB could also provide information on branching ratios of key decays for fundamental symmetries research. Collectively, these experimental approaches are providing a rich array of data for efforts to better understand nuclear structure, astrophysics, and fundamental symmetries. 

The current landscape of nuclear theory was discussed in Section~\ref{sec:many_body}. We discussed first a set of data-driven approaches, which involve determining either energy-density functionals --- as in DFT --- or effective interactions for use in configuration-interaction approaches --- like the QRPA or shell-model --- from global fits to nuclear data and, in some cases, additional local fine-tuning. In addition to these approaches, we discussed what have come to be known as {\it ab initio} many-body methods, which come with varying levels of approximation. Quasi-exact approaches exist in both coordinate space (QMC) and configuration space (NCSM), in which one aims to solve the Sch\"{o}dinger Equation directly stating from the underlying nuclear interaction with minimal approximations. These methods, however, scale exponentially with the size of the system and are limited to the lightest nuclei. The symmetry adapted version of the latter approach, SA-NCSM, uses the symmetries of the nuclear system to perform truncations on the model space, enabling the computation of larger systems. The CC approach and IMSRG family of approaches use similarity and unitary transformations, respectively, to decouple many-body excitations from the Hamiltonian, enabling studies of medium mass nuclei due to polynomial scalings in system size. As facilities confront us with data on more exotic systems, or higher-precision tests of the Standard Model, new challenges will be posed to the current suite of many-body methods. In the following paragraphs, we will summarize these challenges in the context of the relevant scientific problems. 

Away from the valley of stability, nuclei begin to exhibit a rich array of structures and phenomena, as discussed in Sec.~\ref{sec:structure}. For instance, the emergence of halo structures characterized by weakly bound structures emerge near the neutron- and proton-driplines. At present, the effect of halo structures on half-lives is unclear, but comparing theory and experiment along isotopic chains could help to understand the evolution of shell structure as one moves away from stability. As this limit is approached, the role of coupling to the continuum will become more important, and will require many-body methods to properly handle these effects. For neutron rich species, the observables that can be accessed are limited to integral quantities like half-lives, $\beta$-delayed particle emission probabilities, and minimal daughter excited-state spectroscopy. In particular, developing frameworks to study $\beta$-delayed particle emission with state-of-the-art many-body frameworks poses an interesting challenge to theory, and developments in this area would help to maximize the discovery potential of experiments conducted near the limites of stability. 

Another phenomenon that appears away from stability is the breakdown of the spherical shell model of the nucleus. In particular, configuration mixing of nuclear states and deformation occurs. Isomersism arises when the lifetime of a state is significantly increased by mismatches in the initial and final state nuclear configurations. The $\beta$ decay mechanism can be used to populate isomeric states, and one can expect many new isomers to be found at FRIB. In order to connect with experimental capabilities in the future, it will be important for many-body methods to efficiently describe nuclear wave functions with complex intrinsic structures. Hybrid methods that combine {\it ab initio} approaches with explicitly deformed reference states have recently been developed. Combining, for instance, renormalization group techniques with other data-drive approaches that are well suited to study deformed systems could enhance discovery potential in the future. Enhancing the convergence of quasi-exact approaches could also allow for studies of heavier, neutron- and proton-rich nuclei with these approaches. As new data on exotic systems become available, having {\it ab initio} approaches combined with data-analysis techniques will allow us understanding the features of the nuclear interaction that are responsible for capturing rich shape and deformation phenomena, and could potential provide insight that will better constrain interaction models in the future. 

Along with learning about the structure of nuclei at the limits of stability, understanding the origin of elements and the life-cycles of stars is an important goal of low-energy nuclear physics. Elemental abundances and $r$-process nucleosynthesis depend on half-life information and the half lives of the most abundant isotopes in an isotopic chain. Presently, there are global theories used to describe half-lives across the nuclear chart. FRIB will be able to provide data on key isotopes that impact astrophysical processes, and allow us to confront global models with experiment. Identifying key nuclei where {\it ab initio} nuclear theory could help to constrain global models is also an opportunity for synergy between different theoretical approaches. $\beta$ decay and electron capture strength functions are also an important dynamical quantity, and either through direct spectroscopic measurements or charge exchange reactions, it is possible to gain information experimentally with FRIB. Methods like QRPA and DFT are well-suited to study these strength functions globally. There has also been recent progress in studying strength functions from an {\it ab initio} point of view using integral transform techniques or direct computation of the many-body Green's function, and turning the attention of these approaches to key $\beta$ decays for astrophysics could help to supplement global models in the future. Of course, confronting both data-driven and {\it ab initio} approaches with new experimental data will help to refine models and benefit astrophysical modeling. 

Searches for physics beyond the SM are another pillar of FRIB science, and $\beta$ decays can play a pivotal role in this area. For instance, the CKM matrix element $V_{ud}$ can be extracted from pure-Fermi superallowed and mixed mirror $\beta$ decays. With the modeling of nuclear structure radiative and isospin-breaking corrections relevant for the extracting of $V_{ud}$ recently being called into question, a tremendous effort has gone into developing formalisms and performing nuclear structure calculations. Additionally, searches for exotic tensor and scalar currents rely on precise nuclear theory in order to search for deviations between the Standard Model and experiment, and there has been a recent thrust to determine recoil corrections to spectra from many-body calculations in systemically improvable frameworks. These recent efforts in theory would be complemented by the measurement of branching ratios, $Q$-values, half-lives, and $\beta$-decay spectra. In the future, SALER at FRIB is an experimental effort poised to make contributions in this area with studies of mirror $\beta$ decay spectra. Studies of charge radii at FRIB also have the potential to help constrain isospin breaking correction in a data-driven manner, and would provide stringent tests for many-body methods being used to compute these quantities from a theoretical point of view. FRIB could also improve data entering into determination of $ft$ values of beyond $sd$-shell superallowed $\beta$ decaying nuclei. 

Nuclear $\beta$ decays are also used in studies of neutrino properties. In particular, $\beta$ decay spectra are crucial for studying the neutrino mass. Low $Q$-value nuclei are particularly useful for these studies, and isotope harvesting of key nuclei for these studies could be useful for future studies in this area. Techniques developed for these studies could also beneficial for other physics searches. For instance, the CRES technique developed to study $^3$H $\beta$ decay for neutrino mass determinations is now also being used to study the decays of $^6$He and $^{19}$Ne in the $^6$He-CRES experiment. Further, the success of quantum sensing techniques developed to study the $^7$Be electron capture spectrum led to the development of techniques to perform the study of $\beta$ decay spectra for radioisotopes created on-line with SALER. Other quantum sensing techniques based on detecting recoils optomechinically using levitated nano-particles are being developed to search for massive neutrinos. Quantum sensing for $\beta$ decay spectra of on-line produced radioisotopes represents a possibility for cross cutting research at FRIB. 

Neutrino properties are also being probed through rare process searches; namely, supporting experimental searches for $0\nu\beta\beta$ were a recommendation of the nuclear science community in the long-range planning process. The measurement of an event is not sufficient to determine new information about neutrino physics, as the half-life of the decay depends both on a quantity related to new physics and a nuclear matrix element. The computation of nuclear matrix elements with robust uncertainties has been a major challenge for nuclear theory, and efforts to provide results with quantified errors has required leveraging high-performance computing and data-science techniques. Studying processes that are correlated with the $0\nu\beta\beta$ matrix element would provide constraints on nuclear models. For instance, charge-exchange reactions that can be probed at FRIB would benchmark many-body approaches being used to study $0\nu\beta\beta$. 

The $\beta$ decay program possible at FRIB is rich in scientific opportunities. Decay spectroscopy of nuclei from near stability to the dripline will contribute to refining models of nuclear structure, informing astrophysical models, and testing our current understanding of the Standard Model. Further, emerging quantum sensing technologies used to study $\beta$ decay represent an opportunity for cross-cutting research at FRIB. The data being obtained will provide a number of unique challenges to theory. Near stability, precision measurements probing the Standard Model will require precise calculations of corrections to decay rates and spectra. To look for precise deviations from the SM, these efforts will also require more stringent uncertainty quantification in the future so that reliable error bars can be provided. We can look to the recent concerted efforts in $0\nu\beta\beta$ as an example of how to perform such an analysis. Far from stability, our models will be tested by exotic systems like halos and deviations from the spherical shell model of the nucleus. Properly handling the continuum in many-body methods and having approaches that can handle complex intrinsic structures will be important for maximizing the discovery potential at FRIB. Finally, synergy between different many-body approaches will also help to maximize the efficiency of our theoretical efforts. Benchmarking global models with {\it ab initio} approaches for key decays, for instance, could help to better inform nuclear structure and astrophysics research. Coordination between experimentalists and theorists in different areas to identify key nuclei, and will help to maximize the era of discovery in $\beta$ decay that FRIB will enable.

\section*{Acknowledgments}
\noindent 
(J.~B.) Support from NSERC and from NRC through TRIUMF.\\
(F.~B., G. H. S.) This work was supported by the U.S. Department of Energy, Office of Science, Office of Nuclear Physics, under the FRIB Theory Alliance award DE-SC0013617, (F.~B.) and by the U.S. Department of Energy, Office of Science, Office of Advanced Scientific Computing Research and Office of Nuclear Physics, Scientific Discovery through Advanced Computing (SciDAC) program (SciDAC-5 NUCLEI). 
This manuscript has been authored in part by UT-Battelle, LLC, under contract DE-AC05-00OR22725 with the US Department of Energy (DOE). The US government retains and the publisher, by accepting the article for publication, acknowledges that the US government retains a nonexclusive, paid-up, irrevocable, worldwide license to publish or reproduce the published form of this manuscript, or allow others to do so, for US government purposes. DOE will provide public access to these results of federally sponsored research in accordance with the DOE Public Access Plan (http://energy.gov/downloads/doe-public-access-plan).\\
(M.~B.) Is supported by the National Science Foundation under Grant PHY-2310059.\\
(M.~A.~C.)
This material is based upon work supported by the
U.S.~Department of Energy, Office of Science, under Award Nos.~DE-FG02-95ER40934 and DE-SC0023633 (through subaward agreement No.~RC113931-ND between Michigan
  State University and the University of Notre Dame).\\
(H.~L.~C) This work was supported by the U.S. Department of Energy, Office of Science, Office of Nuclear Physics under Contract No. DE-AC02-05CH11231 (LBNL). 
(M.~D.) Acknowledges the direct support of the Nuclear Theory for New Physics Topical collaboration. U.S.~Department of Energy under contract DE-SC0023663 and US Department of Energy under Contracts No. DE-SC0021027. \\
(W.D.) acknowledges support from the U.S.\ DOE under Grant No.\ DE-FG02-00ER41132
(M.~G.) This work was supported in part by the Deutsche Forschungsgemeinschaft (DFG) through the Cluster of Excellence ``Precision Physics, Fundamental Interactions, and Structure of Matter'' (Project ID 390831469).
(P.~G.) This work was supported by the U.S. Department of Energy, Office of Science, Office of Nuclear Physics under Award No. DE-SC0023663 (NTNP Topical Collaboration).\\
(H.~S.~H.) This work was supported by the Center for Experimental Nuclear Physics and Astrophysics at the University of Washington's U.S. Department of Energy grant No. DE-FG02-97ER41020.\\
(H.~H.) This work was supported by the U.S. Department of Energy, Office of Science, Office of Nuclear Physics under Awards No. DE-SC0023516, DE-SC0023175 (SciDAC-5 NUCLEI Collaboration) and DE-SC0023663 (NTNP Topical Collaboration). H.~H. also acknowledges the ``@NDB: Advancing Theory for Nuclear Double-Beta Decay'' Focused Research Hub supported by the National Science Foundation (NSF) FRHTP program under award No.  PHY-2402275.
\\
(L.~J.) Acknowledges support of the LOEWE Top Professorship LOEWE/4a/519/05.00.002(0014)98 by the State of Hesse.\\
(G.~B.~K.) Financial support was provided by the Laboratory Directed Research and Development program of Los Alamos National Laboratory under project number 20240742PRD1 \\
(B.~L., T.~H.~O.) This work was carried out under the auspices of the U.S. Department of Energy by Lawrence Livermore National Laboratory under Contract No.~DE-AC52-07NA27344 and (B.~L.) Argonne National Laboratory under Contract No.~DE-AC02-06CH11357.\\
(R.~S.~L) This material is based upon work supported by the U.S. Department of Energy, Office of Science, Office of Nuclear Physics, and used resources of the Facility for Rare Isotope Beams (FRIB) Operations, which is a DOE Office of Science User Facility under
Award No. DE-SC0023633.
(K.~A.~L.) This work was supported by the Network for Neutrinos, Nuclear Astrophysics and Symmetries (N3AS), through the National Science Foundation Physics Frontier Center award No. PHY-2020275. K.A.L acknowledges the Institute for Nuclear Theory at the University of Washington for its kind hospitality and stimulating research environment. This research was supported in part by the INT's U.S. Department of Energy grant No. DE-FG02- 00ER41132.\\
(D.~C.~M.) This work was supported through the
DOE Office of Science under Grant DE-SC0026367.\\
(A.~M.) This work was supported by the U.S. National Science Foundation under Award No. OIA-2327385.\\
(D.~M.) Financial support was provided by the U.S. Department of Energy under Award No. DEFG02-93ER40773.\\
(A.~M.) This work was supported by the U.S. Department of Energy, Office of Science, Office of Nuclear Physics, under Award Number DE-AC02-06CH11357.\\
(O.~N-C.) This work was supported by the U.S. National Science Foundation under Award No.
PHY-2111185.\\
(W.~C.~P.) Financial support was provided by the National Science Foundation under Grant PHY-2209530. \\
(B.~C.~R.) This material is based upon work supported in part by the U.S. Department of Energy, Office of Science, Office of Nuclear Physics under Contracts No. DE-AC05-00OR22725 and by the Nuclear Data Inter-Agency Working Group (NDIAWG) Funding Opportunity-2440.\\
(A.~R.) This work was supported  by the U.S. Department of Energy under Award No. DOE-DE-NA0004245 (NNSA, the Stewardship Science Academic Alliances program)\\
(Y.~S.) This work was supported by the U.S. National Science Foundation under grant number 21-16686 (NP3M).\\
(C.-Y.~S.) Financial support was provided by the U.S. Department of Energy (DOE) Topical Collaboration ``Nuclear Theory for New Physics'', award No. DE-SC0023663, and by University of Tennessee, Knoxville.\\
(N.~S.) Acknowledges the financial support of CNRS/IN2P3, France, via ENFIA, ABI-CONFI and ABI-CONFI-II Master projects. (G.~C.~W) Acknowledges support by the US Department of Energy under Contracts No. DE-SC0021027.\\
(A.~S.) Acknowledges support from the National Science Foundation under grand PHY-2514797.
(V.~T.) Acknowledges the support of the National Science Foundation under reward No.~2412808.\\
(X.~W.)
This material is based upon work supported by the U.S.~Department of Energy, Office of Science, Office of Nuclear Physics, under Award Number DE-SC0023633, and by Michigan State University.\\


\bibliography{biblio.bib}

\end{document}